\documentclass[oneside,11pt]{article}

\usepackage[utf8]{inputenc}
\usepackage[T1]{fontenc}

\usepackage{times}
\usepackage{bm}
\usepackage{stmaryrd}
\usepackage{changes}

\usepackage{amsthm}
\usepackage{amsmath}
\usepackage{amsfonts}
\usepackage{amssymb}

\usepackage{graphicx} 
\usepackage{float}
\usepackage{booktabs}
\usepackage{multirow}
\usepackage{rotating}
\usepackage{array}
\usepackage{xcolor}
\usepackage{color}
\usepackage{breakcites}

\usepackage{enumitem}

\usepackage[top=1.5cm,bottom=2cm,right=2.5cm,left=2.5cm]{geometry}
\usepackage{titling}

\usepackage{natbib}
\setcitestyle{square}
\setcitestyle{numbers}

\usepackage{ifplatform}

\usepackage{textcomp}

\ifwindows
\fi

\ifwindows
	\usepackage{bbm}
\else
	\iflinux
		\usepackage{bbm}
	\else
		\usepackage{bbm}
	\fi
\fi

\newcommand{\R}{\mathbb{R}}

\newcommand{\bx}{\mathbf{x}}

\newcommand{\bmm}{\mathbf{m}}

\newcommand{\by}{\mathbf{y}}
\newcommand{\bX}{\mathbf{X}}
\newcommand{\bY}{\mathbf{Y}}
\newcommand{\bZ}{\mathbf{Z}}

\newcommand{\bU}{\mathbf{U}}

\newcommand{\bmu}{\boldsymbol\mu}

\newcommand{\bv}{\mathbf{v}}
\newcommand{\bw}{\mathbf{w}}

\newcommand{\bz}{\mathbf{z}}

\newcommand{\mSigma}{\mathrm\Sigma}

\newcommand{\bV}{\mathbf{V}}

\newcommand{\lrp}[1]{\left(#1\right)}

\newcommand{\lrc}[1]{\left[#1\right]}

\newcommand{\hatmuv}{\hat{\bmu}_{\bV}}
\newcommand{\hatsigmav}{\hat{\sigma}_{\bV}}

\DeclareFontFamily{OT1}{pzc}{}
\DeclareFontShape{OT1}{pzc}{m}{it}{<-> s * [1.10] pzcmi7t}{}
\DeclareMathAlphabet{\mathpzc}{OT1}{pzc}{m}{it}

\newcommand{\E}{\mbox{E}}
\newcommand{\Var}{\mathrm{Var}}

\newcommand{\sign}{\mathrm{sign}}

\newcommand{\MAD}{\mathrm{MAD}}
\newcommand{\MADN}{\mathrm{MADN}}

\newcommand{\erf}{\mathrm{erf}}

\newcommand{\Prob}{\mbox{$\mathbf{P}$}}
\newcommand{\Q}{\mbox{$\mathbf{Q}$}}
\newcommand{\convp}{\stackrel{c.p.}{\rightarrow}}
\newcommand{\convs}{\stackrel{a.s.}{\rightarrow}}

\usepackage[pdftex,final,bookmarksnumbered,bookmarksopen=false,breaklinks,colorlinks]{hyperref}
\hypersetup{
	final,
	bookmarksnumbered=true,
	bookmarksopen=true,
	bookmarksopenlevel=0,
	unicode=false,          % non-Latin characters in Acrobat?s bookmarks
	pdftoolbar=true,        % show Acrobat?s toolbar?
	pdfmenubar=true,        % show Acrobat?s menu?
	pdffitwindow=false,     % window fit to page when opened
	pdftitle={High-dimensional outlier detection using random projections},    % title
	pdfdisplaydoctitle=true, %display document title instead of file name in title bar
	pdftoolbar=true,
	pdfmenubar=true,
	pdflang={English},
	pdfauthor={Paula Navarro-Esteban},     % author
	pdfsubject={arXiv paper},   % subject of the document
	pdfcreator={Paula Navarro-Esteban},   % creator of the document
	pdfproducer={Paula Navarro-Esteban}, % producer of the document
	pdfkeywords={Outlier detection}{Multivariate data}{High-dimensional data}{Random projections}{Sequential analysis}, % list of keywords
	pdfnewwindow=true,      % links in new window
	breaklinks=true,
	hidelinks,       
	linkcolor=black,          % color of internal links (change box color with linkbordercolor)
	citecolor=black,        % color of links to bibliography
	filecolor=magenta,      % color of file links
	urlcolor=cyan           % color of external links
}

\newtheorem{defin}{Definition}
\newtheorem{theorem}{Theorem}
\newtheorem{corollary}{Corollary}
\newtheorem{remark}{Remark}
\newtheorem{proposition}{Proposition}
\newtheorem{lemma}{Lemma}

\allowdisplaybreaks

\makeatletter
\def\munderbar#1{\underline{\sbox\tw@{$#1$}\dp\tw@\z@\box\tw@}}
\makeatother

\begin{document}

%---------- TITLE ---------------%
\title{High-dimensional outlier detection using random projections}
\setlength{\droptitle}{-1cm}
\predate{}%
\postdate{}%

%---------- AUTHORS ---------------%
\date{}

\author{P. Navarro-Esteban$^{1,2}$ \and
	J.A. Cuesta-Albertos$^{1}$}
\footnotetext[1]{
Departmento de Matem\'aticas, Estad\'\i stica y Computaci\' on, Universidad de Cantabria, Santander (Spain).}
\footnotetext[2]{Corresponding author, e-mail: \href{mailto:paula.navarro@unican.es}{paula.navarro@unican.es}.}

\maketitle

%-------------------------------------%

\begin{abstract}
There exist multiple methods to detect outliers in multivariate data in the literature, but most of them require to estimate the covariance matrix. The higher the dimension, the more complex the estimation of the matrix becoming impossible in high dimensions. In order to avoid estimating this matrix, we propose a novel random projections-based procedure to detect outliers in Gaussian multivariate data. It consists in projecting the data in several one-dimensional subspaces where an appropriate univariate outlier detection method, similar to Tukey's method but with a threshold depending on the initial dimension and the sample size, is applied. The required number of projections is determined using sequential analysis. Simulated and real datasets illustrate the performance of the proposed method.
\end{abstract}
\begin{flushleft}
\end{flushleft}

%-----------------------------------------------%
\section{Introduction}
\label{intro}
%-----------------------------------------------%
Outliers are often identified as observations obtained from a distribution different from that one producing the bulk of the data set. Notwithstanding the fact that this definition has sometimes been handled (see, for instance, \cite{Febrero2007}
), it is  more convenient in practice  to consider as outliers those points lying at a distance greater than a given threshold from the centre of the sample, independently of the distribution which produced them. Detecting outliers in a sample is one of the first steps when handling data, since they may lead to model misspecification, biased parameter estimation and incorrect results in general, see Aggarwal \cite{Aggarwal2015} for instance. 

Despite the fact that other settings can also be handled with our method, in this paper we focus on testing outlyingness of some vectors $\bx$ in $\R^{d}$ with respect to a sample of iid (independent and identically distributed) rv's (random vectors) $\bX_1,\ldots,\bX_n$ in $\R^{d}$ with normal distribution $N_d(\bmu,\mSigma)$, where $\bmu$ and $\mSigma$ are respectively the mean vector and the covariance matrix. Thus, the hypotheses to be tested are
\begin{align}\label{eq:hypothe_1}
\mathbf{H}_0: {\bx}\text{ is not an outlier } \quad\text{vs.}\quad\mathbf{H}_1:{\bx}\text{ is an outlier.}
\end{align}

Our method can be applied to any combination of sample sizes and dimensions, but our principal interest is in the cases in which $d>n$. This method uses projections as a dimensionality-reduction technique to avoid the estimation of $\bmu$ and mostly $\mSigma$. Our procedure declares $\bx$ to be an outlier if the distance of a one-dimensional projection of $\bx$ to the centre of the projected sample exceeds a data driven threshold. It is noteworthy that we avoid imposing any structure to the covariance matrix thus including high dimensional and/or very correlated data. 

There exists an abundant literature on multidimensional outliers for $d$ low or moderate in comparison with $n$ (see, for instance Barnett and Lewis \cite{Barnett1994} and Aggarwal \cite{Aggarwal2015} and references therein). 
When the dimension is higher than the sample size the literature is not so abundant but we can mention Filzmoser et al. \cite{Filzmoser2008} and Ro et al. \cite{Ro2015}. The first paper is based on the properties of principal components analysis (PCA). However, the principal componentes are difficult to be estimated in very high-dimensional settings, see Johnstone et al. \cite{Johnstone2009}. For instance, it occurs that it is only possible to obtain the asymptotic distribution for $O(n^{1/5})$ coefficients in the linear functional regression model when a PCA-based estimator is used, Cardot et al. \cite{Cardot2007}. On the other hand, the method introduced in Ro et al. \cite{Ro2015} is based on a modification of the Mahalanobis distance which involves only the diagonal elements of $\mSigma$. Thus, it is equivalent to consider uncorrelated marginals and this does not usually occur in practice. 

Our proposal is based on that an outlier is a point lying far away from the centre of a given data set. Then, according to the  Stahel-Donoho estimators, Stahel \cite{Stahel} and Donoho \cite{Donoho}, we   look for a univariate projection that makes an observation outlier, because ``\ldots if a point is a multivariate outlier, then there must be some one-dimensional projection of the data for which the point is a (univariate) outlier'', see Maronna and Yohai \cite{Maronna1995}. Hence we only handle one-dimensional projections and thus we avoid the estimation of $\mSigma$.

The idea that an outlier is a point too separated from the centre of a data set can dated back to 1968 in Healy \cite{Healy1968}. It was made more precise in Davies and Gather \cite{Davies} for dimension $d=1$ and in Becker and Gather \cite{Becker1999} for multidimensional data. Those papers propose computing (robust) estimators of the centre and of the covariance matrix of the data set at hand, and, then declaring outliers those points whose Mahalanobis distances to the estimated centre are greater than a previously fixed threshold. An important characteristic is that the threshold depends on both $d$ and $n$ (see Theorem \ref{theo:conv_Cnd}  below). Some computational problems were reported, for instance,  in Cerioli et al. \cite{Cerioli2009} and Cerioli \cite{Cerioli}, albeit they have not appeared in our implementation here.

A possibility of implementing the idea is using projection pursuit. However this technique, in principle, requires to examine all the possible directions, what is impossible in practice. To overcome this problem, there exist procedures which only involve many finite deterministic data-dependent projections such as Pe\~na and Prieto \cite{Pena} and Serfling and Mazumder \cite{Serfling}, but they require to estimate the covariance matrix, and Pan et al. \cite{Pan} who do not provide the exact number of the required directions.

As an alternative, we propose to use a number of random directions independently chosen from the sample at hand. Johnson and Lindenstrauss' Lemma \cite{Johnson} is the basis of the feasibility of random projections. Their most useful property for us is a result stated in Cuesta-Albertos et al. \cite{Cuesta}. From there, it is known that a.s. just a one-dimensional random projection is enough to distinguish between two distributions defined on a separable Hilbert space if one of them satisfies a certain condition on their moments: if two distributions are given, and a one-dimensional marginal of them is randomly chosen, we have that almost surely, the two distributions are different/equal if and only if the two marginals are different/equal. Thus, this procedure projects the original high-dimensional data into a one-dimensional randomly chosen subspace. Since handling only one random direction gives a low power under the alternative hypothesis, we handle several random directions.

Random projections have been applied to solve other problems such as in goodness of fit (Cuesta-Albertos et al. \cite{Cuesta2006}, \cite{Cuesta2007} and \cite{Cuesta2014}), analysis of variance (Cuesta-Albertos and Febrero-Bande \cite{Cuesta2010}), testing linearity in functional regresion (Cuesta-Albertos et al. \cite{Cuesta2018}), constructing depths (Cuesta-Albertos and Nieto-Reyes \cite{Cuesta2008}), etc. A common problem in those results is that no clear guidance on the number of the required projections was given. We propose the use of the sequential analysis to solve this, the same idea could be used in the above referred papers.

A sequential method is characterized by a stopping rule that decides whether to stop the observation process with $\bX_1,\ldots,\bX_n$ or to get an additional observation $\bX_{n+1}$ for each $n\ge 1$. Therefore, the number of observations needed by the procedure is random. Those methods are a powerful technique because they need on average smaller sample sizes than fixed sample size procedures to achieve the same power, Tartakovski et al. \cite{Tartakovsky2014}. In our case, this criteria leads to select a low number of random directions, $K_n$, what makes the method run quite fast, in $O(K_n n)$ time. For instance, the computations usually require a little less than 2 seconds for $d = 500$ and $n = 100$, albeit some particularly difficult cases could require at most 12 seconds. 

The sketch of the procedure to test \eqref{eq:hypothe_1} is the following:

\begin{itemize}
	\item[1.] Select $a,b \in \mathbb{R}^+$, $a\leq b$. 
	
	\item[2.] Take a rv $\bV$ with $N_d(\mathbf{0},I_d)$ distribution and make $\bV=\bV/\Vert\bV\Vert$.
	
	\item[3.] Project $\bx$ and the sample on the subspace generated by $\bV$, i.e. compute $\bx'\bV$ and $\bX_1'\bV, \ldots, \bX_n'\bV$, and calculate $\hat{\nu}_{\bV}$ and $\hat{\lambda}_{\bV}$ estimators of the centre and of the dispersion of the projections $\bX_1'\bV, \ldots, \bX_n'\bV$.
	
	\item[4.] Compute $y^{\bV}:=(\bx'\bV - \hat{\nu}_{\bV})/\hat{\lambda}_{\bV}$. 
	
	\item[5.] If $|y^{\bV}| \in [a,b]$ go back to Step 2., else:
	\begin{itemize}
		\item[-] The point $\bx$ is declared as an outlier if $|y^{\bV}| > b$.
		
		\item[-] The point $\bx$ is declared as non-outlier if $|y^{\bV}| < a$.
	\end{itemize}
\end{itemize}

The choice of parameters $a$ and $b$ is discussed in Section \ref{sec:method}. It turns out that they depend on the sample size, on the dimension of the space and on  $\mSigma$, see \eqref{eq:metodo3} and Proposition \ref{prop:probabilidad2}. This dependency will be analysed in Section \ref{sec:method} through the expected number of required projections to reach the decision about the point we are classifying. Concerning the estimation of $\hat{\nu}_{\bV}$ and $\hat{\lambda}_{\bV}$ we begin using the sample mean and the sample standard deviation. Next we will replace them by the sample median and the sample median absolute deviation, MAD, respectively.

Despite the fact that we propose some expressions determining $a$, $b$ and $\E(K_n)$, the specific computation of their values has happened to be impossible for us. This has led us to consider numerical approximation, including the asymptotic values as $n\to\infty$ with $d$ fixed and $\mSigma=I_d$. Some theoretical work, now in progress, suggests that this solution could work for many covariance matrices, for large values of $d$ (see Subsection \ref{Sec.a_bDeLaIdentidad} in the Appendix). Moreover, extensive simulations reinforce this feeling because they have provided  empirical evidence that the proposed procedure is rather stable with respect to variations on $d$, $n$ and even on $\mSigma$ and, consequently,  those asymptotic values can be applied in practice for all combinations of $d$ and $n$ (including those with $d>n$) as well as with many different $\mSigma$'s (see Subsection \ref{subsec:comp_a_b}).

The paper is organized as follows. In Section \ref{sec:def} we make the definition of outlier precise and include a result on the asymptotic behaviour (on $d$ and on $n$) of the threshold (Theorem \ref{theo:conv_Cnd}). Section \ref{sec:method} gives the main theoretical results on which our method is based. Guidelines for its practical implementation are given in Section \ref{sec:implementation}. A comprehensive simulation study and two real data applications are presented in Section \ref{sec:num_studies}. An independent technical Appendix contains the proofs of the results obtained in the paper and several tables showing computational results not included in the main text.

All along the paper, we assume that all the rv's are defined on the same, rich enough, probability space $(\Upsilon,\mathcal{A},\Prob)$.

%-----------------------------------------------%
\section{Definition of an outlier}\label{sec:def}
%-----------------------------------------------%

In this section we make the definition of outlier precise and analyse some properties of the threshold involved in such definition. Essentially, the idea is that if a point is outside a certain ball centred at the centre of the sample, then it is an outlier. The shape of the ball should be determined by $\mSigma$. Those ideas lead to Definition \ref{def: outlier}, which is based on the well known fact that if $\bX$ is $N_d(\bmu,\mSigma)$, then the square of its $\mSigma$-based Mahalanobis distance to $\bmu$ follows a chi-squared distribution with $d$ degrees of freedom, $\chi_{d}^2$. Given $0<\delta<1$, denote by $C_n^d(\delta)$ the square root of the $\delta$-quantile of the maximum of a random sample with size $n$ and distribution $\chi_{d}^2,$ i.e. $C_n^d(\delta)$ is the solution of the equation:
\begin{equation}\label{eq:def_C_nd}
\Prob\left( \max \left\{ \left\Vert\bX_1-\bmu\right\Vert_{\mSigma},\ldots, \left\Vert\bX_{n}-\bmu\right\Vert_{\mSigma} \right\}\geq C_n^d(\delta) \right) = \delta,
\end{equation}
where $\Vert\bX-\bmu\Vert_{\mSigma}=\Vert\mSigma^{-1/2}\lrp{\bX-\bmu}\Vert$, with $\Vert\cdot\Vert$ being the Euclidean norm and $\bX_1,\ldots,\bX_n$ iid rv's with distribution $N_d(\bmu,\mSigma)$. Thus, $C_n^d(\delta) $ is the square root of the $(1-\delta)^{1/n}$-quantile of the distribution $\chi^2_d$.
To ease the notation we omit $\delta$ in $C_n^d(\delta)$ when its value is clear from the context or its exact value is irrelevant.

\begin{defin}\label{def: outlier}
	\textit{Let $\bx \in \mathbb{R}^d$ and $\delta \in (0,1)$. We say that $\bx$ is an outlier at the level $\delta$ with respect to a simple random sample with size $n$ and a distribution $N_{d}(\bmu,\mSigma)$, if    
		$\left\Vert\bx-\bmu\right\Vert_{\mSigma}\geq C_n^d(\delta).$
	}
\end{defin}

According to this definition, \eqref{eq:hypothe_1} becomes
\begin{align}\label{eq:hypothe_2}
\mathbf{H}_0: \left\Vert\bx-\bmu\right\Vert_{\mSigma} \leq C_n^d(\delta) \quad\text{vs.}\quad\mathbf{H}_1:\left\Vert\bx-\bmu\right\Vert_{\mSigma} > C_n^d(\delta).
\end{align}

Note that Definition \ref{def: outlier} is easily modified to cover dependent data. The only difference in the dependent case will be the expression for $C_n^d(\delta)$ which will be more complex. Extensions to elliptical non-normal distributions are straightforward.

Theorem \ref{theo:conv_Cnd} gives the asymptotic behaviour of $C_n^d$.
\begin{theorem}\label{theo:conv_Cnd}
	Let $C_n^d$ be as defined in \eqref{eq:def_C_nd}. Then $C_n^d \rightarrow \infty$ as $n \rightarrow \infty$ or $d \rightarrow \infty$ while the other parameter remains fixed with rates $\log(n)$ and $d^{1/2}$ respectively.
\end{theorem}

An illustration of Theorem \ref{theo:conv_Cnd} appears in Table \ref{tab:valores_C_nd} in Subsection \ref{subsubsec:C_n} in the Appendix, which shows the values of $C_n^d(\delta)$ for some values of $d$ and $n$, and $\delta=0.05.$ 

%-----------------------------------------------%
\section{The proposed outlier-detection method}
\label{sec:method}
%-----------------------------------------------%
If $\mSigma$ and $\bmu$ are known it is simple to check if a given point satisfies Definition \ref{def: outlier} or not. However, in practice, $\bmu$ and $\mSigma$ must be estimated and, consequently, it is not possible to be completely sure if the definition holds. 
To test \eqref{eq:hypothe_2}, we propose the procedure sketched in the Introduction, paying attention to the determination of $a$ and $b$. We also provide the expected number of projections required to declare a point as an outlier or as regular. We begin with some results related to statistics based on the sample mean and variance; later, in Subsection \ref{subsec:robust}, we will introduce their robust versions. 

Under $\mathbf{H}_0$, the only relevant quantity is the value of $t=\left\Vert\bx-\bmu\right\Vert_{\mSigma}$, so instead of assuming that we have a fixed point, we will replace the point $\bx$ by a random point in the Mahalanobis sphere associated to $\mSigma$ with centre at $\bmu$ and radius $t$. Being more precise, we will replace the point $\bx \in \mathbb{R}^d$ by a rv $\bX$ whose distribution is $N_d(\bmu,\mSigma)$ given that $\left\Vert\bX-\bmu\right\Vert_{\mSigma}=t$. We begin with two assumptions and some notation:
\begin{align*}
\mathrm{(A1)}&\; \bX \mbox{ and } \bX_1,\ldots,\bX_n \mbox{ are iid rv's with distribution } N_d(\bmu,\mSigma).\\
\mathrm{(A2)}&\; \bV \mbox{ and } \bV_1,\ldots,\bV_n \mbox{ are iid rv's with distribution } N_d(\mathbf{0},I_d) \mbox{ which also are}\\
&\; \mbox{ independent form the rv's in (A1)}.
\end{align*}

\paragraph{Notation} Denote the beta and error function as $\mathrm{B}(a,b)$ and $\erf({\cdot})$ respectively. We define $\Omega^{d-1}_{\mSigma}(t):=\{\bx \in \mathbb{R}^d : \Vert\bx\Vert_{\mSigma}=t\}$, the Mahalanobis hypersphere of radius $t$. With an abuse of notation, $\Omega^{d-1}_t$ denotes such a sphere when $\mSigma=I_d$, $\omega^d_t$ is its surface area; thus, $\omega^{d}_t=2\pi^{d/2}t^d/\Gamma(d/2)$. The sample mean and the covariance matrix are denoted by $\hat{\bmu}$ and $\hat{\mSigma}$. Initially the centre and the dispersion of the projected sample $\bX_1'\bV,\ldots,\bX_n'\bV$ are estimated by its sample mean, $\hat{\mu}_\bV$, and standard deviation $\hat{\sigma}_\bV$. 

Under assumptions (A1) and (A2), denote 
\begin{equation}\label{eq:T_Kestimate}
Y^{\bV}:=\frac{ \bX'\bV-\hat{\bmu}_{\bV}}{\hat{\sigma}_{\bV}}.
\end{equation} 
The denominator in \eqref{eq:T_Kestimate} can be zero for some $\bV$'s, however the subset of $\bV$'s satisfying this has null Lebesgue measure even in the case $d>n$. Although $Y^{\bV}$ depends on the sample, we omit this dependency to ease the notation. On the other hand, the distribution of $Y^{\bV}$ does not depend on $\bmu$ nor on an scale. Since our method relies on $Y^\bV$, we can assume w.l.o.g. that $\bmu=\mathbf{0}$ and that the smaller eigenvalue of $\mSigma$ is one. 

When $\bV=\bV_k$, we simplify the notation writing $Y^k$ instead of $Y^{\bV_k}$. The rv number of random projections which we need to decide if $\bX$ is an outlier or not with respect to a sample of size $n$ is denoted by $K_n^{a,b}(\mSigma)$. Thus, given $0<a\leq b$,
\begin{equation}\label{eq:def_Khat}
K_n^{a,b}(\mSigma)=\inf \left\{k : |Y^{k}| < a \mbox{ or } |Y^{k}|> b \right\}.
\end{equation}
If there is no possibility of confusion, or the  values of $a,b$ or $\mSigma$ are not important, we omit them and simplify to $K_n$. 
Note that if $K_n$ is finite, then $\vert Y^{K_n}\vert$ is well defined. 

For $y,t > 0$, $\bmm$, $\bx \in \mathbb{R}^d$, $S$ and $\mSigma$ two $d\times d$ semi-positive and positive definite matrices, $\bX$ with distribution $N_d(\mathbf{0},\mSigma)$ and $\bZ$ with distribution $N_d(\mathbf{0},I_d)$, denote 
\begin{equation}\label{eq:prob_unknown_sigma}
\begin{array}{rl}
%\vspace{.1cm}
y_{\bmm,S}^{\bV}:=&\lrp{\bx-\bmm}'\bV/\lrp{\bV'S\bV}^{1/2},\\
%
%\vspace{.1cm}
F(y,t):=&\Prob\left( \left.|\bZ'\bV| < y \  \right\vert  \left\Vert\bZ\right\Vert=t \right),\\
F_{\mSigma}(a,b,t):=&\Prob (|Y^{K_n}| > b \ | \ \left\Vert\bX\right\Vert_{\mSigma}=t),\\
\Vert\bx\Vert_S:=&\Vert\lrp{S^+}^{1/2}\bx\Vert
\end{array}
\end{equation}
where $S^+$ is the Moore-Penrose inverse of $S$.  The pdf of $\bX$ given that $\Vert\bX\Vert_{\mSigma}=t$ will be denoted by $f_t$.

%-----------------------------------------------%
\subsection{Some properties of the distribution of $Y^\bV$}
\label{subsec:marginal}
%-----------------------------------------------%
We begin obtaining an explicit expression for the conditional cdf of $Y^\bV$ given $\left\Vert\bX\right\Vert_{\mSigma}$. Then, Proposition \ref{prop: pdf_givenXd2} gives an expression of the cdf of the standardized random projection of a given $d$-dimensional vector. In this proposition we suppose that $S$ is diagonal, which entails no loss of generality, since a  rotation of the coordinates axes allows us to obtain this kind of matrix. Notice that we make no assumption on the number of non-null eigenvalues of $S$ as long as there exist two positive ones at least.

\begin{proposition}\label{prop: prob_dist} 
	Under assumptions (A1) and (A2), the cdf of $Y^{\bV}$ given that $\left\Vert\bX\right\Vert_{\mSigma}=t$, with $t>0$, does not depend on $\mSigma$ and its value is 	
	\begin{align*}
	\Prob\lrp{Y^\bV<y\big|\left\Vert\bX\right\Vert_{\mSigma}=t}=&
	\begin{cases}
	-\tau\displaystyle\int_{-\infty}^y\displaystyle\int_{-\infty}^0 \displaystyle\int_{-t}^t g_t(s,x,z) \,ds \,dx \,dz, \quad  y<0,\\
	\frac{1}{2}+\tau \displaystyle\int_{0}^y\displaystyle\int_0^{\infty} \displaystyle\int_{-t}^t g_t(s,x,z)\,ds \,dx \,dz, \quad y>0,
	\end{cases}
	\end{align*}
	
	\noindent where $g_t(s,x,z)=\frac{x^{n-1}}{z^n}\exp \left\{ \frac{-(n-1)x^2}{2z^2} \right\} (t^2-s^2)^{(d-3)/2}\exp\left\{ \frac{-n}{2}(s-x)^2\right\},$ $\tau:=2^{\frac{3-n}{2}}t^{2-d}\sqrt{\frac{n}{2\pi}}(n-1)^{\frac{n-1}{2}}	/\left(\Gamma\left(\tfrac{n-1}{2}\right)B\left(\tfrac{d-1}{2},\frac{1}{2}\right)\right)$.
\end{proposition}

\begin{proposition}\label{prop: pdf_givenXd2}
	Let $\bx=(x_1,\ldots,x_d)' \in \mathbb{R}^d$. Assume that $\bmm=(m_1,\ldots,m_d)' \in \mathbb{R}^d$, $S$ is diagonal with eigenvalues $0=s_{\ell+1}^2=\ldots=s_d^2$ and $0<s_1^2\leq \ldots\leq s_\ell^2$ with $2\leq\ell\leq d$ and that $t:=\Vert\bx-\bmm\Vert_S>0$. If $\bV$ is uniformly distributed on $\Omega_1^{d-1}$, then the distribution of $y_{\bmm,S}^{\bV} $ is supported by $[-t,t]$ and 
	\begin{align*}
	\Prob(y_{\bmm,S}^{\bV} \leq z)=
	\begin{cases}
	% 	0,& y\leq -t, \\
	%
	\tau\displaystyle\int_{A_{-}^{\bv}} \Delta(z)e^{-\frac{1}{2}\sum_{i=2}^dv_i^2} \,d\bv_{-1},& -t<z<-\frac{|u_1|}{s_1}, \\
	\frac{1}{2}-\sign(z)\tau\displaystyle\int_{A_{+}^{\bv}} \Delta(y)e^{-\frac{1}{2}\sum_{i=2}^dv_i^2} \,d\bv_{-1},& -\frac{|u_1|}{s_1} \leq z\leq\frac{|u_1|}{s_1},  \\
	1-\tau\displaystyle\int_{A_{+}^{\bv}} \Delta(z)e^{-\frac{1}{2}\sum_{i=2}^dv_i^2} \,d\bv_{-1}, & \frac{|u_1|}{s_1} <z<t,\\
	%
	% 	1,& y\geq t, 
	\end{cases}
	\end{align*} 
	with $\tau:=(2^{\frac{d+3}{2}}\pi^{\frac{d-1}{2}})^{-1}$, $\Delta(z):=\erf\left(h_{+}(z)/\sqrt{2}\right)-\erf\left(h_{-}(z)/\sqrt{2}\right)$, $\bv_{-1}:=(v_2, \ldots, v_d)$, $h_{\pm}(z)=\lrp{u_1\psi_{\bv}\pm|z|\sqrt{(u_1)^2\varphi_{\bv}+s_1^2\psi_{\bv}^2-s_1^2z^2\varphi_{\bv}}}/\lrp{s_1^2z^2-(u_1)^2}$,  $A_{+}^{\bv}:=\{\bv_{-1} : \psi_{\bv}>0\}$ and $A_{-}^{\bv}:=\{\bv_{-1} : \psi_{\bv}<0\}$ and where $\psi_{\bv}:=\sum_{i=2}^du_iv_i$, $\varphi_{\bv}:=\sum_{i=2}^\ell s_i^2v_i^2$, and $u_i=x_i-m_i$ for $i=1,\ldots,d$.
\end{proposition}

We obtain now some properties of the distribution of $Y^{K_n}$ which will be key in the practical determination of $a$ and $b$. Given $\alpha \in (0,1),$ the intended error of type I, our goal is to obtain  $0 < a \leq b$ such that $\Prob(K_n < \infty)=1$, and the probability of declaring a point $\bX$ as outlier when it is not  is less or equal than $\alpha$, i.e.

\begin{align}\label{eq:metodo2}
\sup \limits_{t\leq C_n^d} F_{\mSigma}(a,b,t)=\alpha. 
\end{align}

Trivially we need to exclude the case $a=0$, because for any $\bx$ and any sample we have a.s. $\bx\neq\hat{\bmu}$ and $\Prob(|y^{\bV}|\leq 0)=0$ a.s. Thus, Fubini's Theorem implies that the probability of not rejecting $\mathbf{H}_0$ is zero for any $\bx$.

\begin{proposition}\label{prop:probabilidad2}
	Under assumptions (A1) and (A2), if $a,b,t$ are strictly positive constants such that $a \leq b$, then	
	\begin{align*}
	F_{\mSigma}(a,b,t)=
	\int_{\Omega^{d-1}_{\mSigma}(t)}\int_{\mathbb{R}^d} \int_{\mathbb{R}^{d^2}}g_a^b(\bx,\bmm,S) f_t(\bx)\,\Prob_{\hat{\mSigma}}(dS) 
	\Prob_{\hat{\bmu}}(d\bmm) d\bx,
	\end{align*}
	where $g_a^b(\bx,\bmm,S):= \Prob\left( \vert y_{\bmm,S}^{\bV}\vert > b  \right)/\left(\Prob\left( \vert y_{\bmm,S}^{\bV}\vert > b  \right)+\Prob\left( \vert y_{\bmm,S}^{\bV}\vert < a  \right)\right)$, $\Prob_{\hat{\mSigma}}$ is the Wishart distribution with parameters $n$ and $\mSigma$, and $\Prob_{\hat{\bmu}}$ is the $N_d\left(\mathbf{0},n^{-1}\mSigma\right)$.
\end{proposition}

From the proof of Proposition \ref{prop:probabilidad2}, it is clear the following corollary.
\begin{corollary}
	Under the assumptions in Proposition \ref{prop:probabilidad2}, we have that 
	\begin{align*}
	\Prob \left(\left\vert Y^{K_n}\right\vert > b \ \vert \left\Vert\bX\right\Vert_{\mSigma}=t,\bX_1,\ldots,\bX_n\right)=\int_{\Omega^{d-1}_{\mSigma}(t)}g_a^b(\bx,\hat{\bmu},\hat{\mSigma}) f_t(\bx)\,d\bx.
	\end{align*}
\end{corollary}

Proposition \ref{prop:dependence_sigma_unknown}  leads to an easier expression of Proposition \ref{prop:probabilidad2} for $\mSigma=I_d$ provided in Corollary \ref{cor:probabilidad2}. The quantities in such corollary can be computed from Proposition \ref{prop: prob_dist}.

\begin{proposition}\label{prop:dependence_sigma_unknown}
	Under assumptions (A1) and (A2) the rv's $Y^1,\ldots,Y^k$ defined in \eqref{eq:T_Kestimate} are conditionally independent given $\left\Vert\bX\right\Vert_{\mSigma}$ if and only if $\mSigma=I_d$. 
\end{proposition}

\begin{corollary}\label{cor:probabilidad2}
	Under assumptions in Proposition \ref{prop:probabilidad2}. If $\mSigma=I_d$, then
	\begin{align*}
	F_{\mSigma}(a,b,t)=&\frac{\Prob\lrp{|Y^\bV|>b\vert\Vert\bX\Vert=t}}{1-\Prob\lrp{|Y^\bV|\in(a,b) \vert\Vert\bX\Vert=t}}.
	\end{align*}
\end{corollary}

Proposition \ref{prop: decrease2} shows that the probabilities involved in \eqref{eq:metodo2} are monotone on $t$. Thus \eqref{eq:metodo2} can be simplified to 
\begin{align}\label{eq:metodo3}
F_{\mSigma}(a,b,C_n^d)&=\alpha.
\end{align}

\begin{proposition}\label{prop: decrease2}
	Under assumptions (A1) and (A2), if $a$, $b$ and $t$ are positive constants such that $0 < a \leq b$, then, the function $F_{\mSigma}(a,b,t)$ is strictly increasing in $t$. 
\end{proposition}

\subsection{Moments of $K_n$}\label{subsec:momentsKn}

Proposition \ref{prop:esperanza2} gives  an expression of the variance and the expected number of projections that we need to declare
a point as an outlier or as regular.  
Its proof is not included because it is similar to that of Proposition  \ref{prop:probabilidad2}.%, 

\begin{proposition}\label{prop:esperanza2}
	Under assumptions (A1) and (A2), assume that $a,b$ and $t$ are positive numbers such that $a\leq b$ and consider $K_n$ defined as in \eqref{eq:def_Khat}, then
	\begin{align*}
	\E\left(K_n \vert \left\Vert\bX\right\Vert_{\mSigma}=t\right)=&\iiint\limits_{\mathcal{D}}\bar{g}_a^b(\bx,\bmm,S)f_t(\bx)\,\Prob_{\hat{\mSigma}}(dS) 
	\Prob_{\hat{\bmu}}(d\bmm) d\bx,\\
	\Var\left(K_n \vert \left\Vert\bX\right\Vert_{\mSigma}=t\right)=&\iiint\limits_{\mathcal{D}} \bar{g}_a^b(\bx,\bmm,S)(2\bar{g}_a^b(\bx,\bmm,S)-1)f_t(\bx)\,\Prob_{\hat{\mSigma}}(dS) 
	\Prob_{\hat{\bmu}}(d\bmm) d\bx\\
	&-\left(\iiint\limits_{\mathcal{D}} \bar{g}_a^b(\bx,\bmm,S) f_t(\bx)\,\Prob_{\hat{\mSigma}}(dS) 
	\Prob_{\hat{\bmu}}(d\bmm) d\bx\right)^2,
	\end{align*}
	where $\bar{g}_a^b(\bx,\bmm,S)= \left(\Prob\left( \vert y_{\bmm,S}^\bV\vert > b  \right)+\Prob\left( \vert y_{\bmm,S}^\bV\vert < a  \right)\right)^{-1}$, $\mathcal{D}:=\Omega^{d-1}_{\mSigma}(t)\times \mathbb{R}^d \times \mathbb{R}^{d^2}$, and $\Prob_{\hat{\mSigma}}$ and $\Prob_{\hat{\bmu}}$ are the Wishart distribution with parameters $n$ and $\mSigma$, and the $N_d(\mathbf{0},\mSigma/n)$, respectively.  
\end{proposition}

Propositions \ref{prop:dependence_sigma_unknown} and \ref{prop:esperanza2} allow to obtain Corollary \ref{cor:esperanza2}.
\begin{corollary}\label{cor:esperanza2}
	Under assumptions in Proposition \ref{prop:esperanza2}, if $\mSigma=I_d$, then
	\begin{align*}
	\E\left(K_n \vert \left\Vert\bX\right\Vert=t\right)=&\frac{1}{1-\Prob\lrp{|Y^\bV|\in (a,b)\vert\Vert\bX\Vert=t}},\\
	\Var\left(K_n \vert \left\Vert\bX\right\Vert=t\right)=&\frac{\Prob\lrp{|Y^\bV|<b\vert\Vert\bX\Vert=t}-\Prob\lrp{|Y^\bV|<a\vert\Vert\bX\Vert=t}}{1-\lrp{\Prob\lrp{|Y^\bV|\in (a,b)\vert\Vert\bX\Vert=t}}^2}.
	\end{align*}
\end{corollary}

It is clear from Corollary \ref{cor:esperanza2} that $\E(K_n\vert\Vert\bX\Vert=t)$ and $\Var(K_n\vert\Vert\bX\Vert=t)$ do not depend on either the specific value of $t$ or the dimension, but rather, only on the probability $\Prob\lrp{\vert Y^{\bV}\vert\in (a,b) \vert\Vert\bX\Vert=t}$. 
A graphical representation of those functions appears in Figure \ref{fig:EK} in Subsection \ref{subsubsec:graph} in the Appendix.

%-----------------------------------------------%
\subsection{Robust versions of $K_n$ and $Y^\bV$ }
\label{subsec:robust}
%-----------------------------------------------%
The results in Sections \ref{subsec:marginal} and \ref{subsec:momentsKn} fix the problem when we have a clean sample and we want to decide on a point which is not in the sample. However, usually, we are interested in detecting outliers inside the sample, which may  affect the estimation of the mean and the standard deviation. Thus, we propose to replace $\hat{\mu}_\bV$ and $\hat{\sigma}_\bV$ in \eqref{eq:T_Kestimate} by some robust counterparts. Our selections  are the median, $m_{\bV}$, and the MAD, $M^*_{\bV}$. 

It is well known that under normality the MAD overestimates the standard deviation (see Maronna et at. \cite{Maronna2019}). To make it consistent (see ibid), we use the normalized $\MAD$, abridged to $\MADN$: 
$M_{\bV}=M^*_{\bV}/q_3$, where $q_3$ is the third quantile  of a $N_1(0,1)$ distribution. We will denote by $\hat{m}_{\bV}$ and $\hat{M}_{\bV}$ to  the sample  median and MADN respectively. Furthermore, since both of them may not be unique,  the notation $\hat{m_{\bv}}$ and $\hat{M}_{\bv}$ refers to the choice of any  of the available possibilities.
To reflect the change, we replace $Y^{\bV}$ and $K_n$ by $\tilde{Y}^{\bV}$ and ${L}_n$, respectively. Now, \eqref{eq:metodo3} becomes 
\begin{align}\label{eq:metodo4}
\Prob \left(\left.\left\vert\tilde{Y}^{{L}_n}\right\vert > b \ \right\vert \left\Vert\bX\right\Vert_{\mSigma}=C_n^d\right)&=\alpha.
\end{align}
As it usually occurs with robust estimators (see for instance Cerioli et al. \cite{Cerioli} or Becker and Gather \cite{Becker1999}), it is difficult to obtain the conditional exact distribution of $\tilde{Y}^{{L}_n}$. Because of this we prove, in Theorem \ref{theo:assymp3}, that asymptotically on $n$ this distribution coincides with that of $Y^{K_n}$. Afterwards, in Section \ref{sec:num_studies}, we will present simulations suggesting that this approximation gives acceptable results in many cases for small sample sizes and arbitrary values of the dimension. 

Theorem \ref{theo:Juan} is an auxiliary result to obtain Theorem \ref{theo:assymp3}. However, we state it separately because it could have some independent interest.

\begin{theorem}\label{theo:Juan}
	Under assumptions (A1) and (A2), there exists  $A_0\in\mathcal{A}$ with $\Prob(A_0)=1$ such that if $\omega\in A_0$, then
	\begin{align}
	\sup_{\bv\in\Omega_1^{d-1}}\vert\hat{m}_{\bv}-m_{\bv}\vert\to 0\quad \mbox{and } \sup_{\bv\in\Omega_1^{d-1}}\vert\hat{M}_{\bv}-M_{\bv}\vert\to 0.\label{eq:prop_Juan1}
	\end{align}
\end{theorem}

\begin{theorem}\label{theo:assymp3}
	Let us consider $g_{a}^b(\cdot,\cdot,\cdot)$ as defined in Proposition \ref{prop:probabilidad2}. Assume (A1) and (A2). If $a,b$ and $t$ are positive constants such that $ 0 < a \leq b$, and  $\tilde{Y}^{{L}_n}$  is defined as above, then, a.s.
	\begin{align*}
	\Prob\lrp{\left|\tilde{Y}^{{L}_n}\right|>b \left\vert \right. \Vert\bX\Vert_{\mSigma}=t}\to&\;\displaystyle\int_{\Omega^{d-1}_{\mSigma}(t)}g_{a}^b(\bx,\mathbf{0},\mSigma)f_t(\bx)\,d\bx.
	\end{align*}
\end{theorem}

Next proposition gives the asymptotic behaviour of the first two moments of ${L}_n$. Its proof is similar to that one of Theorem \ref{theo:assymp3} and we do not include it. 

\begin{proposition}\label{prop:EK_assym}
	Let us consider $\bar{g}_a^b(\cdot,\cdot,\cdot)$ as defined  in Proposition \ref{prop:esperanza2}. Assume (A1) and (A2). If $a,b$ and $t$ are positive constants such that $ 0 < a \leq b$,  then, a.s.
	\begin{align*}
	\E\left({L}_n \ \vert \ \Vert\bX\Vert_{\mSigma}=t\right)\to&\;\int_{\Omega_{\mSigma}^{d-1}(t)}\bar{g}_a^b\lrp{\bx,\mathbf{0},\mSigma}f_t(\bx)\, d\bx,\\
	\Var({L}_n \ | \left\Vert\bX\right\Vert_{\mSigma}=t)\to&\;\int_{\Omega_{\mSigma}^{d-1}(t)}\bar{g}_a^b\lrp{\bx,\mathbf{0},\mSigma}\lrp{2\bar{g}_a^b\lrp{\bx,\mathbf{0},\mSigma}-1}f_t(\bx)\, d\bx\\
	&-\lrp{\int_{\Omega_{\mSigma}^{d-1}(t)}\bar{g}_a^b\lrp{\bx,\mathbf{0},\mSigma}f_t(\bx)\, d\bx}^2,
	\end{align*}
\end{proposition}

The expressions of Theorem \ref{theo:assymp3} and Proposition \ref{prop:EK_assym} simplify in the case $\mSigma=I_d$ as shown in the following corollary.

\begin{corollary}\label{cor:assymp_EK}
	With the assumptions and the notation in Theorem \ref{theo:assymp3}, consider $F(\cdot,t)$ as defined  in \eqref{eq:prob_unknown_sigma}. If $\mSigma=I_d$, then, as $n \to \infty$,  a.s.,  
	\begin{align*}
	\Prob\lrp{\left|\tilde{Y}^{{L}_n}\right|>b \left\vert\right. \Vert\bX\Vert=t}\to&\;\lrp{1-F(b,t)}/\lrp{1-F(b,t)+F(a,t)},\\
	\E({L}_n  \left\vert\right.\left\Vert\bX\right\Vert=t)\to&\;\lrp{1}/\lrp{1-F(b,t)+F(a,t)},\\
	\Var({L}_n \left\vert\right. \left\Vert\bX\right\Vert=t)\to&\;\lrp{F(b,t)-F(a,t))}/\lrp{\left(1-F(b,t)+F(a,t)\right)^2}.
	\end{align*}
\end{corollary}	

\begin{remark}\label{remark:expectation_lim}
	Denote $\mathbb{X}_n=\{\bX_1,\ldots,\bX_n\}$. From the proofs of Theorem \ref{theo:assymp3} and Proposition \ref{prop:EK_assym} it is clear that the a.s.  limits of the expressions 
	$
	\Prob\lrp{|\tilde{Y}^{{L}_n}|>b  \left\vert\right. \Vert\bX\Vert_{\mSigma}=t,\mathbb{X}_n}$, $
	\E\left({L}_n  \left\vert\right.\left\Vert\bX\right\Vert_{\mSigma}=t, \mathbb{X}_n\right)$ and $
	\Var({L}_n  \left\vert\right. \left\Vert\bX\right\Vert_{\mSigma}=t, \mathbb{X}_n)$ coincide with those shown in Proposition \ref{prop:EK_assym}. 
\end{remark}

%-------------------------------------------------------------------------%
\subsection{Computation of the constants $a$ and $b$}\label{subsec:comp_a_b}
%-------------------------------------------------------------------------%
The explicit computation of $a$ and $b$ requires to find a solution of \eqref{eq:metodo4} satisfying that $\E\lrp{{L}_n\vert\Vert\bX\Vert_{\mSigma}=C_n^d}$ equals to a pre-specified value. This problem has been impossible for us even in the non-robust version  \eqref{eq:metodo3} which handles the function $F_{\mSigma}(a,b,t)$.

Proposition \ref{prop:probabilidad2} gives an explicit expression for $F_{\mSigma}(a,b,t)$; the problem being that the integrand in this expression is so involved that, excepting if $a=b$, we have not been able to  compute the integral even when $\mSigma=I_d$ (note that $a$ and $b$ depend on the covariance matrix). In addition, the  complexity increases when $\mSigma\neq I_d$, because of the dependency of the projections given $\Vert\bX\Vert_{\mSigma}$ as Proposition \ref{prop:dependence_sigma_unknown} showed.

An option to solve \eqref{eq:metodo3} as an approximation to \eqref{eq:metodo4} would be take $a=b=a_{\alpha}$, the conditional $(1-\alpha)$-quantile of $ Y^\bV$ given that $\Vert\bX\Vert_{\mSigma}=C_n^d$; but this does not seem very sensible because this means taking the decision based on one single random projection. However, according to Proposition \ref{prop:worst_case} below, for every $a \in (0,a_\alpha)$, there exists a unique $b_a$ such that the pair $(a, b_a $) gives a test at the level  $\alpha$ for the covariance matrix under consideration. Moreover, the lower the $a$, the larger the number of required projections,  what increases the chances  to take the right decision (at the price of a higher computational time).

\begin{proposition}\label{prop:worst_case}
	Given $a>0$ with $\Prob(\vert Y^\bV\vert<a)\leq \alpha$, there exists a unique $b_a$ such that $F_{\mSigma}(a,b_a,C_n^d)=\alpha$. Moreover,  the map $a\mapsto b_a$ is strictly decreasing on $a$.
\end{proposition}

Proposition \ref{prop:EK_comparison} somehow eases the computation of $a$ and $b$ because it states that given $0<a\leq b$, then asymptotically on $n$ the expected number of observations required to reach a decision  is minimal if the sample comes from a $N_d(\mathbf{0},I_d)$. Therefore, if we use for a general covariance matrix the constants of the identity, then we will make the decision using the pre-specified number of projections or more.

\begin{proposition}\label{prop:EK_comparison}
	Let us assume (A1) and (A2) and let $t>0$ and $0<a\leq b$. Let $\mSigma\neq I_d$ be a positive definite matrix. Let $\mathbb{X}^\mSigma:= \{\bX_n^\mSigma\}$ and $\mathbb{X}^{I_d}:=\{\bX_n^{I_d}\}$ be two random samples taken from the $N_d(\mathbf{0},\mSigma)$ and $N_d(\mathbf{0},I_d)$ respectively. Then, almost surely,
	\[
	\lim_n \E\left( {L}_n^{a,b} \ \bigg| \Vert\bX\Vert_{\mSigma}=t,\mathbb{X}^\mSigma \right) > 
	\lim_n \E\left( {L}_n^{a,b} \ \bigg| \Vert\bX\Vert=t,\mathbb{X}^{I_d}\right).
	\]
\end{proposition}

After Proposition \ref{prop:EK_comparison}, our proposal consists of using  $a$ and $b$ computed for the $N_d(\mathbf{0},I_d)$. However, Proposition \ref{prop:EK_comparison} leaves two open points: the level of the test obtained when using those constants  with $\mSigma \neq I_d$; and some hints on the expected number of observations when $n$ is low, mostly,  when $\mSigma \neq I_d$. We have  obtained no theoretical result on this line, but we have produced  practical evidence suggesting that the situation is reasonably good. Specifically, we have selected several covariance matrices  and we have conducted numerical experiments using pairs $(a,b)$ computed for $\mSigma=I_d$ with the following results (see   Subsections \ref{subsec:a_b_id} and \ref{subsec:a_b_Other}):

\begin{itemize}
	\item[1)]
	The obtained rejection levels with  $\mSigma \neq I_d$ are close to the levels of the identity.
	
	\item[2)]
	The results obtained for sample sizes as low as $n=50$ are similar to those predicted by Proposition \ref{prop:EK_comparison}. I.e., for  sample sizes $n\geq 50$ and covariance matrices $\mSigma\neq I_d$, the mean of the obtained values for $L_n$ are mostly larger  than the expected for the identity and they are seldom only slightly lower.

	\item[3)]
	The mean of the values obtained for ${L}_n$ when $\mSigma\neq I_d$ are generally similar to those obtained when $\mSigma=I_d$ but sometimes they are much higher.
\end{itemize}

\paragraph{Notation} We denote by $l_\mSigma^r $ to   $E(L_n| \|X\|_\mSigma=r  C_n^d(\delta))$, when $\delta=.05$.  Its  sample mean along the simulations we do will be represented by $\hat{ l}_\Sigma^r$. However, to ease the notation, we will write  $l_I^r, \hat l_I^r, l_i^r $ and $\hat l_i^r $ when $\mSigma =I_d$ or $\mSigma =\mSigma_i^d, i=1,\ldots, 4$ for the matrices $\mSigma_1^d,\ldots,\mSigma_4^d$ defined in the next subsection respectively.

\subsubsection{Computation of $(a,b)$, $\mSigma=I_d$}\label{subsec:a_b_id}

In this subsection, given $n\in\mathbb{N}$, we want to compute  the constants $a$ and $b$ giving a power $\alpha$-test with a given value $h\ge 1$ for $l_I^1$. To this, taking into account the expressions in Corollary \ref{cor:assymp_EK}, we could solve the equations 
\begin{align}\label{eq:a_b_sigma}
\begin{cases}
h&=(1-v+u)^{-1}\\
\alpha&=(1-v)(1-v+u)^{-1},
\end{cases}
\end{align}
and then to look for $a$, $b$ satisfying that $u=\tilde {F}(a,C_n^d)$ and $v=\tilde{F}(b,C_n^d)$, where $\tilde{F}(y,t)=\linebreak\Prob\lrp{|\tilde Y^\bV|<y|\Vert\bX\Vert=t}$. 

The solution of \eqref{eq:a_b_sigma} is $u=(1-\alpha)/h$ and $v=1-\alpha/h$. Only remains to find the $u$ and $v$ quantiles of the distribution $\tilde{F}(\cdot,C_n^d)$. Since we have no explicit expressions for them, 
we have decided to begin computing $a$ and $b$ by the Monte Carlo method. 

The computation is done as follows:
we fix $N$ large and for $j=1,\ldots,N$, i) generate $\bX_0^j,\bX_1^j,\ldots,\bX_n^j$ and $\bV^j$ iid rv's with distribution $N_d(\mathbf{0},I_d)$, ii) consider $\bX^j=C_n^d\bX_0^j/\Vert\bX_0^j\Vert$, iii) compute $\tilde Y^j=\linebreak\left|(\bX^j)'\bV^j-\hat{m}_{\bV^j}\right|/\hat{M}_{\bV^j}$, iv) take $a$ and $b$ equal to the quantiles $u$ and $v$ of the sample $\tilde Y^1,\ldots,\tilde Y^N$. 

Regrettably, some simulations have shown that the test associated to the obtained pair has generally power lower than $\alpha$ because the value $b$ is lower than desired. To fix this point we recalculate $b$, keeping $a$ fixed, by simulations with the bisection method. This procedure has proved to give  tests at the right level.

Table \ref{tab:comparison_valores_ab2} shows the values of the constants $a$ and $b$ for different values of $l_{I}^1$. Those values have been computed with the above explained methodology with $N=10^6$. 

The choice of the values of the dimension and the sample size attempts to represent the scenarios when the dimension is higher/smaller than the sample size. From this table, the bigger $l_{I}^1$, the wider the interval $(a,b)$ according to Corollary \ref{cor:esperanza2}. 
However, the larger the sample size, the narrower the interval $(a,b)$. This is due to the fact that the estimation of the parameters is more stable for greater sample sizes.

\begin{table}
	\caption{Obtained values of $(a,b)$ when $\mSigma=I_d$ for different values of $n, d$ and $l_{I}^1$ and $C_n^d\equiv C_n^d(0.05)$.}
	\label{tab:comparison_valores_ab2}
	%\begin{center}
	\begin{tabular}{r@{\hskip .08cm}|c@{\hskip .08cm}c@{\hskip .10cm}c@{\hskip .10cm}c@{\hskip .10cm}|c@{\hskip .08cm}c@{\hskip .10cm}c@{\hskip .10cm}c@{\hskip .10cm}c@{\hskip .10cm}|c@{\hskip .08cm}c@{\hskip .10cm}c@{\hskip .10cm}c@{\hskip .10cm}c}
		\hline\noalign{\smallskip}
		& \multicolumn{4}{c|}{$n=50$} && \multicolumn{4}{c|}{$n=100$} && \multicolumn{4}{c}{$n=500$}\\
		&  \multicolumn{2}{c}{$l_{I}^1=50$} &  \multicolumn{2}{c|}{$l_{I}^1=100$} &&  \multicolumn{2}{c}{$l_{I}^1=50$} &  \multicolumn{2}{c|}{$l_{I}^1=100$} &&  \multicolumn{2}{c}{$l_{I}^1=50$} &  \multicolumn{2}{c}{$l_{I}^1=100$}\\
		$d$& $a$ & $b$ & $a$ & $b$ && $a$ & $b$ & $a$ & $b$ && $a$ & $b$ & $a$ & $b$\\
		\hline
		50  & 0.0325 &  4.9714 & 0.0163 & 5.3212 && 0.0326 & 4.6374 & 0.0163 & 4.9143 && 0.0336 & 4.4525 & 0.0167 & 4.6989\\
		100  & 0.0303 & 4.7184 & 0.0150 & 5.0936  && 0.0303 & 4.3539 & 0.0151 & 4.6495 && 0.0304 & 4.1478 & 0.0156 & 4.3910 \\
		500  & 0.0268 & 4.3039 & 0.0133 & 4.6239 && 0.0267  & 3.9230 & 0.0133 & 4.2078 && 0.0266 & 3.7278 & 0.0132 & 3.9520 \\
		1000 & 0.0263 & 4.1916 & 0.0130 & 4.5217 && 0.0261 & 3.8253 & 0.0128 & 4.0909 && 0.0259 & 3.6096 &  0.0130 & 3.8197\\
		\noalign{\smallskip}
		\hline
	\end{tabular}
	%\end{center}
\end{table} 
%**********************************************%
\subsubsection{Computation of $(a,b)$, $\mSigma\neq I_d$} \label{subsec:a_b_Other}
%**********************************************%
Based on Proposition  \ref{prop:EK_comparison}, our idea is using the values obtained for $\mSigma = I_d$  to handle any covariance matrix. We firstly check if those values are suitable for general matrices. Since we think that  the worst situation with those constants could occur in matrices with sparse eigenvalues, we have chosen three families with large variation among them, while $\mSigma_4^d$'s are a matrices with little variation.  The considered families are: 
\begin{itemize}
	\item[-] $\mSigma_1^d$ is a matrix with the half of their eigenvalues 1's and others $d^2$.
	
	\item[-] $\mSigma_2^d$  is a matrix with equally spaced eigenvalues from 1 to $d^2$. 
	
	\item[-] $\mSigma_3^d$ is a matrix whose eigenvalues are 1's $d-1$ times and one is $d^2$. 
	
	\item[-] The eigenvalues of $\mSigma_4^d$ vary between 1 and 2. They are the ratio between two equispaced sequences between $d^2$ and 2 and between $d^2$ and 1 respectively.
\end{itemize}

From this point,  for each combination of dimension  and sample size,  we have computed a pair $(a_I,b_I)$ giving an $\alpha$-level test for the identity matrix as explained  in Subsection \ref{subsec:a_b_id}. We have kept $a_I$ and, for every $\mSigma=\mSigma_i^d, i=1,\ldots,4$, we have computed (using the same procedure as in Subsection \ref{subsec:a_b_id} with $N = 10^4$ simulations) the value $b_\mSigma$  such that the pair $(a_I, b_\mSigma )$ is an $\alpha$-level test. 

The good news is that in all cases we have considered,  we have found that the values $b_I$ and $b_\mSigma$ are very similar. In fact, we include in Subsection \ref{Sec.a_bDeLaIdentidad} in the Appendix some guidelines of an ongoing research to find conditions allowing to use the constants $a_I,b_I$ with other covariance matrices when $d$ is large. In fact, the reasoning there suggests that this selection could work for matrices fullfilling a not too restrictive condition on their eigenvalues and that the selection could be not too different even for those matrices not satysfying it (see Remarks \ref{Nota.Ctes.Id_1} and \ref{Nota.Ctes.Id_2} in the Appendix).  Moreover, excepting if $\mSigma=\mSigma_3^d$, the expected numbers of projections $l_I^1$ and $l_\mSigma^1$ are also very similar.

For each pair of sample size and dimension,  Table \ref{tab:comparison_valores_worstb} shows the $b_\mSigma$ maximizing the difference $|b_I-b_\mSigma|$ along the four covariance matrices and the  matrix producing it. All obtained $b_\mSigma$'s are in Table \ref{tab:comparison_valores_abs1} in  Subsection \ref{subsubsec:Computing_b} in the Appendix.

\begin{table}
	\caption{Values of $b_\mSigma$ giving the greatest difference $|b_I-b_\mSigma|$ for $\mSigma=\mSigma_i^d, i=1,\ldots,4$, and different values of $d$ and $n$. $a$'s are taken from Table \ref{tab:comparison_valores_ab2}. Columns $\mSigma$ tell the matrices in which $b_\mSigma$ were obtained.}
	\label{tab:comparison_valores_worstb}
	%\begin{center}
	\begin{tabular}{r@{\hskip .45cm}|c@{\hskip .12cm}c@{\hskip .33cm}c@{\hskip .12cm}c@{\hskip .33cm}|c@{\hskip .33cm}c@{\hskip .12cm}c@{\hskip .33cm}c@{\hskip .12cm}c@{\hskip .33cm}|c@{\hskip .33cm}c@{\hskip .12cm}c@{\hskip .33cm}c@{\hskip .12cm}c}
		\hline\noalign{\smallskip}
		& \multicolumn{4}{c|}{$n=50$} && \multicolumn{4}{c|}{$n=100$} && \multicolumn{4}{c}{$n=500$}\\
		&  \multicolumn{2}{c}{$l_{I}^1$=50} &  \multicolumn{2}{c|}{$l_{I}^1$=100} &&  \multicolumn{2}{c}{$l_{I}^1$=50} &  \multicolumn{2}{c|}{$l_{I}^1$=100} &&  \multicolumn{2}{c}{$l_{I}^1$=50} &  \multicolumn{2}{c}{$l_{I}^1$=100}\\
		$d$& $b_\mSigma$ & $\mSigma$ & $b_\mSigma$ & $\mSigma$ && $b_\mSigma$ & $\mSigma$ & $b_\mSigma$ & $\mSigma$ && $b_\mSigma$ & $\mSigma$ & $b_\mSigma$ & $\mSigma$\\
		\hline
		50  &  5.1413 & $\mSigma_3^d$ & 5.4932 & $\mSigma_3^d$ && 4.6194  & $\mSigma_4^d$ & 4.9504  & $\mSigma_3^d$ && 4.4439  & $\mSigma_3^d$ & 4.6858 & $\mSigma_3^d$\\
		100  &  4.8813 & $\mSigma_3^d$  & 5.1857 & $\mSigma_3^d$ && 4.3497  & $\mSigma_4^d$ & 4.6387 & $\mSigma_2^d$ && 4.1399 & $\mSigma_3^d$ & 4.3691 & $\mSigma_3^d$\\
		500  & 4.3244  & $\mSigma_2^d$ & 4.6946 & $\mSigma_3^d$ && 4.0248  & $\mSigma_3^d$ & 4.2460 & $\mSigma_3^d$ &&  3.7509 & $\mSigma_4^d$  & 3.9143 & $\mSigma_3^d$\\
		%			% 			
		1000  &  4.3129 & $\mSigma_3^d$  & 4.6166 & $\mSigma_3^d$ && 3.9221  & $\mSigma_3^d$ & 4.1094 & $\mSigma_3^d$ && 3.6276  & $\mSigma_1^d$ & 3.8363  & $\mSigma_2^d$\\
		\noalign{\smallskip}
		\hline
	\end{tabular}
	%\end{center}
\end{table}

%-----------------------------------------------%
\section{Practical implementation}\label{sec:implementation}
%-----------------------------------------------%

Here we give some advices on the practical implementation of the method. We pay attention to how to fix the number of expected projections (Subsection \ref{Subs:Fixing.ISigma}) and  how many simulated values of $\tilde Y^\bV$ we should produce  to compute $a$ and $b$ (Subsection \ref{Subs:HowLargeN}). Subsection \ref{Subs:AlgorithmAnaliseSamples} contains an algorithm to analyse all points in a sample. Subsection \ref{Subs:IdentifyingOutliers} shows a procedure to reduce the role of the randomness in the process

%---------------------------------------------------------&
\subsection{Which value should we choose for $l_\mSigma^r$?} \label{Subs:Fixing.ISigma}
%---------------------------------------------------------&
In principle, the higher the $l_\mSigma^r$ the higher the power under the alternative, but also the computational effort increases. The simulations we present below show a detectable increment in power  from $l_\mSigma^r=50$ to  $l_\mSigma^r=100$. However, this increment is not too striking and, of course, the improvement slows down for values of $l_\mSigma^r$  above 100.

Hence, our advice is to fix this parameter at 50, or at most at 100. In fact, in Subsection \ref{sec:comparison} we use $l_\mSigma^r=50$, while we choose $l_\mSigma^r=100$  in Subsection \ref{sec:real_data}.

%---------------------------------------------------------&
\subsection{How many simulated values of $\tilde Y^\bV$ are required to compute $a,b$?} \label{Subs:HowLargeN}
%---------------------------------------------------------&
The algorithm we proposed in Subsection \ref{subsec:a_b_id}   to compute $a,b$ requires a large number $N$ of replicas of $\tilde Y^\bV$. 
In this paper we have chosen $N=10^6$, but this is quite time consuming. Some computations suggest that $N=10^4$ could do it depending on the involved percentiles, but it seems that $N=10^5$ offers a reasonable trade-off between time and precision. Table  \ref{tab:ab_N10^5}, in Subsection \ref{subsubsec:Computational times} in the Appendix, shows the  computational times for some combinations of $d,n$ and $l_I^1$. Those times range from 40 seconds to 33 minutes in a four cores processor 3.2 GHz Intel Core i5. 

The results obtained with $N=10^5$ are not so bad. To see this, it is enough to compare the results in Tables \ref{tab:comparison_valores_ab2_100000} and \ref{tab:proof_prob_depend_pto1R_100000} with those in Tables \ref{tab:comparison_valores_ab2} and \ref{tab:proof_prob_depend_pto1R_app}: there are some differences among the parameters (due to greater uncertainty in the estimation of the involved quantiles) but, in our opinion, they are inside  reasonable margins.

\begin{table}
	\caption{Obtained values of $(a,b)$ when $\mSigma=I_d$ for different values of $n, d$ and $l_{I}^1$ and $C_n^d\equiv C_n^d(0.05)$. Only $10^5$ simulated values in the of $\tilde Y^\bV$ in the first step.}
	\label{tab:comparison_valores_ab2_100000}
	%\begin{center}
	\begin{tabular}{r@{\hskip .08cm}|c@{\hskip .08cm}c@{\hskip .10cm}c@{\hskip .10cm}c@{\hskip .10cm}|c@{\hskip .08cm}c@{\hskip .10cm}c@{\hskip .10cm}c@{\hskip .10cm}c@{\hskip .10cm}|c@{\hskip .08cm}c@{\hskip .10cm}c@{\hskip .10cm}c@{\hskip .10cm}c}
		\hline\noalign{\smallskip}
		& \multicolumn{4}{c|}{$n=50$} && \multicolumn{4}{c|}{$n=100$} && \multicolumn{4}{c}{$n=500$}\\
		&  \multicolumn{2}{c}{$l_{I}^1=50$} &  \multicolumn{2}{c|}{$l_{I}^1=100$} &&  \multicolumn{2}{c}{$l_{I}^1=50$} &  \multicolumn{2}{c|}{$l_{I}^1=100$} &&  \multicolumn{2}{c}{$l_{I}^1=50$} &  \multicolumn{2}{c}{$l_{I}^1=100$}\\
		$d$& $a$ & $b$ & $a$ & $b$ && $a$ & $b$ & $a$ & $b$ && $a$ & $b$ & $a$ & $b$\\
		\hline
		50  & 0.0333 & 4.9870 & 0.0168 & 5.3563 && 0.0326 & 4.6579 & 0.0165 & 4.9365 && 0.0340 & 4.4449 & 0.0170 & 4.6927\\
		100  & 0.0297 & 4.7470 & 0.0149 & 5.1214 && 0.0300 & 4.3772 & 0.0150 & 4.6435 && 0.0312 & 4.1760 & 0.0154 & 4.3896\\
		500  & 0.0262 & 4.3144 & 0.0131 & 4.6484 && 0.0269 & 3.9731 & 0.0140 & 4.2024 && 0.0275 & 3.7194 & 0.0136 & 3.9522\\
		1000 &0.0256 & 4.1863 & 0.0122 & 4.5825 && 0.0262 & 3.8331 & 0.0134 & 4.0953 && 0.0257 & 3.6629 & 0.0123 & 3.8329\\
		\noalign{\smallskip}
		\hline
	\end{tabular}
	
	%\end{center}
\end{table}

\begin{table}
	\caption{Estimation of the probability of declaring as an outlier a vector such that $\left\Vert \bX \right\Vert= C_n^d$, when $\Sigma=I_d$, for several values of $n, d$ using $a,b$ obtained in  Table \ref{tab:comparison_valores_ab2_100000}. }
	\label{tab:proof_prob_depend_pto1R_100000}
	%\begin{center}
	\begin{tabular}{r@{\hskip .16cm}|c@{\hskip .12cm}c@{\hskip .12cm}c@{\hskip .12cm}r@{\hskip .12cm}|c@{\hskip .12cm}c@{\hskip .12cm}c@{\hskip .12cm}c@{\hskip .12cm}r@{\hskip .12cm}|c@{\hskip .12cm}c@{\hskip .12cm}c@{\hskip .12cm}c@{\hskip .12cm}r}
		\hline\noalign{\smallskip}
		& \multicolumn{4}{c|}{$n=50$} && \multicolumn{4}{c|}{$n=100$} && \multicolumn{4}{c}{$n=500$}\\
		&  \multicolumn{2}{c}{$l_{I}^1=50$} &  \multicolumn{2}{c|}{$l_{I}^1=100$} &&  \multicolumn{2}{c}{$l_{I}^1=50$} &  \multicolumn{2}{c|}{$l_{I}^1=100$} &&  \multicolumn{2}{c}{$l_{I}^1=50$} &  \multicolumn{2}{c}{$l_{I}^1=100$}\\
		$d$& Prob. & $\hat l_I^1$ & Prob. & $\hat l_I^1$ &&Prob. & $\hat l_I^1$ &Prob. & $\hat l_I^1$ &&Prob. & $\hat l_I^1$ &Prob. & $\hat l_I^1$ \\
		\hline
		50  & 0.0528  &  48   &  0.0440  &  98   &&  0.0518  &  51   &  0.0454 & 100   &&  0.0466  &  48   &  0.0470  &  97\\
		100  & 0.0464  &  51   &  0.0474 & 103   &&  0.0534  &  50   &  0.0516 & 101   &&  0.0460  &  50   &  0.0482 & 100\\
		500  & 0.0530 & 50 &  0.0480 &102 &&  0.0506 & 48 &  0.0508 & 94 &&  0.0492 & 49 &  0.0492 & 99\\
		1000 &0.0540  &  50   &  0.0532 & 110   &&  0.0482  &  49   &  0.0494  &  96   &&  0.0440  &  51   &  0.0524 & 104\\
		\noalign{\smallskip}
		\hline
	\end{tabular}
	
	%\end{center}
\end{table}

%---------------------------------------------------------&
\subsection{Algorithm to analyse a sample} \label{Subs:AlgorithmAnaliseSamples}
%---------------------------------------------------------&

An algorithm to analyse all points in a sample goes as follows: Let $\mathcal X$ be the set containing all points in the sample at hand and fix the set of the regular points, $\mathcal X_R$, equal to the  empty set. Then follow the steps
\begin{enumerate}
	
	\item
	Take a random projection and analyse all points in $\mathcal X$.
	
	\item
	If some points have been declared as outliers, 
	delete them from $\mathcal X$, 
	set $\mathcal X_R=\emptyset$,
	and go to step 1. Else,  add  the points declared as non-outliers to $\mathcal X_R$.
	
	\item
	If $\mathcal X_R \neq \mathcal X$ go to step 1. Else, return $\mathcal X_R$.
\end{enumerate}

Notice that the algorithm always ends. Moreover, some points declared  regular in initial rounds, could later be declared as outliers, because in step 2 we make $\mathcal X_R=\emptyset$ every time a new outlier is identified. This is done so to reduce the masking effect.

%---------------------------------------------------------&
\subsection{How to reduce the role of the randomness in deciding if a point is outlier or not?} \label{Subs:IdentifyingOutliers}
%---------------------------------------------------------&

Some people can feel uncomfortable with the randomness of the  procedure. As stated, the larger $l_\mSigma^r$ the lower the role of the randomness. A possibility to reduce further this role is to repeat a not so large number of times, $T$, the process  using a significance level $\alpha$. 	Thus, since points $\bx$ satisfying that $\|\bx- \bmu\|_\mSigma=C_n^d(\alpha)$ are declared as outliers a proportion $\alpha$ of times, we could resort to declare as outliers those points which have been identified as outliers more than a proportion $\alpha$ of times along the $T$ repetitions.  We have applied this criteria in Subsection \ref{sec:real_data}, with $T=100$.

The criteria can be strengthened (resp. relaxed) identifying as outliers only the points declared as outliers a number of times higher  (resp. lower) than the 0.95 (resp. 0.05) quantile of a binomial with parameters $T$ and $\alpha$.

%-----------------------------------------------%
\section{Numerical studies}\label{sec:num_studies}
%-----------------------------------------------%
In this section we analyse the behaviour of the method thorough simulated experiments and real datasets. Here, only the results for $n=50$ are shown (the complete results are in the Appendix). We also compare our procedure with existing methods. 

The computations of the constants $a$ and $b$ determining the tests are carried out as described in Subsections \ref{subsec:a_b_id} with $N=10^6$ simulated values of $\tilde Y^\bV$.

\subsection{Simulations}\label{sec:simul}
We use the notation introduced at the end of Subsection \ref{subsec:comp_a_b}.  All the results  are obtained from $5000$ replicated simulations.

Table \ref{tab:proof_prob_depend_pto1R_n50}  shows the proportion of times we have declared a point with Mahalanobis norm $C_n^d(\delta)$ with $\delta = 0.05$ as an outlier for $n=50$ and several values of $d$.  More results including the cases $n=100,500$ are in Table \ref{tab:proof_prob_depend_pto1R_app}  in Subsection \ref{subsubsec:Detecting_Outliers} in  the Appendix. The results are not bad because the proportions  are close to the intended: the percentiles 0.025 and 0.975 of the obtained proportions are 0.044 and 0.0562 and the price we pay to achieve robustness seems to be a slightly conservative test, since  we obtain  18 (out of 120) proportions outside those values, all of them in the upper part, but with the maximum (equal to 0.0668) being  close to the target.

\begin{table}[!htb]
	\caption{Estimation of the probability of declaring as an outlier a vector such that $\left\Vert \bX \right\Vert_{\mSigma}= C_n^d$, for $n=50$ and several values of $ d$ and $\mSigma$. We also show the sample means of $L_n$.}
	\label{tab:proof_prob_depend_pto1R_n50}
	%\begin{center}
	\begin{tabular}{cc|rcrcrcrcrc}
		\hline\noalign{\smallskip}
		$d$ & $l_{I}^1$ &$\hat l_{I}^1$ & $I_d$ & $\hat  l_{1}^1$ & $\mSigma_1^d$ & $\hat  l_{2}^1$ & $\mSigma_2^d$ & $\hat  l_{3}^1$ & $\mSigma_3^d$ & $\hat  l_{4}^1$ & $\mSigma_4^d$ \\
		\hline
		$50$ & 50 & 51 & 0.0528 & 49 & 0.0571 & 49 & 0.0541 & 186 & 0.0668 & 50 & 0.0569\\
		& 100 & 98 & 0.0560 & 99 & 0.0558 & 103 & 0.0553 & 366 & 0.0580 & 99 & 0.0572\\
		\hline
		$100$ & 50 & 49 & 0.0507 & 48 & 0.0496 & 50 & 0.0501 & 249 & 0.0628 & 50 & 0.0489\\ 
		& 100 & 100 & 0.0538 & 101 & 0.0519 & 100 & 0.0526 & 526 & 0.0603 & 98 & 0.0494\\
		\hline
		$500$ & 50 & 49 & 0.0481 & 50 & 0.0507 & 50 & 0.0518 & 552 & 0.0628 & 50 & 0.0483\\
		& 100 & 100 & 0.0520 & 102 & 0.0509 & 99 & 0.0545 & 1111 & 0.0589 & 101 & 0.0538\\
		\hline
		$1000$ & 50 & 50 & 0.0496 & 50 & 0.0538 & 49 & 0.0534 & 790 & 0.0586 & 50 & 0.0500\\
		& 100 & 100 & 0.0520 & 101 & 0.0476 & 102 & 0.0507 & 1601 & 0.0549 & 99 & 0.0553\\
		\noalign{\smallskip}
		\hline
	\end{tabular}
	%\end{center}
\end{table}

The mean number of projections $\hat l_1^1, \ldots ,\hat l_4^1$ are always greater or very close to $l_I^1$ (giving support to the fact that the asymptotical result shown in Proposition \ref{prop:EK_comparison} also holds for finite sample sizes), being  $\hat l_3^1$ always the largest one. 

Moreover $\hat l_1^1, \hat l_2^1$ and $\hat l_4^1$  are always reasonably similar to $\hat l_I^1$, which, in turn, are close to the goal $l_I^1$.   The values obtained for $\hat l_3^1$ increase with the dimension and, when $d=500,10^3$, they are an order of magnitude larger than intended.

Table  \ref{tab:proof_prob_depend_pto2R_n50} shows the estimations of the probability of declaring a point as an outlier when its Mahalanobis norm is $1.2C_n^d$ or $2C_n^d$ and $n=50$. Complete results are in Tables \ref{tab:proof_prob_depend_pto1_2R_app} and \ref{tab:proof_prob_depend_pto2R.app} in Subsection \ref{subsubsec:Detecting_Outliers} in the Appendix. 

\begin{table}%[!htb]
	\caption{Estimation of the probability of declaring as an outlier a vector such that $\left\Vert \bX \right\Vert_{\mSigma}= r C_n^d, r=1.2, 2$ with $n=50$. We also show the sample means of $L_n$.}
	\label{tab:proof_prob_depend_pto2R_n50}
	%\begin{center}
	\begin{tabular}{ccr|rcccccrccc}
		\hline\noalign{\smallskip}
		$d$& $\|X\|_\mSigma$&$l_{I}^{r}$ & $\hat l_I^{r}$ & $I_d$ & $\hat l_1^{r}$ & $\mSigma_1^d$ & $\hat l_2^{r}$ & $\mSigma_2^d$ & $\hat l_3^{r}$ & $\mSigma_3^d$ & $\hat  l_4^{r}$ & $\mSigma_4^d$ \\
		\hline
		$ 50$ & $1.2C_n^d$& $50$&  48 & .2378 & 48 & .2247 & 48 & .2338 & 163 & .1752 & 48 & .2333\\
		&&$100$ & 93 & .2729 & 96 & .2412 & 95 & .2617 & 313 & .1867 & 92 & .2639\\
		\cline{2-13}
		&$2C_n^d$ & 50& 12 & .8817 & 13 & .8660 & 12 & .8830 & 47 & .6575 & 12 & .8912\\
		&&$100$& 16 & .9259 & 19 & .9061 & 16 & .9153 & 74 & .6985 & 16 & .9229\\
		\hline
		$ 100$& $1.2C_n^d$ &$50$& 48 & .2235 & 49 & .2093 & 48 & .2146 & 223 & .1729 & 49 & .2191\\
		&&$100$& 97 & .2387 & 97 & .2236 & 95 & .2320 & 460 & .1723 & 96 & .2487\\
		\cline{2-13}
		& $ 2C_n^d$ &$50$ & 13 & .8829 & 13 & .8678 & 13 & .8734 & 70 & .6289 & 13 & .8743\\
		&&$100$& 18 & .9150 & 19 & .9081 & 18 & .9115 & 113 & .6738 & 18 & .9160\\
		%\
		\hline
		$ 500$& $1.2C_n^d$ &$50$ & 50 & .2160 & 48 & .2132 & 49 & .2168 & 518 & .1711 & 50 & .2198\\
		&&$100$& 97 & .2454 & 99 & .2375 & 96 & .2399 & 973 & .1761 & 97 & .2412\\
		\cline{2-13}
		& $ 2C_n^d$ &$50$ & 13 & .8771 & 13 & .8617 & 13 & .8780 & 150 & .6139 & 13 & .8726\\
		&&$100$& 18 & .9166 & 18 & .9185 & 18 & .9075 & 249 & .6513 & 18 & .9090\\
		%\noalign{\smallskip}
		\hline
		%\noalign{\smallskip}
		$ 1000$& $1.2C_n^d$ &$50$ & 49 & .2202 & 51 & .2136 & 49 & .2159 & 700 & .1632 & 49 & .2156\\
		&&$100$& 98 & .2470 & 97 & .2338 & 97 & .2429 & 1383 & .1616 & 96 & .2366\\
		\cline{2-13}
		&$2C_n^d$ & 50 & 13 & .8797 & 13 & .8728 & 13 & .8729 & 214 & .6128 & 13 & .8674\\
		&&$100$& 19 & .9116 & 19 & .9134 & 19 & .9093 & 360 & .6551 & 19 & .9124\\
		\noalign{\smallskip}
		\hline
	\end{tabular}
	%\end{center}
\end{table}

The values corresponding to  $I_d$ and $\mSigma_4^d$ are the highest, being those of the identity slightly better. The worst results (and the highest number of required projections) are obtained for $\mSigma_3^d$; the remaining ones being similar to those corresponding to the identity. Obviously when $l_{I}^1$ increases, so does the probability to detect the outliers. We also see an increase of the power when $n$ becomes larger  and a slight decrease when $d$ becomes larger. This makes sense because for larger values of $n$, the estimation of the parameters is more accurate, while the larger $d$, the greater the noise in the sample.

Other features of the procedure are analysed  later. In particular, the masking and  swamping effects as well as the size of the outliers are analysed in Subsection \ref{Subs.MaskingSwamping} and  the proportion of observations wrongly classified as outliers in Subsection \ref{sec:comparison}. Finally, it is also worth seeing the effect in the analysis of large outliers. This task is done in Subsection \ref{sec:wine} using real data because the data set analysed there contains a quite large outlier and we compare the result of the analysis with and without this point.

\subsection{Masking and Swamping} \label{Subs.MaskingSwamping}

In order to analyse the masking and the swamping effects in our procedure, we have generated samples with sizes $n=50,100$ and dimensions $d=50, 500, 1000$ for the covariance matrices $I_d$ and $\mSigma_i^d, i=1,\ldots, 4$. Those samples contain 10\% outliers.  More precisely, when $n=50$ (resp. $n=100$), we take one point (resp. two) with the distribution of $\bX$ given $\Vert\bX\Vert_\mSigma=rC_n^d$ with $r=1.05,1.25,2,3$. 

Moreover, since the introduced outliers have different sizes, this analysis is also useful to analyse the effect of the size of the outliers in their detection.

The results of those simulations appear in Table  \ref{tab:masking_swamping} in Subsection \ref{subsubsec:Detecting_Outliers} in the Appendix. More exactly, this table shows the proportion of regular observations which are incorrectly identified as outliers (swamping effect) in column $<C_n^d$ and, also, the proportion of genuine outliers which are declared as outliers (masking effect) in the rest of the columns. We see that the swamping effect increases with the dimension and depends on the covariance matrix ($\mSigma_3^d$ is the worst case), but it is always small (bellow 0.03 excepting when $\mSigma=\mSigma_3^d$ in which the value 0.05 happens once). This phenomenom decreases with $n$. Obviously, the proportion of real outliers declared as outliers increases with their Mahalanobis norm and also, there is a slight increase when $d$ or $n$ increase. 

More precisely, when we have points with Mahalanobis norm close to the limit value $C_n^d$, for instance $1.05C_n^d$, we have proportions of detected outliers from 0.061 to 0.1030 (depending on  $\mSigma$, the dimension and the sample size), while when we move away from that, for instance to $3C_n^d$, these proportions increase to values from 0.7840 to 0.9775. In addition, we see that $\mSigma_3^d$ gives the slowest increase.

%--------------------------------------------------------%
\subsection{Comparison with other procedures}\label{sec:comparison}
%--------------------------------------------------------%

Here we  compare our method (denoted RP) with other ones proposed for high-dimensional data, such as the principal component outlier detection  (PCOut), Filzmoser et al. \cite{Filzmoser2008}, and the minimum diagonal product  (MDP), Ro et al. \cite{Ro2015}. 

Main interest in this subsection is twofold: first  to check how $d$ and $\mSigma$ affect  those methods, second  to see the capability of the procedures to detect multiple outliers once the parameters have been fixed to have a similar behaviour under the null.

To this, we use two settings: in the first one we handle clean samples and compute how many points  are declared as outliers. In the second one the  samples contain 10\% outliers and analyse the proportion of them which are detected by the procedures.

In both settings, we have employed $n=50, 100$, $d=50,500,1000$ and seven covariance matrices: first one is the identity, the second one is  $S_2 = (e^{-|i-j|/d})$. Then, we generate a matrix $A$ whose elements are iid $N(0,1)$ and  take $S_3 = A'A$. Remaining  matrices are the $\mSigma_i^d$'s   defined in Subsection \ref{subsec:a_b_Other}. The results  obtained with the $\mSigma_i^d$'s are in Subsection \ref{subsubsec:Comparison_MDP_PCOut} in the  Appendix.  We report here those corresponding to $I_d, S_2$ and $S_3$,  covering a  situation with independent marginals, another one with relatively high correlations and a third one with   randomly chosen correlations.

In the three settings that we handle here, we have generated  data for $\mSigma=I_d$ and we have multiplied them by the appropriate matrix to obtain the desired covariance; thus, somehow, we handle the same data with the three covariance matrices. We have done 500 simulations. Matrix $A$ varies from simulation to simulation.

PCOut and MDP are implemented in the functions {\tt pcout} and  \texttt{rmdp} in the R packages \texttt{mvoutlier} and \texttt{Rfast}, respectively. 
We have kept the default parameters of those functions excepting that when we use {\tt rmdp}, we fix {\tt itertime} = $d^{1.5}$ in order to keep the suggestion of the help that this parameter should be similar to $d$ for sample sizes equal to $50$, from where we have concluded that for higher sample sizes, the number of iterations should be greater than the sample size. Regrettably, this makes MDP quite slow and we do not report their  results when $d=1000$ because it  took 364.18 seconds to compute five values when $n=50$ in the first setting. 

The default options of the functions  {\tt pcvout} and  \texttt{rmdp} lead to a claim of around 10\% of outliers in the clean samples. Thus, for each pair $n,d$, we have fixed the parameters $a,b$ for RP in order to declare around this percentage of outliers. This is achieved taking $a,b$ such that $E\left({L}_n \ \vert \ \Vert\bX\Vert={q_d^n} \right)= 50$ and $\Prob( \bX \mbox{ declared outlier}   |$ $ \|\bX\| = q_d^n) = 0.1$, where $q_d^n$ is the 0.75-quantile of the square roots of a random sample with size $n$ taken from a $\chi^2_d$. Those parameters have been used in both settings.

The results obtained when using the covariance matrices $\mSigma_i^d$  are similar to those obtained when $\mSigma=S_3$. Those cases are handled as described before, excepting for the fact that we have used a randomly chosen basis in order to prevent the matrices $\mSigma_i^d$ being diagonal. We did not this before because RP is invariant against those rotations. However, on the one hand, it seems that MDP may depend on when  $\mSigma$ is diagonal or not; and, on the other hand, the first step in PCOut is to standardise the data, thus making all cases in which $\mSigma$ is diagonal equivalent to $\mSigma=I_d$.

\subsubsection{Handling a clean sample}

Here we generate a sample from a $N_d(\mathbf{0},\mSigma)$ without outliers and compute the proportion of the points in the sample the procedures declare as outliers. Since there are no outliers in the sample, no observation should be declared as outlier. However, the proportion of outliers is not interesting here (because you can get the right proportion tuning appropriately the parameters). Here, we are only interested in detecting the stability of the procedures; more precisely in seeing if the dimension or the  covariance matrix  affect to the capacity of the procedures to detect outliers. 

The conclusion of those simulations (see Tables \ref{tab:Comparison.Other.Procedures.1} and   \ref{tab:Comparison.Other.Procedures.App.1} in  Subsection \ref{subsubsec:Comparison_MDP_PCOut} in the  Appendix) seems to be that the behaviour of MDP is very different depending on when $\mSigma$ is diagonal or not and, when $\mSigma \neq I_d$, the dimension also affects its behaviour. The increment of the sample size decreases the number of wrongly detected outliers.

PCOut and RP are quite stable when the dimension varies, in spite of PCOut tends to declare more outliers when $n=50$. This effect is more noticeable in the results in Table \ref{tab:Comparison.Other.Procedures.App.1}. Additionally, PCOut seems to declare less outliers when the dependence is not too strong while the oposite happens with RP. Overall, results from RP are more stable than those from MDP or PCOut.

\begin{table}
	\caption{Proportion of outliers found in a clean data set for several covariance matrices. }
	\label{tab:Comparison.Other.Procedures.1}
	%\begin{center}
	\begin{tabular}{rr|ccc|ccc|ccc}
		\hline\noalign{\smallskip}
		& &\multicolumn{3}{c}{MDP}&\multicolumn{3}{|c}{PCOut} & \multicolumn{3}{|c}{RP}
		\\
		$n$ & $d$ & $I_d$ & $S_2$ & $S_3$ & $I_d$ & $S_2$ & $S_3$  &  $I_d$ & $S_2$ & $S_3$ 
		\\
		%\noalign{\smallskip}
		\hline
		%\noalign{\smallskip}
		$50$ & $50$ & .1360  & .1190  & .1158&  .1025 & .1377 & .1110  & .1108 &  .0909 &  .1085
		\\
		& $500$& .1404  & .0320  & .0585  & .0950 & .1308 & .1003  & .1149 &  .1003 &  .1101 
		\\
		& $1000$& --- & --- & --- &   .1028 & .1317 & .1018  & .1132 &  .1000 &  .1141 
		\\
		%\noalign{\smallskip}
		\hline
		%\noalign{\smallskip}
		$100$ & $50$& .0735 & .0896 & .0739 & .1022 & .1219 & .1086  &.1044  &  .0790 &  .1020 
		\\
		& $500$& .0827 & .0187 & .0498 &  .0829 & .1235& .0813  & .1104  &  .0810 &  .1104 
		\\
		& $1000$& --- & ---  & --- & .0787 & .1232 & .0808& .1108  &  .0830 &  .1098 
		\\
		\noalign{\smallskip}
		\hline
	\end{tabular}
	%\end{center}
\end{table}

\subsubsection{Handling a sample with 10\% outliers}

Here we generate a clean sample with size $.9n$  from a $N_d(\mathbf{0},\mSigma)$ and we add $n_{out}=.1n$ outliers with distribution  $N_d(\mathbf{0},\mSigma)$ given that $\|\bX\|_\mSigma= p_i, i=1,\ldots, n_{out}$; where we take $q_i, i=1,\ldots, n_{out},$ an equispaced sequence from $.95$ to $.99$ and, then, the $p_i$'s are the square roots of the $q_i$'s-quantiles of the $\chi^2_d$ distribution. 

Tables \ref{tab:Comparison.Other.Procedures.2} and \ref{tab:Comparison.Other.Procedures.App.2} (last one in Subsection \ref{subsubsec:Comparison_MDP_PCOut} in the  Appendix) show the  proportion of outliers which were correctly identified along  500 repetitions; thus, the higher the proportions, the better.  MDP does a good work when $\mSigma =I_d$, with better results than PCOut, but its behaviour seems to deteriorate in the other two situations in Table \ref{tab:Comparison.Other.Procedures.2}, mostly when $d$ increases. In the situations handled in Table \ref{tab:Comparison.Other.Procedures.App.2}  this  method gives the best results when  $\mSigma=\mSigma_3^d$. It is not too bad when $d=50$ with the remaining matrices, but its behaviour deteriorates noticeably when $d$ increases. 

Broadly speaking, we can say that PCOut is the winner when $\mSigma=S_2$ while RP is the choice in the remaining cases with $\mSigma\neq \mSigma_3^d$. Those results suggest that, on highly dependent situations, the user could benefit from using PCOut; while he should use RP in no so dependent ones.

\begin{table}
	\caption{Samples contain 10\% of real outliers. Columns show the proportion of them correctly identified. }
	\label{tab:Comparison.Other.Procedures.2}
	
	\begin{center}
		\begin{tabular}{rr|ccc|ccc|ccc}
			\hline\noalign{\smallskip}
			& &\multicolumn{3}{c}{MDP}&\multicolumn{3}{|c}{PCOut} & \multicolumn{3}{|c}{RP}
			\\
			$n$ & $d$ & $I_d$ & $S_2$ & $S_3$  & $I_d$ & $S_2$ & $S_3$  &  $I_d$ & $S_2$ & $S_3$ 
			\\
			%\noalign{\smallskip}
			\hline
			%\noalign{\smallskip}
			$50$ & $50$ & .1959 &  .1216&  .1357  & .1564 & .2912 & .1684  & .2856  &  .2032 &  .2844 
			\\
			& $500$& .1842 & .0340 & .0688  & .1196 & .1756 & .1040  & .1568  &  .1056 &  .1420 
			\\
			& $1000$& --- & --- & --- &   .1060 & .1552 & .1112 &  .1576  &  .1092 &  .1424 
			\\
			%\noalign{\smallskip}
			\hline
			%\noalign{\smallskip}
			$100$ & $50$& .1301 & .0902 & .0905 & .2112 & .3120 & .2282  & .3076  &  .1816 &  .2864
			\\
			& $500$&.1424 & .0185 & .0610  & .0856 & .1808 & .0928  & .1790  &  .1138 &  .1642 
			\\
			& $1000$& --- & --- & --- &  .0812 & .1598 & .0852  &  .1478  &  .1064 &  .1526 
			\\
			\noalign{\smallskip}\hline
		\end{tabular}
	\end{center}
\end{table}

%--------------------------------------------------------%
\subsection{The procedure in practice: Two real data examples}\label{sec:real_data}
%--------------------------------------------------------%
The practical relevance of the proposed test is illustrated on two well-known real data sets. They have been studied by Hubert et al. \cite{Hubert2015}. Those data are functional; however, all observations in both two  sets have been measured on the same values of the independent variable, and they can be also considered as $d$-dimensional. 

We compute $a$ and $b$  as in Section \ref{subsec:a_b_id} with $\Prob \left(\left\vert\tilde Y^{L_n}\right\vert > b \vert\left\Vert\bX\right\Vert_{\mSigma}=C_n^d\right)=0.05 $, $l_I^1=100$ and $N=10^6$. Consequently, a point $\bx$ such that  $\left\Vert\bx\right\Vert_{\mSigma}=C_n^d$ should be identified as outlier 5\% of times. 

We have applied the method $T=100$ times to every point in the sample at hand and we have declared as outliers those points who were identified as outliers 5\% of times or more, following the procedure described in Subsection \ref{Subs:IdentifyingOutliers}.

In the analysis we show the outliers identified by the procedures introduced in this paper (denoted RP),  in Hubert et al. \cite{Hubert2015} (denoted  Hub), in Filzmoser et al. \cite{Filzmoser2008}, (denoted PCOut), and Ro et al. \cite{Ro2015} (denoted MDP). PCOut and MDP are handled with their default parameters, excepting that we take 
{\tt itertime} = $d$ in MDP according to the suggestion that this value should be similar to the dimension when $n=50$. 
%--------------------------------------------------------%
\subsubsection{Wine Data}\label{sec:wine}
%--------------------------------------------------------%
This dataset contains the proton nuclear magnetic resonance  spectra of 40 different wine samples, Larsen et al. \cite{Larsen}. As in Hubert et al. \cite{Hubert2015}, we select the  region between wavelengths 5.37 and 5.62, on which each sample has $d=397$ measurements. 

\begin{figure}[!htb]
	\begin{center}
		\includegraphics[width=0.49\textwidth]{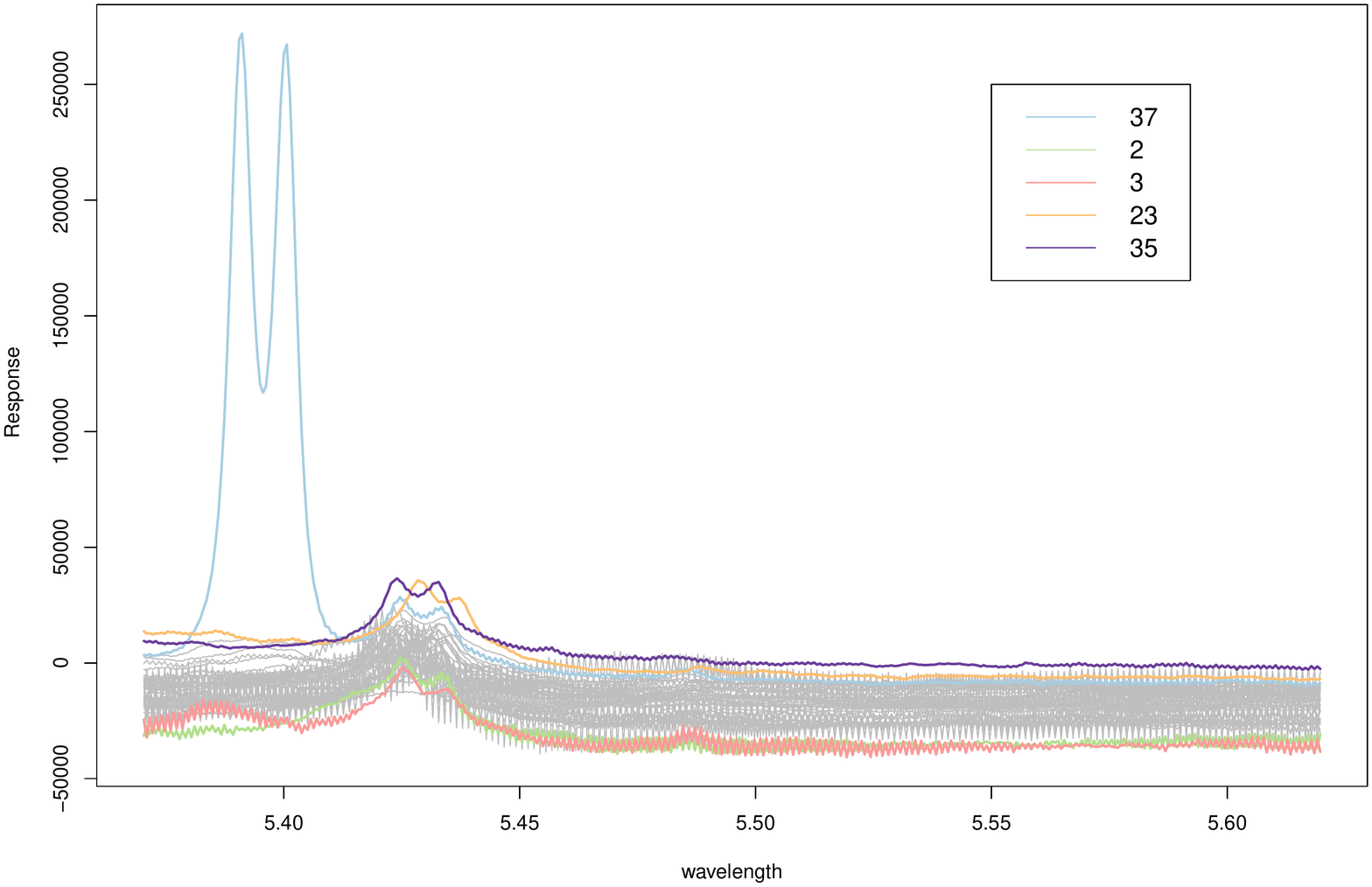}
		\includegraphics[width=0.49\textwidth]{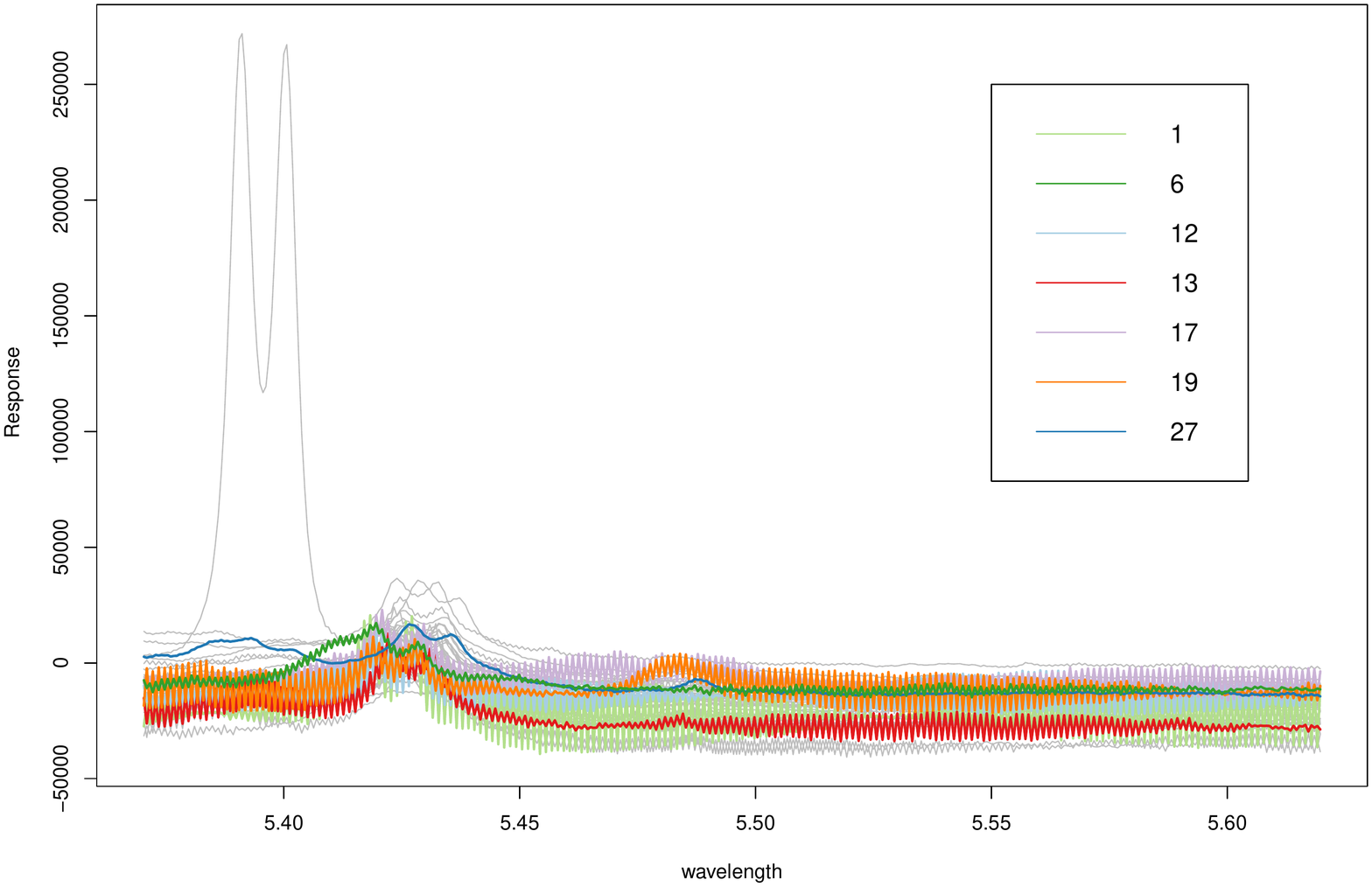}
		\caption{The left panel shows the outliers which are also detected in \cite{Hubert2015}. The right panel shows the outliers detected with the proposed method but not  in \cite{Hubert2015}.}
		\label{fig:out_wine}
	\end{center}
\end{figure}

Table \ref{tab:Wine.Outliers} shows the data identified as outliers by the considered procedures. Those curves are represented in Figure \ref{fig:out_wine} with coloured lines. We see that the curve 37 has large peaks around wavelength 5.4 and may be considered an isolated outlier. To see if this curve has a big effect in the analysis (I.e. to give an idea of the robustness of the method), we have carried out the same procedure eliminating it. The results of RP, Hub and MDP are similar to those in Table \ref{tab:Wine.Outliers} (see Table \ref{tab:Wine.Outliers_no37}), however they are not so for PCOut method. Therefore, in this case, we can conclude that including extreme outliers, as the curve 37, does not affect too much the result of our procedure.

\begin{table}[!htb]
	\caption{Wines identified as outliers. Each number in row RP (resp. PCOut) is the proportion of times this wine was declared  outlier by RP (resp. the  weight of this wine. Low weights identify potential outliers). }
	\label{tab:Wine.Outliers}%set.seed(123875)
	\begin{center}
	\begin{tabular}{r|ccccccccccccc}
		\hline\noalign{\smallskip}
		& 1 & 2  & 3 & 6  & 12  & 13  & 17  & 18 & 19  & 23  & 27  & 35  & 37 
		\\
		\hline
		%\
		RP& 0.89 \hspace*{-2mm}&\hspace*{-2mm} 0.16 \hspace*{-2mm}&\hspace*{-2mm} 0.24 \hspace*{-2mm}&\hspace*{-2mm} 0.05 \hspace*{-2mm}&\hspace*{-2mm} 0.67 \hspace*{-2mm}&\hspace*{-2mm} 0.32 \hspace*{-2mm}&\hspace*{-2mm} 0.63 \hspace*{-2mm}& &\hspace*{-2mm} 0.64 \hspace*{-2mm}&\hspace*{-2mm}  0.61 \hspace*{-2mm}&\hspace*{-2mm}  0.06 \hspace*{-2mm}&\hspace*{-2mm} 0.19 \hspace*{-2mm}&\hspace*{-2mm}  1.00 
		\\
		[2mm]
		Hub & 	& X & X & & & & & & & X& & X & X	
		\\
		[2mm]
		PCOut& 0.04 \hspace*{-2mm}& 0.06 \hspace*{-2mm}& 0.16\hspace*{-2mm} & 0.20\hspace*{-2mm} & 0.04\hspace*{-2mm} & 0.08\hspace*{-2mm} & & 0.16\hspace*{-2mm}& 0.04\hspace*{-2mm} & 0.14\hspace*{-2mm}& & & 0.04\hspace*{-2mm}
		\\
		[2mm]
		MDP& & & & & & & & & & & & &  X
		\\	
		\noalign{\smallskip}\hline
	\end{tabular}
	\end{center}
\end{table}

\begin{table}[!htb]
	\caption{Wines identified as outliers deleting the wine 37. The description of Table \ref{tab:Wine.Outliers} applies. }
	\label{tab:Wine.Outliers_no37}%set.seed(123875)
	\begin{center}
		\begin{tabular}{r|cccccccccccccccc}
			\hline\noalign{\smallskip}
			& 1 & 2  & 3 & 6  & 9 & 12  & 13 & 14 & 17 & 18  &  19 & 20 &  23  & 27 & 29  & 35  
			\\
			%\noalign{\smallskip}
			\hline
			%\noalign{\smallskip}
			RP \hspace*{-2mm}&  0.92 \hspace*{-2mm}&\hspace*{-2mm}0.19 \hspace*{-2mm}&\hspace*{-2mm}0.30 \hspace*{-2mm}&\hspace*{-2mm}0.08 \hspace*{-2mm} & &\hspace*{-2mm}0.57 \hspace*{-2mm}&\hspace*{-2mm}0.21 \hspace*{-2mm}& & \hspace*{-2mm}0.47 \hspace*{-2mm}& & \hspace*{-2mm}0.59 \hspace*{-2mm}& &\hspace*{-2mm}0.57 \hspace*{-2mm}&\hspace*{-2mm}0.08 \hspace*{-2mm}& &\hspace*{-2mm}0.26 \hspace*{-2mm}
			\\
			[2mm]
			Hub& & X & X &  &  &  &  & & &  & & & X &  & & X
			\\
			[2mm]
			PCOut \hspace*{-2mm}&0.04 \hspace*{-2mm}& & &\hspace*{-2mm}0.20 \hspace*{-2mm}&\hspace*{-2mm}0.52\hspace*{-2mm} & \hspace*{-2mm}0.04\hspace*{-2mm} &\hspace*{-2mm}0.08\hspace*{-2mm} &\hspace*{-2mm}0.66\hspace*{-2mm} &  &\hspace*{-2mm}0.16\hspace*{-2mm}&\hspace*{-2mm}0.04\hspace*{-2mm} &\hspace*{-2mm}0.35\hspace*{-2mm}&\hspace*{-2mm}0.14\hspace*{-2mm}&\hspace*{-2mm}0.78\hspace*{-2mm} &\hspace*{-2mm}0.27\hspace*{-2mm} &
			\\
			[2mm]
			MDP& & & & & & & & & & & & & & & &  
			\\	
			\noalign{\smallskip}\hline
		\end{tabular}
	\end{center}
\end{table}

\begin{figure}[!htb]
	\begin{center}
		\includegraphics[width=.5\textwidth]{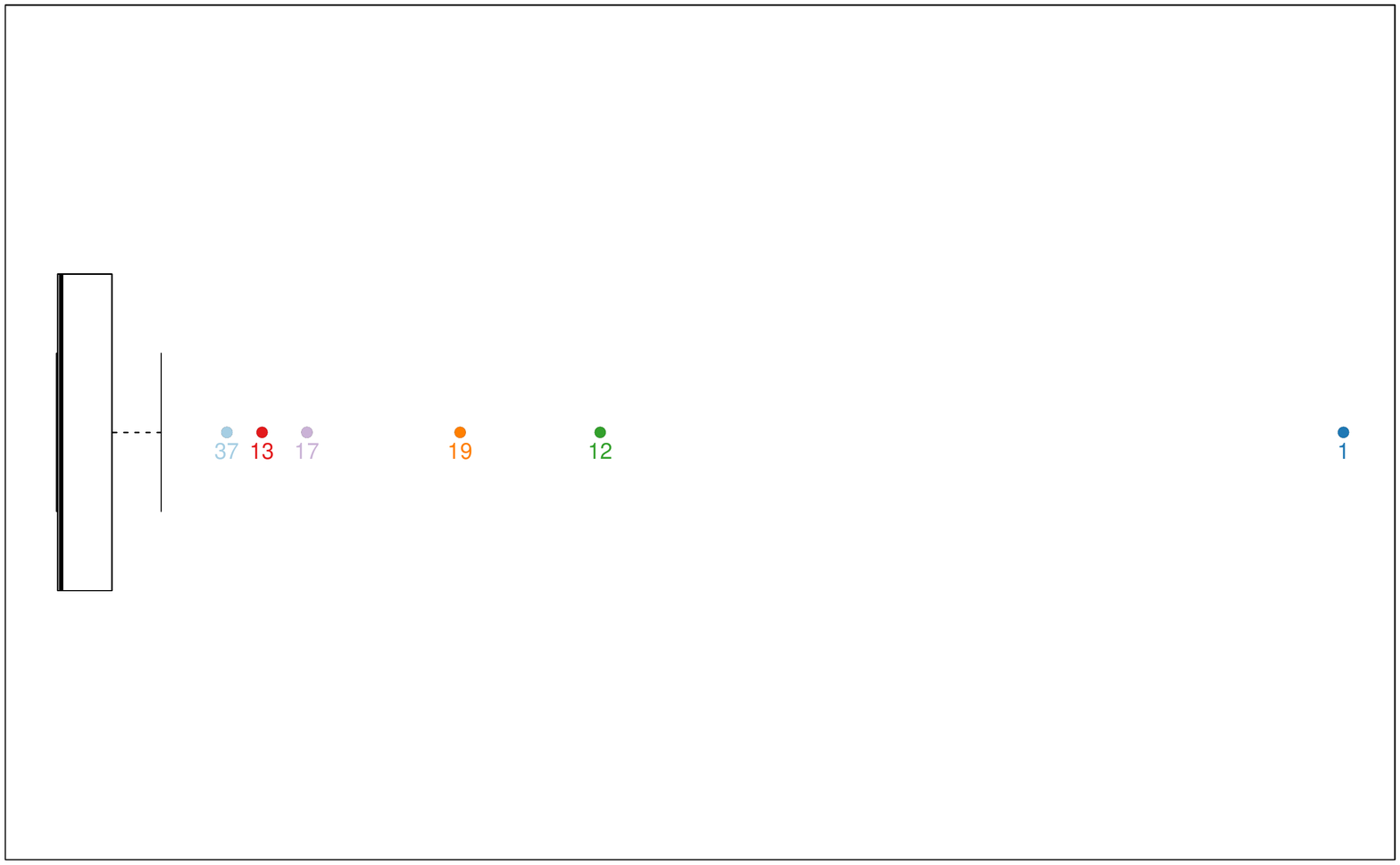}
		\caption{Boxplot of the squared of the differences among the components of each point in the wine data.}
		\label{fig:boxplot}
	\end{center}
\end{figure}

The curves 1, 12, 17 and 19 oscillate too much, as shown in Figure \ref{fig:boxplot}, where  the boxplot of the indicators of the oscillation $\sum_{j=1}^d(X^i_{j+1}-X^i_j)^2$ computed for each point $\bX^i, i=1,\ldots, 40$ appear. Unlike Hub, RP and PCOut declare them as outliers (except for 17 which is not declared by PCOut): RP with probability greater than 0.6 and PCOut with weights 0.04 (weights close to zero indicate potential outliers). 

RP, PCOut and Hub also declare the curves 2, 3 and 23 as outliers. Curve 35 is only identified by RP and Hub. Figure \ref{fig:out_wine} shows that those curves are in the external part of the bulk of the data: 2 and 3 in the bottom and 23 and  35 in the top part. RP and PCOut additionally detect 13 as  outlier;   this curve is in the bottom part of the  data just above of 2 and 3 (see  Figure \ref{fig:out_wine}).  RP detects the curves 6 and 27 (in coloured lines in Figure \ref{fig:out_wine}), PCOut only the curve 6, and  Hub none of them. We see that curve 6  starts to increase before the other curves; while  27 has a similar shape of the curve 3 (which is declared as an outlier by Hub) but in the top part of the data. However, the number of times these curves have been detected by RP (well below the .95-quantile of a binomial with parameters 100 and .05, which is 9) make them doubtful as outliers from the RP point of view. PCOut gives the maximum weight, 0.2, to  6, i.e. among all the outliers that PCOut detects, this curve is the least anomalous.

\begin{figure}[!htb]
	\begin{center}
		\includegraphics[width=0.49\textwidth]{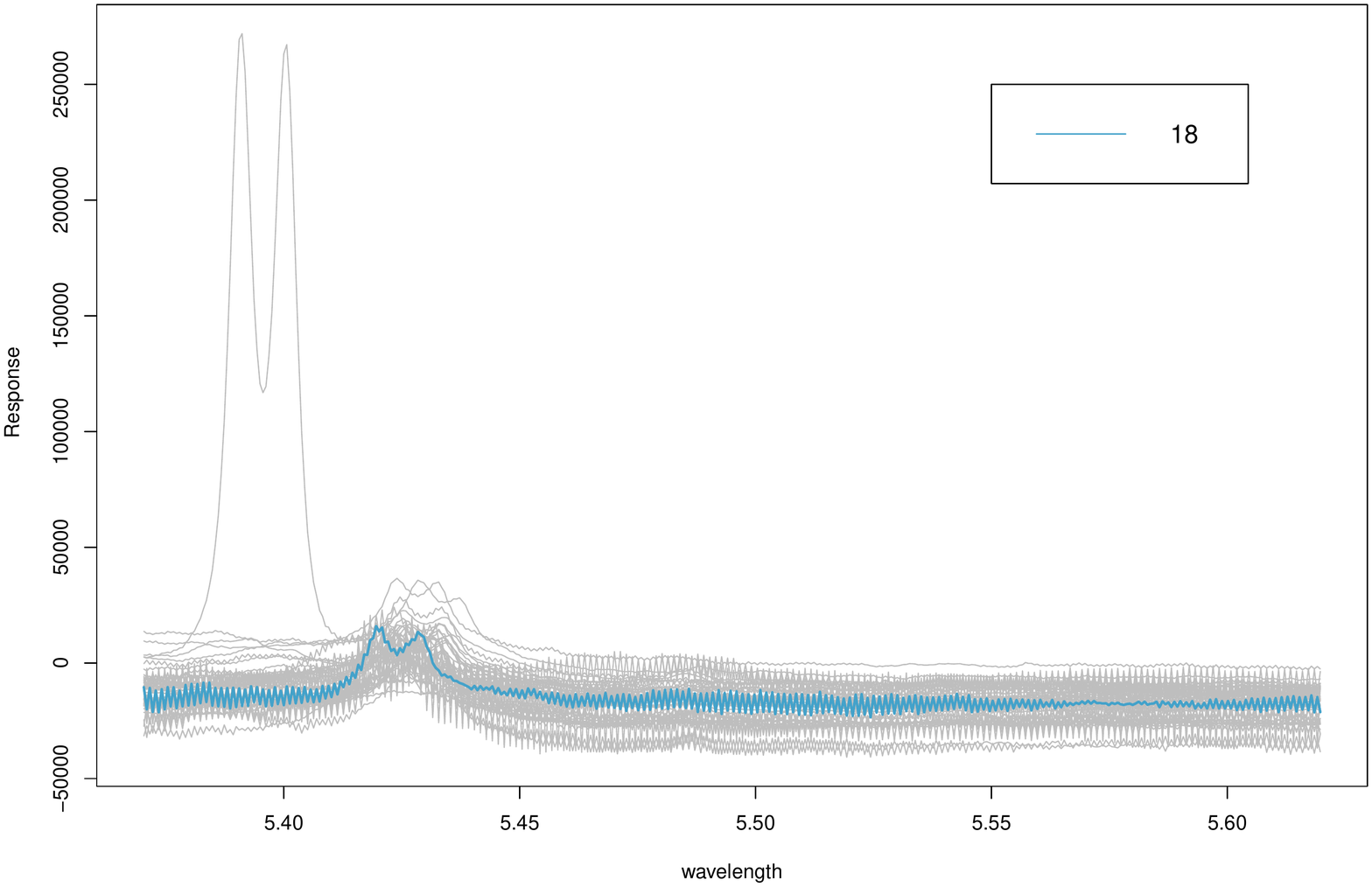}
		\includegraphics[width=0.49\textwidth]{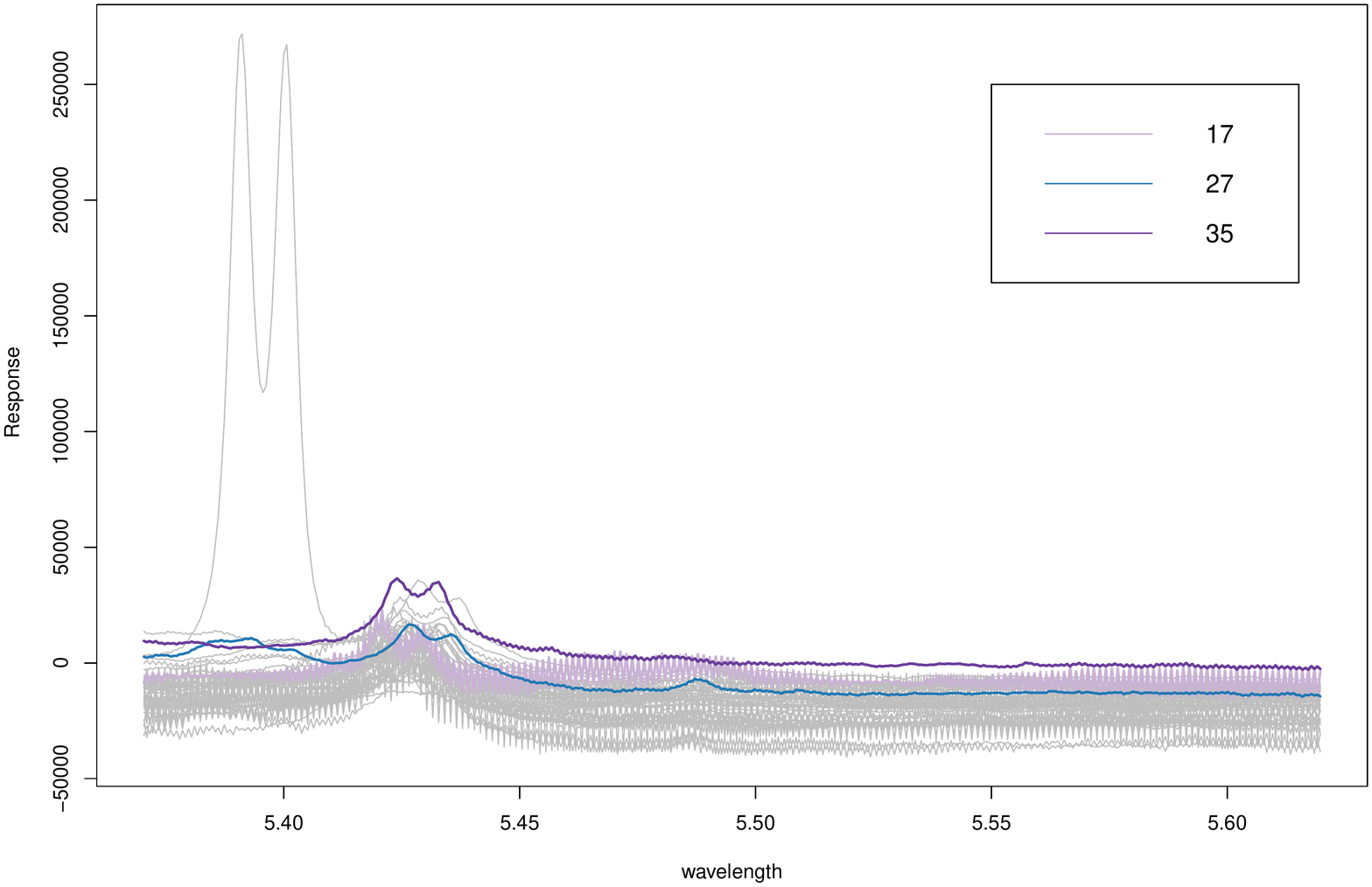}
		\caption{The left panel shows the outlier which was  detect by PCOut but not with our method. The right panel shows the outliers detected  with our method but not with PCOut.}
		\label{fig:out_wineFilz}
	\end{center}
\end{figure}

The difference between the detected curves by PCOut and RP is that PCOut detects the curve 18 (with the same weight that curve 3), and RP detects the curves 17, 27 and 35. Figure \ref{fig:out_wineFilz} shows these curves. It seems the curve 18 has some fluctuation however this curve does not appear as outlier in the boxplot in Figure \ref{fig:boxplot}. 

In conclusion, it seems that PCOut and RP detect better the shape outliers than Hub. RP also detects curves that have little peculiarities or those which are in the border of the bulk of the data albeit with lower probability.

%--------------------------------------------------------%
\subsubsection{Octane data}\label{sec:octane}
%--------------------------------------------------------%
This data set  consists of 39 near infrared  spectra of gasoline samples over $d=226$ wavelengths ranging from 1102 nm to 1552 nm with measurements every two nm. It is known that samples 25, 26 and 36-39 have a very different spectrum because they contain added ethanol, Esbensen et al. \cite{Esbensen}, Rousseeuw et al. \cite{Rousseeuw_2006} and Hubert et al. \cite{Hubert2015}. 
Table \ref{tab:Octane.Outliers} shows the data identified as outliers by the considered procedures.

\begin{table}[!htb]
	\caption{Outliers in the gasolines. Each number in row RP (resp. PCOut) is the proportion of times this wine was declared  outlier by RP (resp. the  weight of this wine. Low weights identify potential outliers). }
	\label{tab:Octane.Outliers}%set.seed(123875)
	%\begin{center}
	\begin{tabular}{r|ccccccccc}
		\hline\noalign{\smallskip}
		&  6 & 23 & 25 & 26 & 34 & 36 & 37 & 38 & 39
		\\
		\hline
		RP&   0.11 &   0.06 &   0.99 &   1.00 &   0.28 &  1.00 &   1.00 &  1 .00 &  0.99
		\\
		[2mm]
		Hub & 	& & X & X & & X &X & X & X
		\\
		[2mm]
		PCOut & & 0.10 & 0.04 & 0.04 & 0.08 & 0.04 & 0.04 & 0.04 & 0.04
		\\
		[2mm]
		MDP& 	& & X & X & & X &X & X & X 
		\\	
		\noalign{\smallskip}
		\hline
	\end{tabular}
	%\end{center}
\end{table}

\begin{figure*}[!htb]
	\begin{center}
		\includegraphics[width=0.49\textwidth]{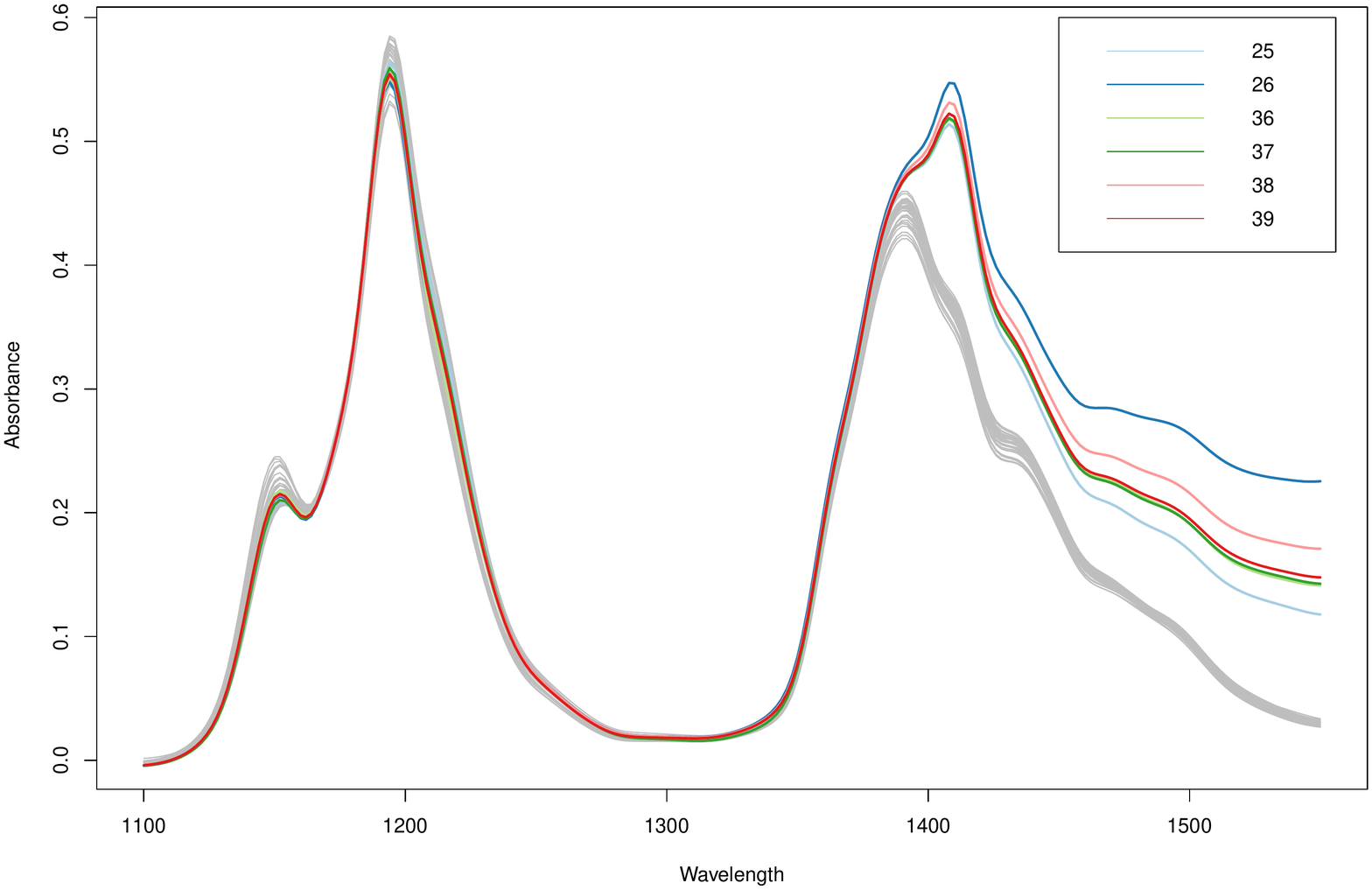}
		%\vspace{0.5cm}
		\includegraphics[width=0.49\textwidth]{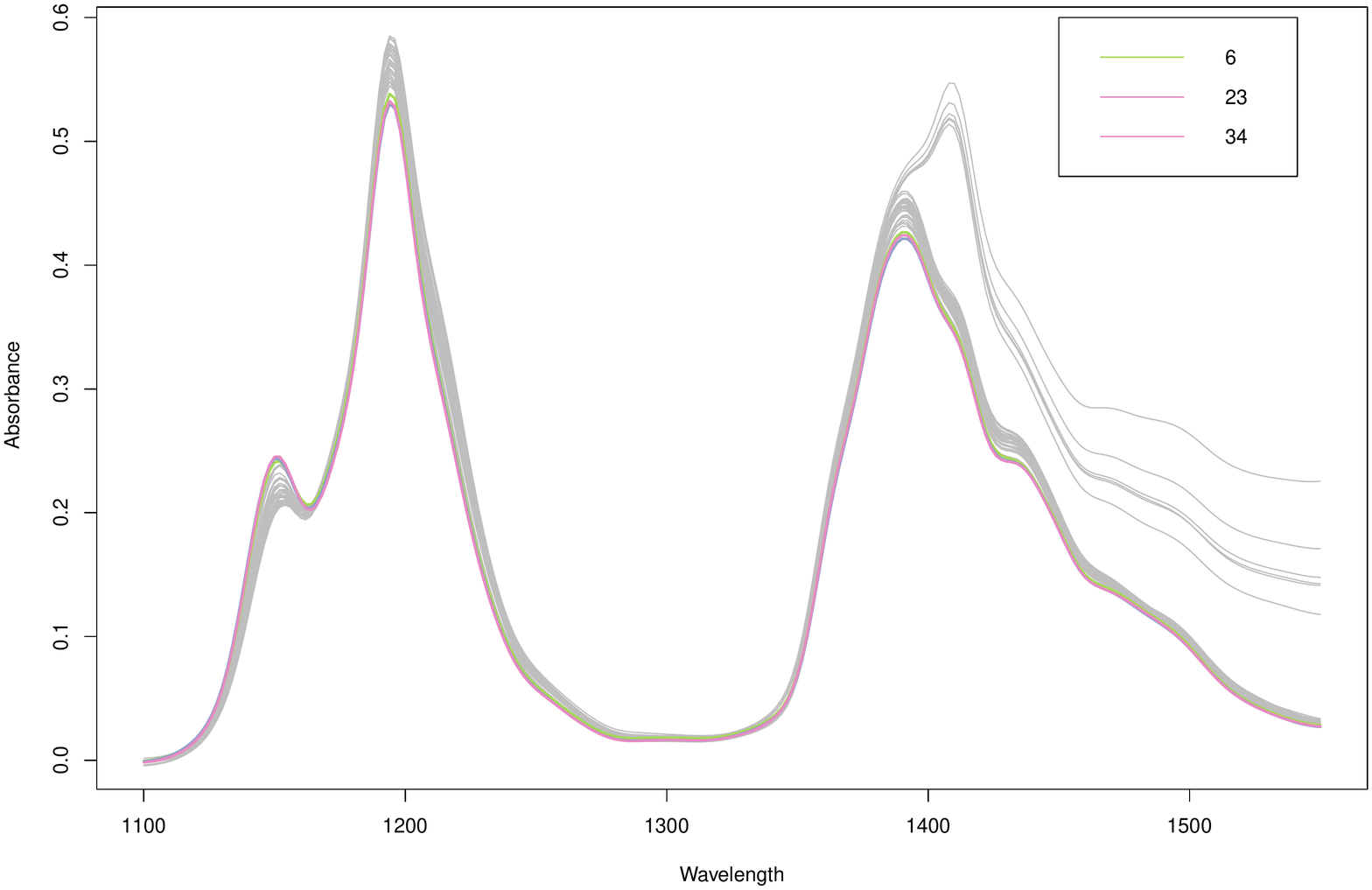}
		\caption{Outliers detected in octane data with the proposed method: the left panel shows the outliers which are also detected in \cite{Hubert2015} and the right panel shows the outliers detected with our method but not with the method proposed in \cite{Hubert2015}.}
		\label{fig:out_octane}
	\end{center}
\end{figure*}

All those curves are plotted with coloured lines in Figure \ref{fig:out_octane}. Curiously, Hub and MDP (resp. PCOut and RP) detect the same curves as outliers.
Clearly the curves 25, 26, 36, 37, 38 and 39, represented in the left panel, are persistently outlying from wavelength 1390 onward and all procedures detect them. The curves 23 and 34, represented in the right panel, are declared outliers by PCOut and RP but not  by Hub and MDP. Additionally, RP detects the curve 6. We see that these three curves are  in the border of the bulk of the data and they are slightly separated from the rest on wavelengths  around 1150,  1195 and 1390. Anyhow, curve 23 is only declared as outlier 6\% of times what makes it doubtful from the point of view of RP. This is the curve with the highest weight, 0.1, when we apply PCOut. 

Similarly to the wine dataset, it seems that PCOut and RP detect the outliers which are far away from the bulk of the data (curves 25, 26 and 36  to 39) and those which always are in the border of the data (23 and 34, and additionally RP detects 6).

\newpage

%-----------------------------------------------%
\section{Appendix I. Proofs}
\label{sec:appendix}
%-----------------------------------------------%
%----------------------------------------------%
\subsection{Proofs of Section \ref{sec:def}}
%----------------------------------------------%

Two lemmas are required to prove Theorem \ref{theo:conv_Cnd}. Here $F^{-1}_{d}$ denotes the quantile function of the distribution $\chi_{d}^2$. Lemma \ref{lemma: Laurent} gives an upper bound of $F^{-1}_{d}$ and is the first part in Lemma 1 of Laurent and Massart \cite{Laurent}. Lemma \ref{lemma: LHopital} follows applying  twice   L'H\^opital's rule.%, 

\begin{lemma}\label{lemma: Laurent}[Laurent and Massart]
	Let $d\geq 1$. Then
	\begin{align*}
	F^{-1}_{d}(s) \leq d+ \log\left(\frac{1}{1-s}\right) + 2\sqrt{d\log\left(\frac{1}{1-s}\right)}, \ s \in (0,1).
	\end{align*}
\end{lemma}

\begin{lemma}\label{lemma: LHopital}
	Let $f$ and $g$ functions such that $\lim \limits_{t\rightarrow \infty} f(t)=\lim \limits_{t\rightarrow \infty} g(t)=0$ and $\lim \limits_{t\rightarrow \infty} \frac{f'(t)}{g'(t)}=c \in \mathbb{R}.$ Then
	$$\lim \limits_{t\rightarrow \infty} \frac{\log(f(t))}{\log(g(t))}=1.$$
\end{lemma}

\begin{proof}[Proof of Theorem \ref{theo:conv_Cnd}]
	Firstly, we obtain the limit when $n\rightarrow \infty$ and $d$ is fixed. From (\ref{eq:def_C_nd}) we have that if $q_{n,d}:=F^{-1}_{d}((1-\delta)^{1/n}),$ then $q_{n,d}=(C_n^d(\delta))^2.$ 
	
	Lower bound: By definition we have $(1-\delta)^{1/n}=\mathcal{P}(d/2,q_{n,d}/2),$ where $\mathcal{P}(a,x):=\gamma(a,x)/\Gamma(a)$ is the regularized lower gamma function, with $\gamma(a,x):=  \int_0^x y^{a-1} e^{-y}\,dy$ the incomplete lower gamma function. For $d=1$, 
	\begin{align*}
	(1-\delta)^{1/n}=\mathcal{P}\left(\tfrac{1}{2},\tfrac{q_{n,1}}{2}\right)=\int_0^{q_{n,1}/2} t^{-1/2}e^{-t}\, dt,
	\end{align*}
	Take $\beta=2$ and $\alpha=\sqrt{e/(2\pi)}$ in Theorem 2 in \cite{Chang} to obtain
	\begin{align*}
	(1-\delta)^{1/n}<1-\sqrt{e/(2\pi)} \exp\{-q_{n,1}\}, 
	\end{align*}
	from where, taking into account that the quantiles of a chi-squared increase with the dimension, we have that for any $d\geq 1$: 
	\begin{align*}
	q_{n,d} > -\log(1-(1-\delta)^{1/n}) +\log(\sqrt{e/(2\pi)}).
	\end{align*}

	Upper bound: By Lemma \ref{lemma: Laurent}, for any $n$ and $d$ we have that 
	\begin{align*}
	q_{n,d} \leq d-2\log\left(1-(1-\delta)^{1/n}\right) + 2\sqrt{-d\log\left(1-(1-\delta)^{1/n}\right)}.
	\end{align*}
	
	As the third term of the above inequality has a lower order than the second one when $n\rightarrow \infty$ and $d$ is fixed, we have that both upper and lower bounds have the same order. Take $f(n):=1-(1-\delta)^{1/n}$ and $g(n):=n^{-1}.$ Both functions $f(n)$ and $g(n)$ trivially go to zero when $n\rightarrow \infty.$ Furthermore
	\begin{align*}
	\lim \limits_{n\rightarrow \infty} \frac{f'(n)}{g'(n)}=-\log(1-\delta).
	\end{align*}
	
	Hence Lemma \ref{lemma: LHopital} gives that $q_{d,n}$ has the same order than $\log(n)$ when $n\rightarrow\infty.$
	
	Secondly, analyse the limit of $C_n^d$ when $d\rightarrow \infty$ and fixed $n \in \mathbb{N}$.  
	Let $Y_d$ be a rv with distribution $\chi_{d}^2.$ Thus, $Y_d$ is the sum of $d$ iid rv's with distribution $\chi_1^2$ whose mean is 1 and whose variance is 2. Then, by the Central Limit Theorem, for $a \in \mathbb{R},$
	\begin{equation}\label{eq:distr_func_C_nd}
	F_{Y_d^{\star}}(a) \rightarrow \Phi(a),
	\end{equation}
	\noindent where $F_{Y_d^{\star}}$ denotes the d.f. of $Y_d^{\star}:=(Y_d-d)/\sqrt{2d}.$ Instead of in a fixed $a,$ we are interested in computing this limit on $a_d:=((C_n^d)^2-d)/\sqrt{2d}.$
	
	Suppose for a contradiction that $\{a_d\}$ is unbounded. Then there exists a subsequence $\{a_{d_k}\}$ such that $\lim_{k \to \infty} a_{d_k}=\infty.$ By \eqref{eq:distr_func_C_nd} since $F_{Y_d^{\star}}$ is increasing, for any $a>0$, we have that 
	\begin{align*}
	1\geq \varlimsup F_{Y_{d_k}^{\star}}(a_{d_k}) \geq \varliminf F_{Y_{d_k}^{\star}}(a_{d_k})\geq \lim F_{Y_{d_k}^{\star}}(a)=\Phi(a).
	\end{align*}
	
	On the other hand, since $\lim_{t \to \infty} \Phi(t)=1,$ we would have that $\lim_{k \to \infty} F_{Y_{d_k}^{\star}}(a_{d_k})=1.$ This is a contradiction because by definition $F_{Y^{\star}_{d_k}}\lrp{\lrp{C_n^{d_k}}^2}=\Prob\lrp{Y^{\star}_{d_k}\leq \lrp{C_n^{d_k}}^2}=(1-\delta)^{1/n}\neq 1$ (remember that $n$ is fixed now). Thus $\{a_d\}$ is bounded.
	
	Suppose now that $\{a_d\}$ does not converge, i.e. suppose that there exist two subsequences $\{d_k^1\}$ and $\{d_k^2\}$ such that $a_{d_k^1}\rightarrow a_1$ and $a_{d_k^2}\rightarrow a_2,$ with $a_1<a_2.$ Let $a_1<x_1<x_2<a_2.$ From an index $k$ onward: 
	\begin{align}\label{eq:sucesiones1}
	(1-\delta)^{1/n}  =  \Prob\left(\chi^2_{d_k^1} \leq \lrp{C_n^{d_k^1}}^2\right)=\Prob\left(Y^{\star}_{d_k^1} \leq a_{d_k^1}\right) \leq \Prob\left(Y^{\star}_{d_k^1} \leq x_1 \right)\rightarrow \Phi(x_1),
	\end{align}
	\begin{align}\label{eq:sucesiones2}
	(1-\delta)^{1/n}  =  \Prob\left(\chi^2_{d_k^2} \leq \lrp{C_n^{d_k^2}}^2\right )=\Prob\left(Y_{d_k^2} \leq a_{d_k^2}\right) \geq \Prob\left(Y^{\star}_{d_k^2}  \leq x_2 \right)\rightarrow \Phi(x_2), 
	\end{align}
	where the convergence follow from \eqref{eq:distr_func_C_nd}. Since \eqref{eq:sucesiones1} and \eqref{eq:sucesiones2} are simultaneously impossible, because $ \Phi(x_1)\neq \Phi(x_2)$, we conclude that   $\{a_d\}$ does converge. 
	
	Let $a:=\lim_{d \to \infty} a_d,$ then $0<(1-\delta)^{1/n} = \lim_{d \to \infty} F_{Y_d^{\star}}(a_d)=\Phi(a)$ and we have that $a\neq 0.$ 
	Then, the result follows from the fact that for any $\epsilon>0$, from an index onward
	\begin{align*}%\label{eq:sucesiones3}
	(a-\epsilon)\sqrt{2d}+d \leq (C_n^d)^2 \leq (a+\epsilon)\sqrt{2d}+d.
	\end{align*}
\end{proof}

%----------------------------------------------%
\subsection{Proofs of Section \ref{subsec:marginal}}
%----------------------------------------------%
To prove Proposition \ref{prop: prob_dist}, the equality $Y^{\bV}=\left(\frac{\bX'\bV}{\mSigma_{\bV}} - \frac{\hatmuv}{\mSigma_{\bV}}\right)\frac{\mSigma_{\bV}}{\hatsigmav}$ leads us to consider the following rv's:
\begin{align}\label{def:rv_separadas}
Y_1 & := \frac{\bX'\bV}{\mSigma_{\bV}},\quad Y_2 :=  \frac{\hatmuv}{\mSigma_{\bV}}, \quad Y_3 :=  \frac{\mSigma_{\bV}}{\hatsigmav}. 
\end{align}

We next obtain the pdf's of those rv's given that $\left\Vert\bX\right\Vert_{\mSigma}=t$, with $t>0$. Since we compute the conditional pdf's given the norm of the point, the rv $Y_1$ does not follow a standard normal distribution. Recall also that the sample mean and the sample variance are calculated using only the sample and therefore, $Y_1$ is the only rv which depends on $\bX$. We need  Lemma \ref{lemma:prop_hyperg} which gives the cdf of the marginal of a uniform distribution on $\Omega_1^{d-1}$.

\begin{lemma}\label{lemma:prop_hyperg}
	Let $\bU=(U_1,\ldots,U_d)'$ be a rv with distribution uniform on $\Omega_1^{d-1}.$ The cdf of $U_1$, denoted by $F_{U_1},$ is given by the following expression:
	\begin{align*}
	F_{U_1}(u)=\sign(u) \frac{1}{2}\mathrm{I}_{u^2}\left(\frac{1}{2},\frac{d-1}{2}\right)+\frac{1}{2}, u \in [-1,1].
	\end{align*}
\end{lemma}

\begin{proof} 	
	By (2.5.11) in Fang and Zhang \cite{Fang}, 
	\begin{align*}
	F_{U_1}(u)=\frac{\Gamma\left(\frac{d}{2}\right)}{\Gamma\left(\frac{1}{2}\right)\Gamma\left(\frac{d-1}{2}\right)}\displaystyle \int_{-1}^u(1-y^2)^{(d-3)/2}\, dy.
	\end{align*}	
	The change of variable $y^2=s$, gives
	
	\begin{align*}
	F_{U_1}(u)=\frac{1}{2\mathrm{B}\left(\frac{1}{2},\frac{d-1}{2}\right)}\left(\sign(u) \mathrm{B}\left(u^2;\frac{1}{2},\frac{d-1}{2}\right)+\mathrm{B}\left(\frac{1}{2},\frac{d-1}{2}\right) \right).
	\end{align*}
	The result is deduced from the definition of the incomplete beta function.
\end{proof}

\begin{lemma}\label{prop: pdf_separadas}
	Under assumptions (A1) and (A2), the pdf's of the rv's $Y_1$, $Y_2$, $Y_3$, defined in \eqref{def:rv_separadas} given that $\left\Vert\bX\right\Vert_{\mSigma}=t$ with $t>0$, are
	\begin{align*}
	f_{Y_1}^t(u)& = \left(\mathrm{B}\left(\tfrac{d-1}{2},\tfrac{1}{2}\right)\right)^{-1}t^{2-d} (t^2-u^2)^{(d-3)/2}, \mbox{ if } u \in [-t,t] \mbox{ and null otherwise, } \\
	f_{Y_2}(u) & =  \left(\frac{n}{2\pi}\right)^{1/2} \exp \{-nu^2/2 \}, u \in \mathbb{R}, \\
	f_{Y_3}(u) & =  \frac{(n-1)^{(n-1)/2}}{2^{(n-3)/2}\Gamma\left(\tfrac{n-1}{2}\right)} u^{-n}\exp\left\{-\tfrac{n-1}{2u^2}\right\}, \mbox{ for } u\in [0,\infty) \mbox{ and null otherwise. } 
	\end{align*}
	
\end{lemma}

\begin{proof}
	Firstly, fix $\bV=\bv$. Using Lemma \ref{lemma:prop_hyperg},  it  is easily seen that the pdf of $Y_1$ given $\bv$ and that $\Vert\bX\Vert_{\mSigma}=t$  coincide with the expression we propose for $f_{Y_1}^t$.
	Secondly, for $Y_2$, since $\hat{\mu}_{\bv}$ follows a $N_1(0,\mSigma_{\bv}^2/n)$ distribution, then the rv $\hat{\mu}_{\bv}/\mSigma_{\bv}$ follows a $N_1(0,1/n)$ distribution. For $Y_3$, it is known that $\hat{\mSigma}_{\bv}^2 (n-1)/\mSigma_{\bv}^2$ follows a  $\chi^2_{n-1}$ distribution. Then, a change of variable gives that the pdf of $Y_3$ given $\bv$ is $f_{\chi^2_{n-1}}((n-1)u^{-2})2(n-1)u^{-3}$, which writing the expression of $f_{\chi^2_{n-1}}$ gives the function we propose for $f_{Y_3}$. 
	The result follows because none of those distributions depend on the chosen $\bv$.% 
\end{proof}

\begin{proof}[Proof of Proposition \ref{prop: prob_dist}]
	The rv's $Y_1$, $Y_2$, $Y_3$ defined in \eqref{def:rv_separadas} are conditionally independent given $\bV$. 
	If $y<0$, then the pdf of the rv $Y^{\bV}$ given $\Vert\bX\Vert_{\mSigma}=t$ is:
	\begin{equation*}
	\begin{split}
	f^t_{Y^{\bV}}(y)&= f^t_{(Y_1-Y_2)Y_3}(y)\\
	& = \displaystyle\int_{\mathbb{R}} f^t_{Y_1-Y_2}(x)f_{Y_3}(y/x)|x|^{-1}\,dx\\
	& = \displaystyle\int_{\mathbb{R}} f_{Y_3}(y/x)|x|^{-1} \left(\displaystyle\int_{\mathbb{R}}f^t_{Y_1}(s)f_{Y_2}(s-x)\,ds \right) \,dx.\\
	\end{split}
	\end{equation*}
	Write the expressions of the pdf's of the rv's $Y_1$, $Y_2$, $Y_3$, given by Lemma \ref{prop: pdf_separadas} and obtain the first equality of this proposition. The reasoning when $y$ is positive is identical.
\end{proof}

Lemma \ref{lemma:bijectiond2} and Corollary \ref{corol:bijectiond2}, which are proved next, will be used in the proof of Proposition \ref{prop: pdf_givenXd2}. In the sets $\mathcal{R}_i$ in Lemma \ref{lemma:bijectiond2}, we assume $v_2\neq 0$ just to simplify the writing. It is enough that $v_i\neq 0$ for some $i\in\{2,\ldots, \ell\}$.

\begin{lemma}\label{lemma:bijectiond2}
	With the assumptions and the notation of Proposition \ref{prop: pdf_givenXd2}, 
	the map $\mathcal{H}:\mathbb{R}^{d} \longrightarrow \mathbb{R}^{d}$ given by
	\begin{align*}
	\mathcal{H}(v_1,\ldots,v_d):=\left(\frac{u_1v_1+\psi_{\bv}}{(s_1^2v_1^2+\varphi_{\bv})^{1/2}},v_2,\ldots,v_d\right)'
	\end{align*}
	is injective when restricted to each of the following regions:
	\begin{align*}
	\mathcal{R}_1&:=\left\{\bv : v_1 < \tfrac{u_1\varphi_{\bv}}{s_1^2\psi_{\bv}}, \psi_{\bv} > 0, v_2\neq 0\right\}; \mathcal{R}_2:=\ \left\{\bv : v_1 <  \tfrac{u_1\varphi_{\bv}}{s_1^2\psi_{\bv}}, \psi_{\bv} < 0, v_2\neq 0\right\}\\
	\mathcal{R}_3&:=\left\{\bv : v_1 >  \tfrac{u_1\varphi_{\bv}}{s_1^2\psi_{\bv}}, \psi_{\bv} > 0, v_2\neq 0\right\}; \mathcal{R}_4:=\ \left\{\bv : v_1 >  \tfrac{u_1\varphi_{\bv}}{s_1^2\psi_{\bv}}, \psi_{\bv} < 0, v_2\neq 0\right\}. 
	\end{align*}
\end{lemma}
\begin{proof}%[ of Lemma \ref{lemma:bijectiond2}]
	To ease the notation, we omit the subindex $\bv$ in $\psi_{\bv}$ and $\varphi_{\bv}$. Note firstly that the last $d-1$ components of $\mathcal{H}$ coincide with the identity function, which is obviously injective. Therefore we assume that $v_2,\ldots,v_d$ are fixed and study the monotonicity of the function $\mathcal{H}^1(v_1)=\left( u_1v_1+\psi\right)/(s_1^2v_1^2+\varphi)^{1/2}$. We have  
	
	\begin{equation*}
	\begin{split}
	\frac{d \mathcal{H}^1(v_1)}{d v_1}& =\frac{u_1\left(s_1^2v_1^2+\varphi\right)^{1/2}-(u_1v_1+\psi)s_1^2v_1(s_1^2v_1^2+\varphi)^{-1/2}}{s_1^2v_1^2+\varphi}\\
	&=\frac{u_1\varphi-\psi s_1^2v_1}{(s_1^2v_1^2+\varphi)^{3/2}}.\\
	\end{split}
	\end{equation*}
	Then $\frac{d \mathcal{H}^1(v_1)}{d V_1}=0$ if and only if $v_1=\frac{u_1\varphi}{s_1^2 \psi},$ or $\psi=0$ and $u_1=0.$ It is easy to check that $\mathcal{H}^1$ is strictly increasing on $\mathcal{R}_1$ and $\mathcal{R}_4$ while it is strictly decreasing on $\mathcal{R}_2$ and $\mathcal{R}_3.$ Consequently, the result is proven.
\end{proof}

For the sake of brevity we denote the function $\mathcal{H}$ restricted to the regions $\mathcal{R}_1, \mathcal{R}_2, \mathcal{R}_3$ and $\mathcal{R}_4$ as  $\mathcal{H}_1, \mathcal{H}_2, \mathcal{H}_3,$ and $\mathcal{H}_4,$ respectively.

\begin{corollary}\label{corol:bijectiond2}
	With the notation above introduced, the inverse of $\mathcal{H}_i$, $i=1,\ldots,4$ are 
	\begin{itemize}
		\item On $\mathcal{H}(\mathcal{R}_1)=\left(-\frac{u_1}{s_1},\sqrt{\frac{u_1^2}{s_1^2}+\frac{\psi_{\bv}^2}{\varphi_{\bv}}}\right)\times \{(v_2,\ldots,v_d)' : \psi_{\bv}>0\}$:  
		\begin{align*}
		\mathcal{H}^{-1}_1(y,v_2,\ldots,v_d)=
		\begin{cases}
		\left(h_{+}(y),v_2,\ldots,v_d\right)', & y<\frac{|u_1|}{s_1},\\
		\left(h_{-}(y),v_2,\ldots,v_d\right)',& \frac{|u_1|}{s_1}<y<\sqrt{\frac{u_1^2}{s_1^2}+\frac{\psi_{\bv}^2}{\varphi_{\bv}}},
		\end{cases}
		\end{align*} 
		
		\item On $\mathcal{H}(\mathcal{R}_2)=\left(-\sqrt{\frac{u_1^2}{s_1^2}+\frac{\psi_{\bv}^2}{\varphi_{\bv}}},-\frac{u_1}{s_1}\right)\times \{(v_2,\ldots,v_d)' : \psi_{\bv}<0\}$:
		\begin{align*}
		\mathcal{H}^{-1}_2(y,v_2,\ldots,v_d)=\left(h_{+}(y),v_2,\ldots,v_d\right)' ,\quad -\sqrt{\tfrac{u_1^2}{s_1^2}+\tfrac{\psi_{\bv}^2}{\varphi_{\bv}}}<y<-\tfrac{|u_1|}{s_1}
		\end{align*}
		
		\item On $\mathcal{H}(\mathcal{R}_3)=\left(\frac{u_1}{s_1},\sqrt{\frac{u_1^2}{s_1^2}+\frac{\psi_{\bv}^2}{\varphi_{\bv}}}\right)\times \{(v_2,\ldots,v_d)' : \psi_{\bv}>0\}$:
		\begin{align*}
		\mathcal{H}^{-1}_3(y,v_2,\ldots,v_d)=\left(h_{+}(y),v_2,\ldots,v_d\right)' ,\quad \tfrac{|u_1|}{s_1}<y< \sqrt{\tfrac{u_1^2}{s_1^2}+\tfrac{\psi_{\bv}^2}{\varphi_{\bv}}}
		\end{align*}
		
		\item On $\mathcal{H}(\mathcal{R}_4)=\left(-\sqrt{\frac{u_1^2}{s_1^2}+\frac{\psi_{\bv}^2}{\varphi_{\bv}}},\frac{u_1}{s_1}\right)\times \{(v_2,\ldots,v_d)' : \psi_{\bv}<0\}$:
		\begin{align*}
		\mathcal{H}^{-1}_4(y,v_2,\ldots,v_d)=
		\begin{cases}
		\left(h_{-}(y),v_2,\ldots,v_d\right)', &  -\sqrt{\frac{u_1^2}{s_1^2}+\frac{\psi_{\bv}^2}{\varphi_{\bv}}}<y<-\frac{|u_1|}{s_1}\\
		\left(h_{+}(y),v_2,\ldots,v_d\right)', & y<\frac{|u_1|}{s_1}
		\end{cases}
		\end{align*}
	\end{itemize}
	\noindent where $h_{\pm}(y):= \lrp{u_1\psi_{\bv}\pm|y|\sqrt{u_1^2\varphi_{\bv}+s_1^2\psi_{\bv}^2-y^2s_1^2\varphi_{\bv}}}/\lrp{s_1^2y^2-u_1^2}$.
\end{corollary}

\begin{proof}%[of Corollary \ref{corol:bijectiond2}]
	From Lemma \ref{lemma:bijectiond2}, we write explicitly the inverse of $\mathcal{H}:$ 
	\begin{align*}
	\mathcal{H}^{-1}(y,v_2,\ldots,v_d)=\left( h_{\pm}(y),v_2,\ldots,v_d\right)'.
	\end{align*}

	It remains to determine when the first coordinate of $\mathcal{H}^{-1}(y,v_2,\ldots,v_d)$ is $h_+$ or $h_-.$ Suppose $\psi_{\bv}>0$ and $y>|x_1|,$ (the rest of the cases are analogous), then  
	\begin{align*}
	\frac{u_1\psi_{\bv}+|y|\sqrt{u_1^2\varphi_{\bv}+s_1^2\psi_{\bv}^2-s_1^2y^2\varphi_{\bv}}}{s_1y^2-u_1^2}> \frac{x_1\psi_{\bv}-|y|\sqrt{u_1^2\varphi_{\bv}+s_1^2\psi_{\bv}^2-s_1^2y^2\varphi_{\bv}}}{s_1y^2-u_1^2}.
	\end{align*}
	Hence, by the definition of the regions $\mathcal{R}_i,$ we have that $(h_-(y),v_2,\ldots,v_d) \in \mathcal{H}_1^{-1}$ and $(h_+(y),v_2,\ldots,v_d) \in \mathcal{H}_3^{-1}.$
\end{proof}

\begin{proof}[Proof of Proposition \ref{prop: pdf_givenXd2}]
	To ease the notation, we often omit the sub-indices $\bv$, $\bmm$ and $S$ of the functions. Due to the symmetry of the distribution of $\bV$, we assume $x_i\geq 0$ for $i=1,\ldots,d$. It is clear that $y^{\bV}_{\bmm,S} \in [-t,t]$. 
	Take the transformation $\mathcal{H}$ defined on Lemma \ref{lemma:bijectiond2}. 
	
	We have that if $B$ is a Borel set on $\mathbb{R}$ and $\bV^1,\ldots,\bV^4$ are rv's such that the distribution of $\bV^i$ is that of $\bV$ given that $\bV\in \mathcal{R}_i$ for $i=1,\ldots,4$, then:
	\begin{align}
	\Prob(y^{\bV}_{\bmm,S}  \in B)%&=\Prob(\mathcal{H}(V)\in B) \nonumber\\
	%&=\sum_{i=1}^4\Prob(V\in \mathcal{R}_i)\Prob(\mathcal{H}(V)\in B | V\in \mathcal{R}_i) \nonumber \\
	&=\sum_{i=1}^4\Prob(\bV\in \mathcal{R}_i)\Prob(\mathcal{H}_{i,1}(\bV^i) \in B ),  \label{eq:proof_jacob}
	\end{align}
	where, as stated, $\mathcal{H}_i=(\mathcal{H}_{i,1},\ldots,\mathcal{H}_{i,d})'$ is the restriction of $\mathcal{H}$ to the set $\mathcal{R}_i.$ Since all $\mathcal{H}_i$ are injective and derivable we have that
	\begin{equation}\label{eq:proof_jac2}
	\Prob(\mathcal{H}_{i,1}(\bV^i) \in B)=\displaystyle\int_{B}\int_{\mathbb{R}^{d-1}} f_{V^i}(\mathcal{H}_{i}^{-1}(z,\bv_{-1}))|J_{\mathcal{H}_{i}}(z,\bv_{-1})| \,d\bv_{-1} \,dz,
	\end{equation}
	where $f_{\bV^i}$ is the pdf of the rv $\bV^i$. We trivially have that 
	\begin{align*}
	|J_{\mathcal{H}_{i}}(z,\bv_{-1})|=\left|\frac{\left(\partial \mathcal{H}_i\right)_1^{-1}(z)}{\partial z}\right|
	\end{align*}
	and
	\begin{align*}
	f_{\bV^i}(\mathcal{H}_i^{-1}(z,\bv_{-1}))=\frac{1}{\Prob( \bV \in \mathcal{R}_i)} f_{\bV}(\mathcal{H}_i^{-1}(z,\bv_{-1}))1_{\mathcal{R}_i}(\mathcal{H}_i^{-1}(z,\bv_{-1})),
	\end{align*}
	this expression jointly with \eqref{eq:proof_jacob} and \eqref{eq:proof_jac2} gives
	\begin{equation}\label{eq:proof_jacob2}
	\Prob(y^{\bV}_{\bmm,S} \in B)=\displaystyle\int_{B}\sum_{i=1}^4\int_{\mathbb{R}^{d-1}}f_{\bV}(\mathcal{H}_i^{-1}(z,\bv_{-1}))|J_{\mathcal{H}_i}(z,\bv_{-1})|\,d\bv_{-1} \,dz,
	\end{equation}
	where we have used the fact that, by definition, $1_{\mathcal{R}_i}(\mathcal{H}_i^{-1}(z,\bv_{-1}))=1.$

	We study now the sign of the determinant of the Jacobian. By Corollary \ref{corol:bijectiond2},
	\begin{align*}
	\left\vert\tfrac{\left(\partial \mathcal{H}_i\right)_1^{-1}(z)}{\partial z}\right\vert=&\left\vert\tfrac{\partial h_{\pm}(z)}{\partial z}\right\vert\\
	=&\tfrac{|\mp u_1^4\varphi+u_1^2s_1^2(\pm z^2\varphi\mp\psi^2)\mp\psi^2z^2s_1^4-2 \psi s_1^2 u_1 z \left(\psi^2 s_1^2+\varphi \left(u_1^2-s_1^2 z^2\right)\right)^{1/2}|}{\left(u_1^2-s_1^2 z^2\right)^2 \left(\psi^2 s_1^2+\varphi \left(u_1^2-s_1^2 z^2\right)\right)^{1/2}},
	\end{align*}
	where the signs depend on the particular index and $z$. 
	
	We have that $\left|\frac{\partial h_{\pm} (z)}{\partial z}\right|= 0$ only when $z \in \{0,\pm u_1/s_1\}$.   %
	As those values are not in the mentioned regions, we state
	\begin{equation*}
	\begin{array}{llll}
	\mbox{ If } z>0,& \mbox{ then } \partial h_{+}(z)/\partial z < 0 & \mbox{ and } & \partial h_{-}(z)/\partial z > 0.\\
	\mbox{ If } z<0,& \mbox{ then } \partial h_{+}(z)/\partial z > 0 & \mbox{ and } & \partial h_{-}(z)/\partial z < 0.
	\end{array} 
	\end{equation*}
	
	Take $B=(-\infty,r]$ in \eqref{eq:proof_jacob2} with $r \in (-t,-|u_1/s_1|)$, then
	\begin{align}
	\vspace{.2cm}
	\Prob\lrp{y^{\bV}_{\bmm,S} <r}=&\displaystyle\int_{A_{+}^{\bv}}\left(\int_{-t}^{r}f_{\bV}(\mathcal{H}_1^{-1}(z,\bv_{-1}))|J_{\mathcal{H}_1}(z,\bv_{-1})| \,dz\right. \nonumber\\
	& \phantom{{}+\int_{A_{-}^{\bv}}} +\left.\displaystyle\int_{-t}^{r}f_{\bV}(\mathcal{H}_3^{-1}(z,\bv_{-1}))|J_{\mathcal{H}_3}(z,\bv_{-1})|\,dz\right)\,d\bv_{-1} \nonumber\\
	&+\displaystyle\int_{A_{-}^{\bv}}\left(\int_{-t}^{r}f_{\bV}(\mathcal{H}_2^{-1}(z,\bv_{-1}))|J_{\mathcal{H}_2}(z,\bv_{-1})| \,dz\right. \nonumber\\
	& \phantom{{}++\int_{A_{-}^{\bv}}} +\displaystyle\left.\int_{-t}^{r}f_{\bV}(\mathcal{H}_4^{-1}(z,\bv_{-1}))|J_{\mathcal{H}_4}(z,\bv_{-1})| \,dz\right)\,d\bv_{-1}.\nonumber% \label{eq:proof_jac3}
	\end{align} 
	
	We have $f_{\bV}(\mathcal{H}_i^{-1}(z,\bv_{-1}))|J_{\mathcal{H}_i}(z,\bv_{-1})|=0$ when $r \in \left(-t,-\sqrt{u_1^2/s_1^2+\psi^2/\varphi}\right)$ for $i=1,\ldots,4$ and, using Corollary \ref{corol:bijectiond2},
	\begin{align*}
	\Prob\lrp{y^{\bV}_{\bmm,S} <r}=&\displaystyle\int_{A_{-}^{\bv}}\left(\int_{-\sqrt{u_1^2/s_1^2+\psi^2/\varphi}}^{r}e^{-\frac{1}{2}(h_{+}^2(z)+v_2^2+\cdots+ v_d^2)} \frac{\partial h_{+}(z)}{\partial z} \,dz\right.\\
	& \phantom{{}-\int_{A^{-}}} -\left.\displaystyle\int_{-\sqrt{u_1^2/s_1^2+\psi^2/\varphi}}^{r}e^{-\frac{1}{2}(h_{-}^2(z)+v_2^2+\cdots+ v_d^2)} \frac{\partial h_{-}(z)}{\partial z} \,dz\right)\,d\bv_{-1}, \\
	\end{align*}
	
	From $h_{-}\left(-\sqrt{u_1^2/s_1^2+\psi^2/\varphi}\right)=h_{+}\left(-\sqrt{u_1^2/s_1^2+\psi^2/\varphi}\right),$ the result is obtained. The case $-|u_1/s_1|<r<0$ is analogous and the cases when $r>0$ are deduced by symmetry.
\end{proof}

%----------------------------------------------%	
Lemma \ref{lemma:independ_givenX_unknown} is obvious and it is stated for further reference.

\begin{lemma}\label{lemma:independ_givenX_unknown}
	Let $\bV_1,\ldots,\bV_k$ be  iid rv's, then $Y^1,\ldots,Y^k$ defined in \eqref{eq:T_Kestimate} are conditionally iid given the $d$-dimensional vectors $\bX$ and $\bX_1,\ldots,\bX_n.$ 
\end{lemma}

\begin{proof}[Proof of Proposition \ref{prop:probabilidad2}]
	Denote $\mathbb{X}_n:=(\bX_1,\ldots,\bX_n)'$. Taking into account Lemma \ref{lemma:independ_givenX_unknown} and that $\hat{\bmu}_{\bV}=\hat{\bmu}'\bV$ and $\hat{\mSigma}_{\bV}=\sqrt{\bV'\hat{\mSigma}\bV}$, the result follows from the  reasoning: 
	\begin{align*}
	F_{\mSigma}(a,b,t)&=\; \sum\limits_{k=1}^{\infty} \Prob( \mbox{ declare } \bX \mbox{ as an outlier with } k \mbox{ proy } \ | \left\Vert \bX\right\Vert_{\mSigma}=t) \\
	&=\; \sum\limits_{k=1}^{\infty}  \E\bigg(\Prob\left( \left.\vert Y^{\bV}\vert >b \ \right\vert \bX,\mathbb{X}_n\right)\\
	&\phantom{=\; \sum\limits_{k=1}^{\infty}k \left [ \E \left(\right.\right.} \Prob\left( \left.\vert Y^{\bV}\vert \in (a,b) \ \right\vert \bX,\mathbb{X}_n\right)^{k-1}\bigg| \Vert\bX\Vert_{\mSigma}=t\bigg)\\
	&=\;\E\left( \left. \tfrac{1-\Prob\lrp{|Y^{\bV}|<b}}{1-\Prob\lrp{|Y^{\bV}|<b}+\Prob\lrp{|Y^{\bV}|<a}} \right\vert \ \left\Vert\bX\right\Vert_{\mSigma}=t \right) \\
	&=\;\E\left( \left. g_{a,b}(\bx,\bmm,S) \right\vert \ \left\Vert\bX\right\Vert_{\mSigma}=t \right).% 
	%&=\; \displaystyle\int_{\Omega^{d-1}_{\mSigma}(t)}\int_{\mathbb{R}^d} %\int_{\mathbb{R}^{d^2}} g_{a,b}(\bx,\bmm,S)f_{\hat{\mSigma}} (S)f_{\hat{\bmu}}(\bmm)f_{t} %(\bx) \, dS\, d\bmm\, d\bx. 
	\end{align*}
\end{proof}

\begin{proof}[Proof of Proposition \ref{prop:dependence_sigma_unknown}]
	Let $t>0$ and $\bw:=(\bx,\bmm,S)\in \Omega_\mSigma^{d-1}(t)\times\mathbb{R}^d\times\mathbb{R}^{d^2}$. Define $\delta(\bw):=\linebreak \Prob\lrp{\left\vert Y^{\bV}\right\vert < a | \mathbb{W}=\bw}-\Prob\left(\left\vert Y^{\bV}\right\vert < a \ | \Vert\bX\Vert_{\mSigma}=t\right)$, where $\mathbb{W}=(\bX,\hat{\bmu},\hat{\mSigma})$. Proposition \ref{prop: pdf_givenXd2} gives that the map $\bw\mapsto\Prob\left(\left\vert Y^{\bV}\right\vert < a | (\bX,\hat{\bmu},\hat{\mSigma})=\bw\right)$ is continuous and not constant on $\bx$ for $a \in (0,t)$ if $\mSigma\neq I_d$. Thus $\displaystyle\int\delta^2(\bw)\Prob_{\mathbb{W} | \Vert\bX\Vert_{\mSigma}=t}(d\bw)>0$. However, by definition of $\delta(\bw)$,
	\begin{equation*}
	\begin{split}
	\Prob\lrp{\left\vert Y^{\bV}\right\vert < a | \Vert\bX\Vert_{\mSigma}=t} &=\displaystyle\int\Prob\left(\left\vert Y^{\bV}\right\vert < a | \mathbb{W}=\bw\right)\Prob_{\mathbb{W} | \Vert\bX\Vert_{\mSigma}=t}(d\bw)\\
	&=\Prob\left(\left\vert Y^{\bV}\right\vert< a\ | \Vert\bX\Vert_{\mSigma}=t\right)+\displaystyle\int\delta(\bw)\Prob_{\mathbb{W} | \Vert\bX\Vert_{\mSigma}=t}(d\bw),
	\end{split}
	\end{equation*}
	and, consequently, $\displaystyle\int\delta(\bw)\Prob_{\mathbb{W} | \Vert\bX\Vert_{\mSigma}=t}(d\bw)=0.$ Denote $g(a,t):=\Prob(Y^{\bV_1}< a,Y^{\bV_2}<a \ | \Vert\bX\Vert_{\mSigma}=t )$, then, by Lemma \ref{lemma:independ_givenX_unknown}, 
	\begin{equation*}
	\begin{split}
	g(a,t) =&\displaystyle\int\Prob\left(\left\vert Y^{\bV_1}\right\vert< a, \left\vert Y^{\bV_2}\right\vert<a | \mathbb{W}=\bw \right)\Prob_{\mathbb{W} | \Vert\bX\Vert_{\mSigma}=t}(d\bw)\\
	=&\displaystyle\int\Prob\left(\left\vert Y^{\bV}\right\vert < a | \mathbb{W}=\bw\right)^2\Prob_{\mathbb{W} | \Vert\bX\Vert_{\mSigma}=t}(d\bw)\\
	=&\displaystyle\int\bigg(\Prob\left(\left\vert Y^{\bV}\right\vert < a  \ | \Vert\bX\Vert_{\mSigma}=t\right)^2+\delta^2(\bw)\\
	&\;\phantom{\int()}+2\delta(\bw)\Prob\left(\left\vert Y^{\bV}\right\vert < a\ | \Vert\bX\Vert_{\mSigma}=t \right)\bigg) \Prob_{\mathbb{W} | \Vert\bX\Vert_{\mSigma}=t}(d\bw)\\
	=&\;\Prob\left(\left\vert Y^{\bV}\right\vert < a \ | \Vert\bX\Vert_{\mSigma}=t\right)^2+\displaystyle\int\delta^2(\bw) \Prob_{\mathbb{W} | \Vert\bX\Vert_{\mSigma}=t}(d\bw)\\
	>&\;\Prob\left(\left\vert Y^{\bV}\right\vert < a \ | \Vert\bX\Vert_{\mSigma}=t\right)^2.
	\end{split}
	\end{equation*}
	However, if $\mSigma=I_d$, then $\Prob\lrp{\left\vert Y^\bV\right\vert<a | \mathbb{W}=w}$ is constant on $\bX$ and the same reasoning shows the independence in this case.
\end{proof}

Before proving Proposition \ref{prop: decrease2}, we need some previous results.

\begin{lemma}\label{lemma1}
	Let $d>1$ and let $S$ and $\mSigma$ be $d\times d$ semi-positive symmetric matrices and $\bZ$ be a $d$-dimensional rv. The function $r\mapsto f(r)$ is increasing on $r$, where 
	\begin{align*}
	f(r):= \Prob( \mSigma^{1/2}\bZ \mbox{ be declared outlier  w.r.t. } N_d(\mathbf{0}, S) \ | \ \Vert \mSigma^{1/2}\bZ\Vert=r ).
	\end{align*}
\end{lemma}
\begin{proof}%[of Lemma \ref{lemma1}]
	Let $\bz \in \mathbb{R}^d$ and let $r=\Vert \mSigma^{1/2}\bz\Vert$. Since the distribution of $\bV=(V_1,\ldots,V_d)'$ is rotation invariant, we can compute $F_{\bZ}(\cdot)$ in any basis we choose. Thus, let us consider a basis on $\mathbb{R}^d$ such that $\mSigma^{1/2}\bz/\Vert\mSigma^{1/2}\bz\Vert = (1,0,\cdots,0)'$, then
	\begin{align*}
	\Prob\left(\frac{|(\mSigma^{1/2}\bz)'\bV|}{\mSigma_{\bV}} >b\right)= \Prob\left(\frac{\Vert\mSigma^{1/2}\bz\Vert|V_1|}{\mSigma_{\bV}} >b\right)=\Prob\left(\frac{|V_1|}{\mSigma_{\bV}} > \frac{b}{r}\right),
	\end{align*}
	which is increasing on $r$ and that does not depend on $\bz$. A similar reasoning implies that the map $r\mapsto\Prob\left(\frac{|(\mSigma^{1/2}\bz)'\bV|}{\mSigma_{\bV}} <a\right)$ decreases on $r$ and the result follows from.
	\begin{align*}
	f(r)=&\displaystyle\int\frac{\Prob\left(\frac{|(\mSigma^{1/2}\bz)'\bV|}{\mSigma_{\bV}} >b\right)}{\Prob\left(\frac{|(\mSigma^{1/2}\bz)'\bV|}{\mSigma_{\bV}} >b\right)+\Prob\left(\frac{|(\mSigma^{1/2}\bz)'\bV|}{\mSigma_{\bV}} <a\right)} \Prob_{\bZ | \ \Vert\mSigma^{1/2}\bZ\Vert}(d\bz)\\
	=&\frac{\Prob\left(\frac{|V_1|}{\mSigma_{\bV}} > \frac{b}{r}\right)}{\Prob\left(\frac{|V_1|}{\mSigma_{\bV}} > \frac{b}{r}\right)+\Prob\left(\frac{|V_1|}{\mSigma_{\bV}} < \frac{a}{r}\right)}.
	\end{align*}
\end{proof}

\begin{lemma}\label{lemma2}
	Let $\delta <1$, $c<1$, $\mSigma$ be a semi positive definite symmetric matrix and $\bZ$ be a rv with distribution $N_d\left(\mathbf{0},\delta I_d\right).$ If $\bx\neq 0,$ then for any $g$ increasing function 
	\begin{align*}
	\E[g(\Vert\mSigma^{1/2}(\bZ+\bx)\Vert )] \geq \E[g(\Vert\mSigma^{1/2}(\bZ+c\bx)\Vert )].
	\end{align*}    
\end{lemma}
\begin{proof}%[of Lemma \ref{lemma2}]
	The second part in Corollary 2 in \cite{Anderson1955} gives that if $\bx\neq \mathbf{0}$ and $h(w)=w$, then 
	\begin{align*}
	\Prob(  \Vert\mSigma^{1/2}(\bZ+c\bx)\Vert \leq r) \geq \Prob(\Vert\mSigma^{1/2}(\bZ+\bx)\Vert \leq r ).
	\end{align*}
	From here, the lemma trivially follows.
\end{proof}

\begin{proof}[Proof of Proposition  \ref{prop: decrease2}]
	Given $\bz \in \Omega_1^{d-1}$ and $S\in \mathbb{R}^{d^2}$, let us consider
	\begin{align*}
	G_{\bz,S}(t)=\displaystyle\int_{\mathbb{R}^{d}} \frac{\Prob\left( \frac{\vert (t\mSigma^{1/2}\bz-\by)'\bV\vert}{\Vert S^{1/2}\bV\Vert} > b \right)}{\Prob\left( \frac{\vert (t\mSigma^{1/2}\bz-\by)'\bV\vert}{\Vert S^{1/2}\bV\Vert} > b\right)+\Prob\left( \frac{\vert (t\mSigma^{1/2}\bz-\by)'\bV\vert}{\Vert S^{1/2}\bV\Vert} < a\right)}f_{\hat{\bmu}} (\by)\,d\by,
	\end{align*}
	where $\hat{\bmu}$ follows a $N_d(\mathbf{0},\mSigma/n)$ distribution. The proposition will be proved if we show that $G_{\bz,S}(t)$ is increasing because
	\begin{align*}
	F_{\mSigma}(a,b,t)=\tfrac{1}{\omega^d_1}\displaystyle\int_{\Omega^{d-1}_1}\int_{\mathbb{R}^{d^2}} G_{\bz,S}(t) \,\Prob_{\hat{\mSigma}}(dS) \, d\bz,
	\end{align*}
	where $\hat{\mSigma}$ follows a Wishart distribution with parameters $n$ and $\mSigma$.
	Given the rv $\bZ$, let $\{\bz \mbox{ outlier wrt } N_d(\mathbf{0}, S)\}$ denote the set where $\bZ$ is declared  outlier with respect to $N_d(\mathbf{0}, S)$. Then
	\begin{align*}
	\Prob\{\bZ \mbox{ outlier wrt } N_d(\mathbf{0}, S)\}=\displaystyle\int\frac{\Prob\left(\frac{|\bz'\bV|}{\Vert S^{1/2}\bV\Vert} >b\right)}{\Prob\left(\frac{|\bz'\bV|}{\Vert S^{1/2}\bV\Vert} >b\right)+\Prob\left(\frac{|\bz'\bV|}{\Vert S^{1/2}\bV\Vert} <a\right)} \Prob_{\bZ}(d\bz).
	\end{align*}
	If we take $\bY_n=\mSigma^{-1/2}\hat{\mu}$ and $f$ is the function defined in Lemma \ref{lemma1},	
	\begin{align*}
	G_{\bz,S}(t)&=\Prob( \mSigma^{1/2}(\bY_n + t\bz) \mbox{ outlier wrt } N_d(\mathbf{0}, S))\\
	&=\displaystyle\int_0^{\infty}\Prob( \mSigma^{1/2}(\bY_n + t\bz) \mbox{ outlier wrt } N_d(\mathbf{0},S) \vert \Vert\mSigma^{1/2}(\bY_n+t\bz)\Vert=r)\Prob(dr)\\
	&= \E[f(\Vert\mSigma^{1/2}(\bY_n+t\bz)\Vert)], 
	\end{align*}
	and the result is deduced from Lemmas \ref{lemma1} and \ref{lemma2}.
\end{proof}

%----------------------------------------------%
\subsection{Proofs of Section \ref{subsec:robust}}
%----------------------------------------------%

We first state some additional notation which is needed to prove Theorem \ref{theo:Juan}. %
Under assumptions (A1) and (A2), denote $\Q_{\bV}$ and $\bar{\Q}_{\bV}$ the probability distribution of $\bX'\bV$ and of $|\bX'\bV|$, respectively, and let 
\begin{align*}
R_{\bV}^n&:=\{\bX_1'\bV,\ldots,\bX_n'\bV\}\\
T_{\bV}^{n}&:=\{|\bX_1'\bV|,\ldots,|\bX_n'\bV|\}\\
S_{\bV}^{n}&:=\{|\bX_1'\bV-m_{\bV}|,\ldots,|\bX_n'\bV-m_{\bV}|\}\\
\hat{S}_{\bV}^{n}&:=\{|\bX_1'\bV-\hat{m}_{\bV}|,\ldots,|\bX_n'\bV-\hat{m}_{\bV}|\}.\end{align*}
Given $S\subset\mathbb{R}$ finite (respectively the real rv $X$) and $\alpha\in (0,1)$, $m(S)$ and $M(S)$ (resp. $m(X)$ and $M(X)$) denote the sets of its medians and $\MADN$s,  $[\munderbar{q}_{\alpha}(S),\bar{q}_{\alpha}(S)]$ (resp. $[\munderbar{q}_{\alpha}(X),\bar{q}_{\alpha}(X)]$) is  the interval of the $\alpha$-quantiles of $S$ (resp. $X$);  we define $[\munderbar{M}_{\alpha}(S),\bar{M}_{\alpha}(S)]:=\cup_{m\in m(S)}\lrc{\munderbar{q}_{\alpha}(|S-m|),\bar{q}_{\alpha}(|S-m|)}$, and similarly for $[\munderbar{M}_{\alpha}(X),\bar{M}_{\alpha}(X)]$.
Thus, $m(S)=[\munderbar{q}_{\frac{1}{2}}(S),\bar{q}_{\frac{1}{2}}(S)]$ and $M(S)=[\munderbar{M}_{\frac{1}{2}}(S),\bar{M}_{\frac{1}{2}}(S)]$. 

Since, by assumption, all random quantities we handle are defined on $(\Upsilon,\mathcal{A},\Prob)$, all of them will depend on some $\omega\in\Upsilon$. Very often this dependence is not made explicit, however, when required, $\omega$ will appear as super-index as in $\hat{m}_{\bV}^{\omega}$, or in $S_{\bV}^{n,\omega}$. 

\begin{lemma}\label{lemma:Juan}
	Let $U$ and $V$ be two real rv's such that there exist $\delta$ and $\gamma$ and $\Prob\{|U-V|\leq \delta\}\geq 1-\gamma$. Then for every $\alpha\in[\gamma,1-\gamma]$,
	\begin{align}
	[\munderbar{q}_{\alpha}(U),\bar{q}_{\alpha}(U)]&\subset [\munderbar{q}_{\alpha-\gamma}(V)-\delta,\bar{q}_{\alpha+\gamma}(V)+\delta]\label{eq:lemma9_1}\\
	[\munderbar{M}_{\alpha}(U),\bar{M}_{\alpha}(U)]&\subset [\munderbar{M}_{\alpha-\gamma}(V)-(2\delta+\delta_{\gamma}^*),\bar{M}_{\alpha+\gamma}(V)+(2\delta+\delta_{\gamma}^*)]\label{eq:lemma9_2},
	\end{align}
	where $\delta_{\gamma}^*=\max\{\munderbar{q}_{\frac{1}{2}}(V)-\munderbar{q}_{\frac{1}{2}-\gamma}(V),\bar{q}_{\frac{1}{2}+\gamma}(V)-\bar{q}_{\frac{1}{2}}(V)\}$.
\end{lemma}	
\begin{proof}
	Let $q\in \lrc{\munderbar{q}_\alpha(U),\bar{q}_\alpha(U)}$. Then, by definition of quantile:
	\begin{align*}
	\alpha&\leq \Prob\{U\leq q\}\\
	&\leq \Prob [|U-V|\leq \delta,U\leq q]+\Prob\{|U-V|>\delta\}\\
	%&\leq\Prob[|U-V|\leq \delta,V\leq q+\delta]+\Prob\{|U-V|>\delta\}\\
	&\leq \Prob\{V\leq q+\delta\}+\gamma.
	\end{align*}
	Hence $\alpha-\gamma\leq \Prob\{V\leq q+\delta\}$, which implies $q+\delta\geq \munderbar{q}_{\alpha-\gamma}(V)$. And then $\munderbar{q}_{\alpha}(U)\geq \munderbar{q}_{\alpha-\gamma}(V)-\delta$. Analogously, we can prove $\bar{q}_{\alpha}(U)\leq \bar{q}_{\alpha+\gamma}(V)+\delta$ and \eqref{eq:lemma9_1} is shown.
	
	To prove \eqref{eq:lemma9_2}, consider $m^U\in m(U)$. Take $\alpha=1/2$ in \eqref{eq:lemma9_1}. There exits $m^V\in m(V)$ such that $|m^U-m^V|\leq \delta+\delta_{\gamma}^*$. Hence, if $|U-V|\leq \delta$, then 
	\begin{align*}
	\bigg||U-m^U|-|V-m^V|\bigg|\leq |U-V|+|m^U-m^V|\leq 2\delta+\delta_{\gamma}^*,
	\end{align*}
	and \eqref{eq:lemma9_2} follows from the definition of MAD and \eqref{eq:lemma9_1}.
\end{proof}

\begin{corollary}\label{cor:Juan2}
	Under hypotheses in Lemma \ref{lemma:Juan}, $m(U)\subset \left[\munderbar{q}_{\frac{1}{2}-\gamma}(V)-\delta,\bar{q}_{\frac{1}{2}+\gamma}(V)+\delta\right].$
\end{corollary}	

If we apply Lemma \ref{lemma:Juan} to rv's uniformly distributed on finite sets with the same cardinal, we obtain the following corollary.
\begin{corollary}\label{cor:Juan1}
	If $S=\{s_1,\ldots,s_n\}\subset\mathbb{R}$ and $R=\{r_1,\ldots,r_n\}\subset\mathbb{R}$ satisfy that there exist $\delta,\gamma$ such that $\#\{i: |s_i-r_i|\leq \delta\}\geq n(1-\gamma)$, then for every $\alpha\in[\gamma,1-\gamma]$,
	\begin{align}
	[\munderbar{q}_{\alpha}(S),\bar{q}_{\alpha}(S)]&\subset [\munderbar{q}_{\alpha-\gamma}(R)-\delta,\bar{q}_{\alpha+\gamma}(R)+\delta]\label{eq:lemma9_1b}\\
	[\munderbar{M}_{\alpha}(S),\bar{M}_{\alpha}(S)]&\subset [\munderbar{M}_{\alpha-\gamma}(R)-(2\delta+\delta_{\gamma}^*),\bar{M}_{\alpha+\gamma}(R)+(2\delta+\delta_{\gamma}^*)]\label{eq:lemma9_2b},
	\end{align}
	where $\delta_{\gamma}^*=\max\left\{\munderbar{q}_{\frac{1}{2}}(R)-\munderbar{q}_{\frac{1}{2}-\gamma}(R),\bar{q}_{\frac{1}{2}+\gamma}(R)-\bar{q}_{\frac{1}{2}}(R)\right\}$.
\end{corollary}

\begin{lemma}\label{lemma:uax_theoJ}
	For every $\bv\in\Omega_1^{d-1}$, there exists a probability one set $A$ such that for every $\omega\in A$ and $\gamma\in(0,1/2)$
	\begin{align*}
	\begin{split}
	\sup_{\alpha\in(\gamma,1-\gamma)}\lrp{\max\left\{\left\vert\munderbar{q}_{\alpha}(R^{n,\omega}_{\bv})-q_{\alpha}(\Q_{\bv})\right\vert,\left\vert\bar{q}_{\alpha}(R^{n,\omega}_{\bv})-q_{\alpha}(\Q_{\bv})\right\vert\right\}}&\to 0,\\
	\sup_{\alpha\in(\gamma,1-\gamma)}\lrp{\max\left\{\left\vert\munderbar{q}_{\alpha}(T^{n,\omega}_{\bv})-q_{\alpha}(\bar{\Q}_{\bv})\right\vert,\left\vert\bar{q}_{\alpha}(T^{n,\omega}_{\bv})-q_{\alpha}(\bar{\Q}_{\bv})\right\vert\right\}}&\to 0.
	\end{split}
	\end{align*}
\end{lemma}	
\begin{proof}
	Since $\bX'\bv$ is a normal rv, then the assumptions in Corollary 1.4.3 in \cite{Csorgo1983} are satisfied. Therefore, (1.4.24) in \cite{Csorgo1983} holds and first statement here is verified. A similar reasoning leads to the second one.
\end{proof}

\begin{proof}[Proof of Theorem \ref{theo:Juan}]
	We first apply the Glivenko-Cantelli Theorem to the iid rv's $\{\Vert\bX_i\Vert\}$ and we have that a.s.
	\begin{align}\label{eq:proof_propGC}
	\sup_{r>0}\left\vert \frac{\#\{i\leq n:\Vert\bX_i\Vert\leq r\}}{n}-\Prob(\Vert\bX_1\Vert\leq r)\right\vert\to 0.
	\end{align}
	Given $h\in\mathbb{N}$, since $\Omega_1^{d-1}$ is compact, there exist $\bv_1^h,\ldots,\bv_{J_h}^h \in \Omega_1^{d-1}$ such that for every $\bv\in\Omega_1^{d-1}$ there exists $i_\bv\in\{1,\ldots,J_h\}$ such that $\Vert\bv-\bv^h_{i_h}\Vert\leq h^{-1}$. Lemma \ref{lemma:uax_theoJ} gives that there exists $A_h \in \mathcal{A}$ such that $\Prob(A_h)=1$ and for every $\omega\in A_h$, \eqref{eq:proof_propGC} is satisfied and for every $\gamma\in(0,1/2)$, 
	\begin{align}\label{eq:proof_propGC2}
	\begin{split}
	\sup_{\alpha\in(\gamma,1-\gamma)}\lrp{\max_{i\leq J_h}\left\{\left\vert\munderbar{q}_{\alpha}(R^{n,\omega}_{\bv_i^h})-q_{\alpha}(\Q_{\bv_i^h})\right\vert,\left\vert\bar{q}_{\alpha}(R^{n,\omega}_{\bv_i^h})-q_{\alpha}(\Q_{\bv_i^h})\right\vert\right\}}&\to 0,\\
	\sup_{\alpha\in(\gamma,1-\gamma)}\lrp{\max_{i\leq J_h}\left\{\left\vert\munderbar{q}_{\alpha}(T^{n,\omega}_{\bv_i^h})-q_{\alpha}(\bar{\Q}_{\bv_i^h})\right\vert,\left\vert\bar{q}_{\alpha}(T^{n,\omega}_{\bv_i^h})-q_{\alpha}(\bar{\Q}_{\bv_i^h})\right\vert\right\}}&\to 0.
	\end{split}
	\end{align}
	Denote $A_0=\cap_{h\in \mathbb{N}}A_h$. Obviously $A_0\in\mathcal{A}$ and $\Prob(A_0)=1$. Let $\omega\in A_0$ be a point which will remain fixed along the proof. We begin proving the first statement in \eqref{eq:prop_Juan1}. Let $\varepsilon>0$. Let $\lambda_d$ be the largest eigenvalue of $\mSigma$. Given $\bv\in\Omega_1^{d-1}$ and $\gamma\in\lrp{0,1/2}$, we have that 
	\begin{align*}
	q_{\frac{1}{2}+\gamma}(\Q_{\bv})-q_{\frac{1}{2}-\gamma}(\Q_{\bv})&=(\bv'\mSigma\bv)\lrp{q_{\frac{1}{2}+\gamma}(N_1(0,1))-q_{\frac{1}{2}-\gamma}(N_1(0,1))}\\
	&\leq \lambda_d \lrp{q_{\frac{1}{2}+\gamma}(N_1(0,1))-q_{\frac{1}{2}-\gamma}(N_1(0,1))}.
	\end{align*}
	
	Therefore, there exists $\gamma_1\in(0,1/2)$ such that 
	\begin{align}\label{eq:proof_Juan2s}
	\sup_{\bv\in\Omega_1^{d-1}}\lrp{q_{\frac{1}{2}+\gamma_1}(\Q_{\bv})-q_{\frac{1}{2}-\gamma_1}(\Q_{\bv})}<\frac{\varepsilon}{3}.
	\end{align}
	Analogously, we can prove that there exits $\gamma_2\in(0,1/2)$ such that, 
	\begin{align}\label{eq:proof_Juan3s}
	\sup_{\bv\in\Omega_1^{d-1}}\lrp{q_{\frac{1}{2}+\gamma_2}(\bar{\Q}_{\bv})-q_{\frac{1}{2}-\gamma_2}(\bar{\Q}_{\bv})}<\frac{\varepsilon}{3}.
	\end{align}
	Take $\gamma=\inf\{\gamma_1,\gamma_2,\varepsilon\}$. Let $r>0$ such that $\Prob(\Vert\bX_1\Vert\leq r)>1-\gamma$ and let $h\in\mathbb{N}$ be such that $r/h<\varepsilon/3$ and $2d\lambda_dM_1/h<\varepsilon/3$, where $M_1$ is the MADN of a $N_1(0,1)$. 
	
	By \eqref{eq:proof_propGC} and \eqref{eq:proof_propGC2}, there exists $N^{\omega}$ such that if $n\geq N^{\omega}$, then $\#\{i\leq n: \Vert\bX_i(\omega)\Vert< r\}>n(1-\gamma)$ and
	\begin{align}\label{eq:proof_propGC3}
	\begin{split}
	\sup_{\alpha\in\lrp{\frac{1}{2}-\gamma,\frac{1}{2}+\gamma}}\lrp{\max_{i\leq J_h}\left\{\left\vert\munderbar{q}_{\alpha}(R^{n,\omega}_{\bV_i^h})-q_{\alpha}(\Q_{\bv_i^h})\right\vert,\left\vert\bar{q}_{\alpha}(R^{n,\omega}_{\bv_i^h})-q_{\alpha}(\Q_{\bv_i^h})\right\vert\right\}}&< \frac{\varepsilon}{3}\\
	\sup_{\alpha\in\lrp{\frac{1}{2}-\gamma,\frac{1}{2}+\gamma}}\lrp{\max_{i\leq J_h}\left\{\left\vert\munderbar{q}_{\alpha}(T^{n,\omega}_{\bv_i^h})-q_{\alpha}(\bar{\Q}_{\bv_i^h})\right\vert,\left\vert\bar{q}_{\alpha}(T^{n,\omega}_{\bv_i^h})-q_{\alpha}(\bar{\Q}_{\bv_i^h})\right\vert\right\}}&< \frac{\varepsilon}{3}.
	\end{split}
	\end{align}
	Let $\bv\in\Omega_1^{d-1}$, 
	if $\Vert\bX_j(\omega)\Vert\leq r$,
	\begin{align}\label{eq:proof_propGC4}
	\left|(\bX_j(\omega))'\bv-(\bX_j(\omega))'\bv_{i_{\bv}}^h\right|\leq \Vert\bX_j(\omega)\Vert\Vert\bv-\bv_{i_{\bv}}^h\Vert\leq rh^{-1}<\frac{\varepsilon}{3},
	\end{align}
	and therefore, by Corollary \ref{cor:Juan1} with $\alpha=1/2$, we have that,
	\begin{align*}
	\hat{m}_{\bv}^{\omega}\in&\left[\munderbar{q}_{\frac{1}{2}-\gamma}\lrp{R^{n,\omega}_{\bV^h_{i_\bV}}}-\frac{\varepsilon}{3},\bar{q}_{\frac{1}{2}+\gamma}\lrp{R^{n,\omega}_{\bV^h_{i_\bV}}}+\frac{\varepsilon}{3}\right],
	\end{align*}
	and \eqref{eq:proof_propGC3} % 
	gives
	\begin{align}\label{eq:proof_Juan3}
	\hat{m}_{\bv}^{\omega}\in\left[q_{\frac{1}{2}-\gamma}\lrp{\Q_{\bv_{i_{\bv}}^h}}-\frac{2\varepsilon}{3},q_{\frac{1}{2}+\gamma}\lrp{\Q_{\bv_{i_{\bv}}^h}}+\frac{2\varepsilon}{3}\right].
	\end{align}
	On the other hand, we have that
	\begin{align}\label{eq:proof_Juan4}
	|\hat{m}_{\bv}^{\omega}-m_{\bv}|\leq |\hat{m}_{\bv}^{\omega}-m_{\bv_{i_{\bv}}^h}|+|m_{\bv_{i_{\bv}}^h}-m_{\bv}|.
	\end{align}
	
	Moreover, $\bmm_{\bv}=0$ for every $\bv$ because all probabilities $\Q_{\bv}$ are normal with mean zero. Thus, the second addend in \eqref{eq:proof_Juan4} is null. However, \eqref{eq:proof_Juan3} and \eqref{eq:proof_Juan2s} imply
	\begin{align}\label{eq:proof_Juan5} |\hat{m}_{\bv}^{\omega}-m_{\bv_{i_{\bv}}^h}|<\varepsilon.
	\end{align}
	Then, the first item in \eqref{eq:prop_Juan1} is proved because by \eqref{eq:proof_Juan4} and \eqref{eq:proof_Juan5} we have that, if $n>N^{\omega}$
	\begin{align}\label{eq:proof_Juan4s}
	\sup_{\bv}|\hat{m}_{\bv}^{\omega}-m_{\bv}|<\varepsilon.
	\end{align} 
	
	Concerning to the second item in \eqref{eq:prop_Juan1}, notice that if $\bv\in\Omega_1^{d-1}$ and $h\in\mathbb{N}$, then
	\begin{align}\label{eq:proof_Juan6}
	|\hat{M}_{\bv}^\omega-M_{\bv}|\leq|\hat{M}_{\bv}^\omega-\hat{M}^\omega_{\bv_{i_{\bv}}^h}|+|\hat{M}^\omega_{\bv_{i_{\bv}}^h}-M_{\bv_{i_{\bv}}^h}|+|M_{\bv_{i_{\bv}}^h}-M_{\bv}|.
	\end{align}
	If $n\geq N^\omega$, $i=1,\ldots,n$, and $\bv\in\Omega_1^{d-1}$, from \eqref{eq:proof_Juan4s} (remember that $\bmm=\mathbf{0}$) we have 
	\begin{align}\label{eq:proof_Juan8}
	\bigg|\;\left|\lrp{\bX_i(\omega)}'\bv-\hat{m}_\bv\right|-\left|\lrp{\bX_i(\omega)}'\bv\right|\;\bigg|\leq \left|\hat{m}_\bv\right|<\varepsilon.
	\end{align}
	Therefore, we can apply Corollary \ref{cor:Juan1} with $\alpha=1/2$, $\delta=\varepsilon$ and $\gamma=0$ to obtain that
	\begin{align*}
	\lrc{\munderbar{M}_{\frac{1}{2}}\lrp{R_\bv^{n,\omega}},\bar{M}_{\frac{1}{2}}\lrp{R_\bv^{n,\omega}}}&= \lrc{\munderbar{q}_{\frac{1}{2}}\lrp{\hat{S}_\bv^{n,\omega}},\bar{q}_{\frac{1}{2}}\lrp{\hat{S}_\bv^{n,\omega}}}\\
	&\subset\lrc{\munderbar{q}_{\frac{1}{2}}\lrp{T_\bv^{n,\omega}}-2\varepsilon,\bar{q}_{\frac{1}{2}}\lrp{T_\bv^{n,\omega}}+2\varepsilon}
	\end{align*}
	which joined to \eqref{eq:proof_propGC3} and the fact that $M_\bv=m(\bar{Q}_\bv)$ gives that if $n\geq N^\omega$
	\begin{align*}
	\left|\hat{M}^\omega_{\bV^h_{i_\bV}}-M_{\bV^h_{i_\bV}}\right|<2\varepsilon+\frac{\varepsilon}{3}.
	\end{align*}
	Concerning the third addend in \eqref{eq:proof_Juan6}, notice that $M_{\bv}=m(\bar{\Q}_{\bv})$ coincides with $\bv'\mSigma\bv M_1$. Thus, if we write $\bv=(v^1,\ldots,v^d)'$, then
	\begin{align*}
	|M_{\bv_{i_{\bv}}^h}-M_{\bv}|&=\left\vert\bv'\mSigma\bv-(\bv_{i_{\bv}}^h)'\mSigma\bv_{i_{\bv}}^h\right\vert M_1\\
	&=\left\vert\sum_{j=1}^d \lrp{v^j}^2\lambda_j-\sum_{j=1}^d\lrp{(\bv^h_{i_{\bv}})^j}^2\lambda_j\right\vert M_1\\
	&\leq \lambda_d M_1\sum_{j=1}^d \left\vert\lrp{v^j}^2-\lrp{(\bv^h_{i_{\bv}})^j}^2\right\vert\\
	&\leq 2\lambda_d M_1\sum_{j=1}^d \left\vert v^j-(\bv^h_{i_{\bv}})^j\right\vert\\
	&\leq 2d\lambda_d M_1\left\Vert \bv-\bv^h_{i_j}\right\Vert
	\leq \frac{2d\lambda_d M_1}{h}<\frac{\varepsilon}{3}.
	%&\leq d\lambda_d M_1\left|\vert\Vert\bv\Vert^2-\Vert\bv^h_{i_{\bv}}\Vert^2\Vert\right|\\
	%&=2d\lambda_d M_1\left|\Vert\bv\Vert-\Vert\bv^h_{i_{\bv}}\Vert\right|\\
	%&\leq 2d\lambda_d M_1h^{-1}<\varepsilon/3.
	\end{align*}
	
	Now, let us pay attention to the first addend in \eqref{eq:proof_Juan6}. According to \eqref{eq:proof_propGC4} and \eqref{eq:lemma9_2b} in Corollary \ref{cor:Juan1}, we have that 
	\begin{align}\label{eq:proof_Juan7}
	|\hat{M}_{\bv}-\hat{M}_{\bv_{i_{\bv}}^h}|\leq \bar{M}_{\frac{1}{2}+\alpha}\lrp{R_{\bv_{i_{\bv}}^h}^{n,\omega}}-\munderbar{M}_{\frac{1}{2}-\alpha}\lrp{R_{\bv_{i_{\bv}}^h}^{n,\omega}}+\frac{2\varepsilon}{3}+\delta^*_{\gamma}.
	\end{align}
	
	First, \eqref{eq:proof_propGC3} and \eqref{eq:proof_Juan2s} give that
	\begin{align*}
	\delta_{\gamma}^*\leq \bar{q}_{\frac{1}{2}+\gamma}\lrp{R^{n,\omega}_{\bv^h_{i_\bv}}}-\munderbar{q}_{\frac{1}{2}-\gamma}\lrp{R^{n,\omega}_{\bv^h_{i_\bv}}}\leq q_{\frac{1}{2}+\gamma}\lrp{\Q_{\bv^h_{i_k}}}-q_{\frac{1}{2}-\gamma}\lrp{\Q_{\bv^h_{i_k}}}+\frac{2\varepsilon}{3}<\varepsilon.
	\end{align*}
	
	For the first addend in \eqref{eq:proof_Juan7}, 
	by Corollary \ref{cor:Juan1} with $\gamma=0$, we conclude that
	\begin{align*}
	\bar{M}_{\frac{1}{2}+\gamma}\lrp{R_{\bv_{i_{\bv}}^h}^{n,\omega}}-\munderbar{M}_{\frac{1}{2}-\gamma}\lrp{R_{\bv_{i_{\bv}}^h}^{n,\omega}}&=\bar{q}_{\frac{1}{2}+\gamma}\lrp{S^{n,\omega}_{i_\bv^h}}-\munderbar{q}_{\frac{1}{2}+\gamma}\lrp{S^{n,\omega}_{i_\bv^h}}\\
	&\leq \bar{q}_{\frac{1}{2}+\gamma}(T^{n,\omega}_{\bv_{i_\bv}^h})-\munderbar{q}_{\frac{1}{2}-\gamma}(T^{n,\omega}_{\bv_{i_\bv}^h})+2\left|\hat{m}_{\bv_{i_{\bv}}^h}\right|,
	\end{align*}
	and from \eqref{eq:proof_propGC3}, \eqref{eq:proof_Juan3s} and \eqref{eq:proof_Juan4s} we have that
	\begin{align*}
	\bar{M}_{\frac{1}{2}+\gamma}\lrp{R_{\bv_{i_{\bv}}^h}^{n,\omega}}-\munderbar{M}_{\frac{1}{2}-\gamma}\lrp{R_{\bv_{i_{\bv}}^h}^{n,\omega}}&< 3\varepsilon. 
	%&<\frac{\varepsilon}{8}+\frac{\varepsilon}{8},
	\end{align*}

	And the proof ends because \eqref{eq:proof_Juan7} and previous inequalities give that if $\omega\in A_0$ and $n\geq N^{\omega}$, then
	\begin{align*}
	\left| \hat{M}_{\bv}-\hat{M}{\bv_{i_{\bv}}^h}\right|<6\varepsilon.
	\end{align*}
\end{proof}

\begin{proof}[Proof of Theorem \ref{theo:assymp3}]
	Let $\omega\in\Upsilon$ and denote
	\begin{align*}
	g_{a,b}^{n,\omega}(\bx):=\frac{\Prob\lrp{\bv:  \frac{\left|\bx'\bv-\hat{m}_\bv^{n,\omega}\right|}{\hat{M}_\bv^{n,\omega}}>b}}{\Prob\lrp{\bv:  \frac{\left|\bx'\bv-\hat{m}_\bv^{n,\omega}\right|}{\hat{M}_\bv^{n,\omega}}>b}+\Prob\lrp{\bv:  \frac{\left|\bx'\bv-\hat{m}_\bv^{n,\omega}\right|}{\hat{M}_\bv^{n,\omega}}<a}}.
	\end{align*}
	Notice that the probabilities involved in this expression are conditioned given the sample $\bX_1,\ldots,\bX_n$. It is clear that if we integrate on the samples 
	\begin{align*}
	\Prob\lrp{|\tilde{Y}^{{L}_n}|>b \ | \Vert\bX\Vert_{\mSigma}=t}=&\displaystyle\int_{\Omega^{d-1}_{\mSigma}(t)}\lrp{\int_\Upsilon g_{a,b}^{n,\omega}(\bx)\,d\Prob(\omega)}f_t(\bx)\,d\bx.
	\end{align*}
	
	Let us prove that the map $g_{a,b}^{n,\omega}$ is well defined. As before we denote by $M_1$, the $\MADN$ of the $N_1(0,1)$. If $\bv \in \Omega_1^{d-1}$, then $M_\bv=\lrp{\bv'\mSigma \bv}^{1/2}M_1\geq \lambda_1 M_1$.
	
	According to Theorem \ref{theo:Juan}, there exists a set $A_0\in \mathcal{A}$ with $\Prob(A_0)=1$ such that for every $\omega\in A_0$, there exits $N^\omega$ such that if $n\geq N^\omega$ then for every $\bv\in \Omega_1^{d-1}$, $\hat{M}^{n,\omega}_\bv>\lambda_1M_1/2$ and $\left\vert\hat{m}_{\bv}^{n,\omega}\right\vert<a\lambda_1M_1/4$. Then, if $\bx\in \Omega_{\mSigma}^{d-1}(t)$
	\begin{align*}
	\Prob\lrp{\bv: \tfrac{\left|\bx'\bv-\hat{m}_\bv^{n,\omega}\right|}{\hat{M}_\bv^{n,\omega}}<a}&=\Prob\lrp{\bv: \bx'\bv\in \lrp{\hat{m}_\bv^{n,\omega}-a\hat{M}_\bv^{n,\omega},\hat{m}_\bv^{n,\omega}+a\hat{M}_\bv^{n,\omega}}}\\
	&\geq \Prob\lrp{\bv: \bx'\bv\in\lrp{\hat{m}_\bv^{n,\omega}-\frac{a\lambda_1M_1}{2},\hat{m}_\bv^{n,\omega}+\frac{a\lambda_1M_1}{2}}}\\
	&\geq \Prob\lrp{\bv:\bx'\bv\in\lrp{-\frac{a\lambda_1M_1}{4},\frac{a\lambda_1M_1}{4}}}>0,
	\end{align*}
	where the last inequality follows from the fact that $\left\{\bv: |\bx'\bv|< a\lambda_1M_1/4\right\}\neq\emptyset$.

	Additionally, 
	for every $\omega\in A_0$, $\bv\in\Omega_1^{d-1}$ and $\bx\in \Omega_{\mSigma}^{d-1}(t)$
	\begin{align}\label{eq:proof_prop6}
	1_{\left\{\frac{\left|\bx'\bv-\hat{m}_\bv^{n,\omega}\right|}{\hat{M}_\bV^{n,\omega}}>b\right\}}\to 1_{\left\{\frac{\left|\bx'\bv\right|}{M_\bv}>b\right\}}
	\end{align}
	unless $\bv$ satisfies that $|\bx'\bv|=bM_\bv$, but this equality only happens for $\bv$ in a set (depending on $\bx$) with Lebesgue measure equal to zero. Consequently, for every $\omega\in A_0$ and $\bx\in \Omega_{\mSigma}^{d-1}(t)$, the convergence in \eqref{eq:proof_prop6} holds for almost every $\bv\in\Omega_1^{d-1}$. Since the involved functions are bounded, \eqref{eq:proof_prop6} gives that, for every $\omega\in A_0$,
	\begin{align*}
	g_{a,b}^{n,\omega}(\bx)\to \tilde{g}_{a,b}(\bx).
	\end{align*}
	
	The fact that $0\leq g_{a,b}^{n,\omega}(\bx)\leq 1$ for every $\bx$, allows to apply the dominated convergence theorem and the result is proven.
\end{proof}

%***********************************
\subsection{Proofs of Section \ref{subsec:comp_a_b}}
%***********************************
\begin{proof}[Proof of Proposition \ref{prop:worst_case}]
	We know $a\leq b$. If $a=b$, then 
	\begin{align}\label{eq:proof_unique_b}
	\alpha=\displaystyle\int_{\Omega_{\mSigma}^{d-1}(C_n^d)}\int_{\mathbb{R}^d}\int_{\mathbb{R}^{d^2}}\Prob\lrp{|y_{\bmm,S}^\bV|>b }f_{\hat{\bmu}} (\bmm) f_t(\bx)\,\Prob_{\hat{\mSigma}}(dS)\,d\bmm\,d\bx.
	\end{align}
	This condition determines $b$ because the function $b\mapsto \Prob\lrp{|y_{\bmm,S}^\bV|>b }$ is strictly decreasing and continuous for any $\bmm$, $S$, and $\bx$. Let $b_0$ be the unique solution of \eqref{eq:proof_unique_b}. If $a<b_0$, there exists a unique $b_a$ such that 
	\begin{align*}
	\alpha=F_{\mSigma}(a,b_a,C_n^d),
	\end{align*}
	because the integrand which implicitly appears in $F_{\mSigma}(a,b,C_n^d)$ (see Proposition \ref{prop:probabilidad2}) is strictly increasing on $b$ and continuous.
	
	If $a_1<a_2$, then $g_{a_2}^{b_{a_1}}(\bx,\bmm,S)< g_{a_1}^{b_{a_1}}(\bx,\bmm,S)$ since $\Prob\lrp{|y_{\bmm,S}^\bV|<a_1}<\Prob\lrp{|y_{\bmm,S}^\bV|<a_2}$. Hence $F_{\mSigma}(a_2,b_{a_1},t)>F_{\mSigma}(a_1,b_{a_1},t)$, then $b_{a_2}<b_{a_1}$.
\end{proof}

\begin{proof}[Proof of Proposition \ref{prop:EK_comparison}]
	Remark \ref{remark:expectation_lim} gives that, for  a general $\mSigma$, a.s.
	\begin{equation}\label{Eq.UltimaProp.1}
	E\left( {L}_n^{a,b} \ | \Vert\bX\Vert_{\mSigma}=t,\mathbb{X}_n^\mSigma\right) 
	\to
	\int_{\Omega_{\mSigma}^{d-1}(t)} \frac{1}{\Prob(|y^\bv_{\mathbf{0},\mSigma}|>b)+\Prob(|y^\bv_{\mathbf{0},\mSigma}|<a)}f_t(\bx)\,d\bx,
	\end{equation}
	which, in the case $\mSigma=I_d$, (see Corollary \ref{cor:esperanza2}) becomes that a.s.
	\begin{equation}\label{Eq.UltimaProp.2}
	E\left( {L}_n^{a,b} \ | \Vert\bX\Vert=t,\mathbb{X}_n^{I_d}\right) 
	\to
	\frac{1}{1-F(b,t)+F(a,t)}.
	\end{equation}
	However, Jensen's inequality gives
	\begin{align*}
	\frac{1}{1-F(b,t)+F(a,t)}
	&=%\frac{1}{1-\int_{\Omega_{\mSigma}^{d-1}(t)} (\Prob(|Y_{\mathbf{0},\mSigma}|<b)-\Prob(|Y_{\mathbf{0},\mSigma}|<a)) f_t(\bx) \,d\bx}\\
	\frac{1}{\int_{\Omega_{\mSigma}^{d-1}(t)} \Prob(|y^\bv_{\mathbf{0},\mSigma}|>b)+\Prob(|y^\bv_{\mathbf{0},\mSigma}|<a) f_t(\bx) \,d\bx}\\
	&< \int_{\Omega_{\mSigma}^{d-1}(t)} \frac{1}{\Prob(|y^\bv_{\mathbf{0},\mSigma}|>b)+\Prob(|y^\bv_{\mathbf{0},\mSigma}|<a)}f_t(\bx)\,d\bx,
	\end{align*}
	where the inequality comes from the fact that the map  $\bx\mapsto \Prob(|y^\bv_{\mathbf{0},\mSigma}|>b)+\Prob(|y^\bv_{\mathbf{0},\mSigma}|<a)$ is not a.s. constant.
\end{proof}

%*****************************************************
\section{Appendix II. Additional material} \label{App:AdditionalMaterial}
%*****************************************************
\subsection{Behaviour of $C_n^d(\delta)$}\label{subsubsec:C_n}
%*****************************************************
Table \ref{tab:valores_C_nd} illustrates the variation of  $C_n^d$. In this table,    it is evident that, even for small sizes, the value of $C_n^d$ grows faster on the dimension than on the sample size.

\begin{table}[!htb]
	\begin{center}
		\caption{Values of $C_n^d(\delta)$ for different dimensions, sample sizes and $\delta=0.05$.}
		\label{tab:valores_C_nd}
		\begin{tabular}{@{}rrrrrrr@{}}
			\hline\noalign{\smallskip}
			%& \multicolumn{6}{c}{$n$}\\ 
			$d$ & $n=10$ & $n =20$ & $n=50$ & $n=100$  & $n=200$ & $n=1000$ \\
			\noalign{\smallskip}\hline\noalign{\smallskip}			
			50 & 8.91 & 9.09 & 9.30 & 9.46 & 9.79  & 9.93\\
			
			200 & 15.97 & 16.14 & 16.35 & 16.50 & 16.64 & 16.95\\
			
			500 & 24.19 & 24.35 & 24.56 & 24.71 & 24.85 & 25.15\\
			
			1000 & 33.44 & 33.61 & 33.82  & 33.96 & 34.10 & 34.40\\
			\noalign{\smallskip}\hline
		\end{tabular}
	\end{center}
\end{table}

%*****************************************************
\subsection{Graphical representation of the formulae in Corollary \ref{cor:esperanza2}}\label{subsubsec:graph}
\begin{figure}
	\centering
	\includegraphics[width=.6\textwidth]{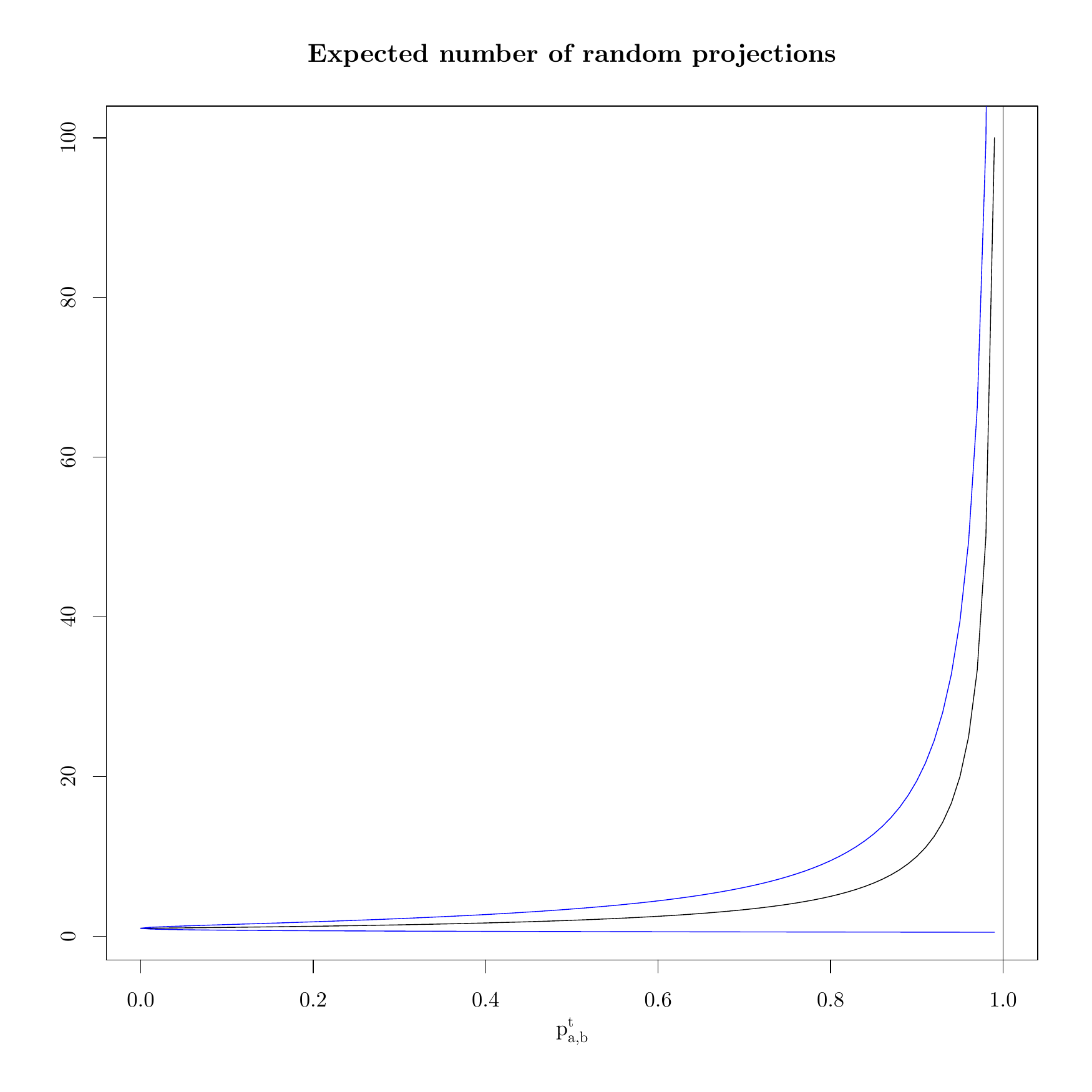}
	\caption{Curves $\E(K_n\vert\Vert\bX\Vert=t)$ (black) and $\E(K_n\vert\Vert\bX\Vert=t)\pm\lrp{\Var(K_n\vert\Vert\bX\Vert=t)}^{1/2}$ (blue)}
	\label{fig:EK}       
\end{figure}

%*****************************************************
Figure \ref{fig:EK} shows the curves $t \to \E(K_n\vert\Vert\bX\Vert=t)\pm\lrp{\Var\lrp{K_n\vert\Vert\bX\Vert=t}}^{1/2}$ and $t \to \E(K_n\vert\Vert\bX\Vert=t)$. Since  those curves only depend on $p_{a,b}^t:=\Prob\lrp{\vert Y^\bV\vert\in (a,b)\vert\Vert\bX\Vert=t}$, those are the values that we represent in the axis of abscissas .

%*****************************************************
\subsection{Some notes on the use of the constants $a_I$ and $b_I$ when we handle general $\mSigma$'s}\label{Sec.a_bDeLaIdentidad}
%*****************************************************
In this section we provide some guidelines of a research now in progress to find conditions allowing to use the constants $a,b$ associated to the identity with other covariance matrices when the dimension is large.

An important piece will be the following simple  lemma.

\begin{lemma}\label{Lemm.SLLN}
	Let $\{\mSigma^d\}_d$ be a sequence of diagonal $d\times d$-dimensional matrices with eigenvalues $0<\sigma_1^d\leq \ldots \leq \sigma_d^d$. 
	Let us assume
	
	\begin{itemize}
		\item[A.1]\label{Eq.SLLN}
		The sequence $\{d\sigma_d^d/\mbox{trace}(\mSigma^d)\}_d$ is bounded.  
	\end{itemize} 
	
	If $\{Z_d\}$ is a sequence of integrable iid real  rv's, then 
	\[
	\frac 1{\mbox{trace}(\mSigma^d)} \sum_{i=1}^d \sigma_i^d Z_i 
	\convs
	\E[X_1].
	\]
\end{lemma}
\begin{proof}
	Obviously
	\[
	\left(\frac 1{\mbox{trace}(\mSigma^d)} \sum_{i=1}^d \sigma_i^d Z_i\right) -\E[Z_1]
	=
	\frac 1 d \sum_{i=1}^d 
	\frac{d\sigma_i^d}{\mbox{trace}(\mSigma^d)} (
	Z_i-\E[Z_1])
	.
	\]
	Since the rv's in the sum in the right hand side are stochastically bounded by \linebreak $\sup_d
	\left(\frac{d\sigma_d^d}{\mbox{trace}(\mSigma^d)}
	\right)(|Z_1|+\E[|Z_1|])$ which is integrable, we can apply  Theorem 2 in \cite{Huetal89} and the result follows.
\end{proof}

Let us assume that the parameters are known. Therefore, the statement in Proposition \ref{prop:probabilidad2} becomes
\[
F_{\mSigma^d}(a,b,t)=
\int_{\Omega^{d-1}_{\mSigma}(t)}  g_a^b(\bx_d,\mathbf{0},\mSigma^d) f_t(\bx_d)\,
d\bx_d,
\]
where we write $\bx_d$ and $\mSigma^d$ instead of $\bx$ and $\mSigma$ to emphasize their dependence on the dimension. We will denote with $\sigma_1^d,\ldots,\sigma_d^d$ to the eigenvalues of $\mSigma^d$.

Let us analyse the numerator of $g_a^b(\bx_d,\mathbf{0},\mSigma^d) $. Let $\{X_d\}$ and $\{V_d\}$ be two iid sequences of one-dimensional standard normal variables. Let us denote \linebreak $\bX_d:=(X_1,\ldots,X_d)'$ and $\bV_d:=(V_1,\ldots,V_d)'$. Obviously, $\bX_d$ and $\bV_d$ are two independent  rv's with distribution standard normal $d$-dimensional.  We also define 
\begin{equation}\label{Eq.LaMadreDelCordero}
{\cal F}_d(\bX_d,V_d) :=\frac1 {\|\bX_d\|} \frac{((\mSigma^d)^{1/2}\bX_d)'\bV_d}{(\bV_d'\mSigma^d \bV_d)^{1/2}}
= 
\frac1 {\|\bX_d\|} \frac{\sum_{i=1}^d (\sigma_i^d)^{1/2}X_iV_i}{\left(\sum_{i=1}^d \sigma_i^d V_i^2\right)^{1/2}}.
\end{equation}

With this notation, we have that
\[
\Prob\left( \vert y_{\mathbf{0},\mSigma^d}^{\bV_d}\vert > b  \right)
=^d \Prob\left( t\; {\cal F}_d(\bX_d,V_d) > b \ | \ \bX_d =\bx_d\right).
\]

Next we divide the analysis depending on when Assumption A.1 holds or not.

\subsubsection{Case 1: Assume that assumption A.1 holds} 

Since A.1 holds, Lemma \ref{Lemm.SLLN} allows us to conclude that
\begin{equation}\label{Eq.SLLN_2}
\frac{\sum_{i=1}^d \sigma_i V_i^2}{\mbox{trace}(\mSigma^d)}
\convs \E[V_1^2] = 1.
\end{equation}

However, the assumption on the vector $\bV_d$ gives that the conditional distribution of the real rv $\sum_{i=1}^d \sigma_i^{1/2}X_iV_i/\linebreak (\mbox{trace}(\mSigma^d))^{1/2}$ given $\bX_d$ is centred normal with variance 
$\frac 1 {\mbox{trace}(\mSigma^d)}\sum_{i=1}^d \sigma_iX_i^2$. Now, Lemma \ref{Lemm.SLLN} gives that, for almost every sequence $\bX_d$, it happens that
$$
\frac 1 {\mbox{trace}(\mSigma^d)}\sum_{i=1}^d \sigma_iX_i^2
\to \E[X_1^2]=
1.$$

The Strong Law of Large Numbers allows to conclude that, for almost every sequence $\bX_d$, it happens that
$
\frac 1 {d^{1/2}} \|\bX_d\| \to 1.
$
Therefore we have that, for almost every sequence $\bX_d$, the expression \eqref{Eq.LaMadreDelCordero}, asymptotically, behaves like a one-dimensional centred normal with variance equal to $d^{-1}$.

The following lemma follows form Tchebichev's inequality and allows to generalise assumption A.1 at the price of replacing the a.s. convergence by convergence in probability. This would allow to obtain similar conclusions as before with a more elaborated reasoning.

\begin{lemma}\label{Lemm.SLLN_2}
	If we replace A.1 in Lemma \ref{Lemm.SLLN} by
	\[
	\lim_d \frac{(\sigma_1^d)^2+\ldots+(\sigma_d^d)^2}{(\sigma_1^d+\ldots+\sigma_d^d)^2}
	=0.
	\]
	and we assume that $\mbox{Var}(Z_1)<\infty$, then 
	\[
	\frac 1{\mbox{trace}(\mSigma^d)} \sum_{i=1}^d \sigma_i^d Z_i 
	\convp
	\E [Z_1].
	\]
\end{lemma}

\begin{remark}\label{Nota.Ctes.Id_1}
	Since $\sigma_1^d\geq 1$, the fact that $\{d\sigma_d^d/\mbox{trace}(\mSigma^d)\}_d$ is bounded holds every time the sequence $\{\sigma_d^d\}$ is bounded, but it also holds for the matrices $\mSigma_1^d$ and $\mSigma_2^d$ (notice that in the last case, $\mbox{trace}(\mSigma_2^d)\geq \frac d 2 \frac{d^2}2$). This shows that all those cases behave similarly and, in particular, that the term in the numerator of $g_a^b(\bx_d,\mathbf{0},\mSigma^d)$ behaves similarly to the case in which $\mSigma^d$ is the identity.
	
	Since the second term in the denominator is similar to that one in the numerator, we can make an educated guessing that, when the dimension is large, if we handle matrices which satisfy A.1, then the constants $a,b$ chosen for the identity should be very similar to the required for the specific matrix at hand.	
\end{remark}

\subsubsection{Case 2: Assume that assumption A.1 fails} 
This case is clearly more tricky than the previous one and here some situations appear in which the equivalence with the identity fails.

For an example, let us consider the case of the sequence of matrices $\mSigma_3^d$. Remember that here $1=\sigma_1^d=\ldots = \sigma_{d-1}^d$ and $\sigma_d^d=d^2$. Thus, $\mbox{trace}(\mSigma_3^d)=d-1 + d^2$, and consequently,
\begin{eqnarray*}
	\frac{\sum_{i=1}^d \sigma_i^d V_i^2}{\mbox{trace}(\mSigma^d)}
	&\approx& V_d^2,
	\\
	\frac 1 {(\mbox{trace}(\mSigma^d))^{1/2}}\sum_{i=1}^d \sigma_i^{1/2}X_iV_i &\approx& \bX_d V_d,
\end{eqnarray*}
therefore, in this case, the behaviour of the three terms in ${\cal F}_d(\bX_d,V_d)$ asymptotically on $d$ is determined by the distribution  of the real rv $d^{-1/2}\frac{X_1}{|V_1|}$, which is of the same order as the normal appearing in Case 1 but different from that.

It seems plausible this being the reason of the different behaviour of this covariance matrix w.r.t. the others.

\begin{remark}
	Obviously many different combinations of asymptotical distributions can be obtained in this scheme, however, we conjecture that all of them will involve random variables with order $d^{-1/2}$ which may explain why the constants obtained for the identity are not so different from the real ones when we consider $\mSigma_3^d$.
	
\end{remark}

\begin{remark}\label{Nota.Ctes.Id_2} 
	Notice that the precise value of $\sigma_d^d$ is not so important as the number of times in which it is reached, because  $\sigma_d^d=d^2$ for $\mSigma_i^d, i=1,2,3$, the difference being that the order of the maximum is reached only in a value for $\mSigma_3^d$ and in a half (or more) of them for $\mSigma_i^d,i=1,2$.
	
	Thus, we conjecture that the constants of the identity could be used for every covariance matrix, excepting  those included in a sequence such that $\sigma_d^d/d \to \infty$  and that, additionally, satisfies something similar to the existence of  $\{p_d\}\subset \mathbb{N}$ such that
	\[
	p_d \leq d, \ \frac {p_d}d \to 1 \ \mbox{ and } \ \frac 1 {\sigma_d^d}\sum_{i=1}^{p_d} \sigma_i^d \to 0.
	\]
	
\end{remark}

%*****************************************************
\subsection{Additional tables}\label{subsubsec:tables}
%*****************************************************
%*****************************************************
\subsubsection{Computation of $a,b$ for general covariance matrices}
%*****************************************************
\label{subsubsec:Computing_b}

Table \ref{tab:comparison_valores_abs1} shows the values of $b$ for some covariance matrices $\mSigma$, $n$ and $d$.

\begin{table}[h]
	\caption{Approximated values of $b_\mSigma$ for $\mSigma=\mSigma_1^d$ and different values of $n$ and $d$. The $a$'s are the values obtained in Table \ref{tab:comparison_valores_ab2} for $l_{I}^1=50,100$. The values of $\hat l_1^1$ are also shown.}
	\label{tab:comparison_valores_abs1}
	\begin{center}
		\begin{tabular}{rl|c@{\hskip .12cm}c@{\hskip .23cm}c@{\hskip .12cm}r@{\hskip .23cm}|c@{\hskip .23cm}c@{\hskip .12cm}c@{\hskip .23cm}c@{\hskip .12cm}r@{\hskip .23cm}|c@{\hskip .23cm}c@{\hskip .12cm}c@{\hskip .23cm}c@{\hskip .12cm}r}
			\hline\noalign{\smallskip}
			& & \multicolumn{4}{c|}{$n=50$} && \multicolumn{4}{c|}{$n=100$} && \multicolumn{4}{c}{$n=500$}\\
			\noalign{\smallskip}\hline\noalign{\smallskip}
			& &  \multicolumn{2}{c}{$l_{I}^1$=50} &  \multicolumn{2}{c|}{$l_{I}^1$=100} &&  \multicolumn{2}{c}{$l_{I}^1$=50} &  \multicolumn{2}{c|}{$l_{I}^1$=100} &&  \multicolumn{2}{c}{$l_{I}^1$=50} &  \multicolumn{2}{c}{$l_{I}^1$=100}\\
			\noalign{\smallskip}\hline\noalign{\smallskip}
			$d$& $\mSigma$ & $b_\mSigma$ & $\hat l_{1}^1$ & $b_\mSigma$ & $\hat l_{1}^1$ && $b_\mSigma$ & $\hat l_{1}^1$ & $b_\mSigma$ & $\hat l_{1}^1$ && $b_\mSigma$ & $\hat l_{1}^1$ & $b_\mSigma$ & $\hat l_{1}^1$\\
			\noalign{\smallskip}\hline\noalign{\smallskip}
			50 & $\mSigma_1$ & 5.0236 & 49 & 5.3996 & 99 && 4.6351 & 51 & 4.9236 &97 &&  4.4517 & 48 & 4.7039 & 100\\
			& $\mSigma_2$  & 4.9995 & 50 & 5.4076 & 100 && 4.6284 & 48 & 4.9289 & 99 && 4.4522 & 48 & 4.6932 & 101\\
			& $\mSigma_3$  &  5.1413 & 189 & 5.4932 & 389 && 4.6374 & 192 & 4.9504 & 384 && 4.4439 & 180 & 4.6858 & 381\\
			& $\mSigma_4$  &  5.0236 & 51 & 5.3798 & 125 && 4.6194 &  49 & 4.9075 & 103 && 4.4520 & 51 & 4.7080 & 99\\
			\noalign{\smallskip}\hline\noalign{\smallskip}
			100 & $\mSigma_1$ &4.7425 & 51 & 5.0891 & 101 && 4.3529 & 51 & 4.6494 & 93 &&  4.1477 & 50 & 4.4083 & 99\\
			& $\mSigma_2$  & 4.7425 & 50 & 5.1213 & 100 && 4.3538 & 52 & 4.6387 & 99 && 4.1471 & 49 &  4.4056 & 98\\
			& $\mSigma_3$  & 4.8813 & 260 & 5.1857 & 541 && 4.3539 & 269 & 4.6494 & 525 && 4.1399 & 260 & 4.3691 & 530\\
			& $\mSigma_4$  & 4.7523 & 48 & 5.1052 & 110 && 4.3497 & 51 & 4.6467 & 98 && 4.1458 & 49 & 4.3922 & 100\\
			\noalign{\smallskip}\hline\noalign{\smallskip}
			500 & $\mSigma_1$ & 4.3012 & 51 & 4.6556 & 100 && 3.9438 & 49 & 4.2325 & 99 && 3.7279 & 48 & 3.9533 & 102\\
			& $\mSigma_2$ & 4.3244 & 48 & 4.6361 & 99 && 3.9701 & 49 &  4.2069 & 99 && 3.7421 & 51 & 3.9509 & 97\\
			& $\mSigma_3$  &4.3244 & 580 & 4.6946 & 1162 && 4.0248 & 590 & 4.2460 & 1180 && 3.7421 & 580 & 3.9143 & 1195\\
			& $\mSigma_4$  & 4.3219 & 49 & 4.6166 & 100 && 3.9613 & 51 & 4.2216 & 101 && 3.7509 & 49 & 3.9475 & 102\\
			\noalign{\smallskip}\hline\noalign{\smallskip}
			1000 & $\mSigma_1$ & 4.2058 & 50 & 4.5254 & 100 && 3.8623 & 48 & 4.1146 & 101 &&  3.6276 & 49 & 3.8192& 98\\
			& $\mSigma_2$  & 4.2272 & 50 & 4.5385 & 100 && 3.8442 & 48 & 4.1094 & 101 && 3.6236 & 50 & 3.8363 & 101\\
			& $\mSigma_3$  & 4.3129 & 830 & 4.6166 & 1678 && 3.9221 & 805 & 4.1094 & 1610 && 3.6080 & 834 & 3.8168 & 1620\\
			& $\mSigma_4$  & 4.2272 & 49 & 4.5385 & 102 && 3.8442 & 51 & 4.1094 & 100 && 3.6080 & 50 &  3.8326 & 101\\
			\noalign{\smallskip}\hline
		\end{tabular}
		
	\end{center}
\end{table}

%*****************************************************
\subsubsection{Computational times}
%*****************************************************
\label{subsubsec:Computational times}
Table \ref{tab:ab_N10^5} shows the required time to compute the values of $a$ and $b$ for several combinations of dimensions, sample sizes and expected number of projections. The computation was carried out in a four cores processor 3.2 GHz Intel Core i5. The decrement observed in the cases $n=100$, $l_I^1=100$ are due to the fact that those cases required a very sort bisection step.

\begin{table}[!htb]
	\caption{Computation times (in seconds) of $a,b$ with $N=10^5$ simulated values of $\tilde Y^\bV$.}
	\label{tab:ab_N10^5}%set.seed(123875)
	\begin{center}
		\begin{tabular}{r|rr|rr|rr}
			\hline\noalign{\smallskip}
			& \multicolumn{2}{c}{$n=50$} & \multicolumn{2}{|c}{$n=100$}& \multicolumn{2}{|c}{$n=500$}
			\\
			$d$ & $l_I^1=50$ & $l_I^1=100$ & $l_I^1=50$ & $l_I^1=100$ & $l_I^1=50$ & $l_I^1=100$\\ 
			\noalign{\smallskip}
			\hline
			\noalign{\smallskip}
			$50$ &37.972 &142.930 &74.400 & 55.347 &92.132 & 321.988
			\\
			$1000$ & 184.780 & 624.712 & 282.090 & 535.981 & 1047.797 & 1982.664
			\\
			\noalign{\smallskip}\hline    
		\end{tabular}
	\end{center}
\end{table}

%*****************************************************
\subsubsection{Detecting outliers}
%*****************************************************
\label{subsubsec:Detecting_Outliers}

Table \ref{tab:proof_prob_depend_pto1R_app} is  Table \ref{tab:proof_prob_depend_pto1R_n50} expanded to $n=100,500$. It shows the values of $\hat l_\Sigma$ and the proportion of times that  a point with Mahalanobis norm $C_n^d$ was identified as an outlier.

% ROBUST

\begin{table}[!htb]
	\caption{Estimation of the probability of declaring as an outlier a vector such that $\left\Vert \bX \right\Vert_{\mSigma}= C_n^d$, for several values of $n, d$ and $\mSigma$. We also show the sample means of $L_n$.}
	\label{tab:proof_prob_depend_pto1R_app}
	\begin{center}
		\begin{tabular}{cc|rcrcrcrcrc}
			\hline\noalign{\smallskip}
			& & \multicolumn{10}{c}{$d=50$}\\
			\cmidrule(r){2-12}
			$n$& $l_{I}^1$ &$\hat l_{I}^1$ & $I_d$ & $\hat  l_{1}^1$ & $\mSigma_1^d$ & $\hat l_{2}^1$ & $\mSigma_2^d$ & $\hat l_{3}^1$ & $\mSigma_3^d$ & $\hat l_{4}^1$ & $\mSigma_4^d$ \\
			\noalign{\smallskip}\hline\noalign{\smallskip}
			$50$ & 50 & 51 & 0.0528 & 49 & 0.0571 & 49 & 0.0541 & 186 & 0.0668 & 50 & 0.0569\\
			& 100 & 98 & 0.0560 & 99 & 0.0558 & 103 & 0.0553 & 366 & 0.0580 & 99 & 0.0572\\
			\noalign{\smallskip}\hline\noalign{\smallskip}
			$100$ & 50  & 50 & 0.0497 & 52 & 0.0528 & 50 & 0.0511 & 186 & 0.0566 & 49 & 0.0477\\
			& 100 & 102 & 0.0478 & 101 & 0.0535 & 100 & 0.0535 & 382 & 0.0510 & 101 & 0.0539\\
			\noalign{\smallskip}\hline\noalign{\smallskip}
			$500$ & 50 & 50 & 0.0501 & 50 & 0.0537 & 50 & 0.0484 & 194 & 0.0537 & 50 & 0.0558\\
			& 100 &99 & 0.0533 & 99 & 0.0484 & 99 & 0.0492 & 387 & 0.0499 & 98 & 0.0487\\
			\noalign{\smallskip}\hline\noalign{\smallskip}
			& & \multicolumn{10}{c}{$d=100$}\\
			\cmidrule(r){2-12}
			$n$& $l_{I}^1$ &$\hat l_{I}^1$ & $I_d$ & $\hat l_{1}^1$ & $\mSigma_1^d$ & $\hat l_{2}^1$ & $\mSigma_2^d$ & $\hat l_{3}^1$ & $\mSigma_3^d$ & $\hat l_{4}^1$ & $\mSigma_4^d$ \\
			\noalign{\smallskip}\hline\noalign{\smallskip}
			$50$ & 50  & 49 & 0.0507 & 48 & 0.0496 & 50 & 0.0501 & 249 & 0.0628 & 50 & 0.0489\\ 
			& 100 & 100 & 0.0538 & 101 & 0.0519 & 100 & 0.0526 & 526 & 0.0603 & 98 & 0.0494\\
			\noalign{\smallskip}\hline\noalign{\smallskip}
			$100$ & 50  & 50 & 0.0535 & 51 & 0.0496 & 48 & 0.0548 & 251 & 0.0560 & 49 & 0.0528\\
			& 100 & 99 & 0.0479 & 99 & 0.0495 & 99 & 0.0490 & 508 & 0.0540 & 99 & 0.0481\\
			\noalign{\smallskip}\hline\noalign{\smallskip}
			$500$ &50 & 50 & 0.0547 & 49 & 0.0538 & 50 & 0.0538 & 264 & 0.0573 & 50 & 0.0536\\
			& 100 & 97 & 0.0557 & 98 & 0.0572 & 100 & 0.0523 & 516 & 0.0494 & 97 & 0.0507\\
			\noalign{\smallskip}\hline\noalign{\smallskip}
			& & \multicolumn{10}{c}{$d=500$}\\
			\cmidrule(r){2-12}
			$n$& $l_{I}^1$ &$\hat l_{I}^1$ & $I_d$ & $\hat l_{1}^1$ & $\mSigma_1^d$ & $\hat l_{2}^1$ & $\mSigma_2^d$ & $\hat l_{3}^1$ & $\mSigma_3^d$ & $\hat l_{4}^1$ & $\mSigma_4^d$ \\
			\noalign{\smallskip}\hline\noalign{\smallskip}
			$50$ & 50  & 49 & 0.0481 & 50 & 0.0507 & 50 & 0.0518 & 552 & 0.0628 & 50 & 0.0483\\
			&100 & 100 & 0.0520 & 102 & 0.0509 & 99 & 0.0545 & 1111 & 0.0589 & 101 & 0.0538\\
			\noalign{\smallskip}\hline\noalign{\smallskip}
			$100$ & 50  & 49 & 0.0543 & 50 & 0.0532 & 50 & 0.054 & 571 & 0.0588 & 50 & 0.0527\\
			&100 & 101 & 0.0522 & 102 & 0.0518 & 100 & 0.0511 & 1136 & 0.0549 & 99 & 0.0550\\
			\noalign{\smallskip}\hline\noalign{\smallskip}
			$500$& 50   & 51 & 0.0508 & 50 & 0.0530 & 50 & 0.0502 & 596 & 0.0538 & 50 & 0.0506\\
			& 100 & 103 & 0.0520 & 100 & 0.0540 & 102 & 0.0470 & 1199 & 0.0489 & 102 & 0.0505\\
			\noalign{\smallskip}\hline\noalign{\smallskip}
			&  &\multicolumn{10}{c}{$d=1000$}\\
			\cmidrule(r){2-12}
			$n$& $l_{I}^1$ &$\hat l_{I}^1$ & $I_d$ & $\hat l_{1}^1$ & $\mSigma_1^d$ & $\hat l_{2}^1$ & $\mSigma_2^d$ & $\hat l_{3}^1$ & $\mSigma_3^d$ & $\hat l_{4}^1$ & $\mSigma_4^d$ \\
			\noalign{\smallskip}\hline\noalign{\smallskip}
			$50$ & 50  & 50 & 0.0496 & 50 & 0.0538 & 49 & 0.0534 & 790 & 0.0586 & 50 & 0.0500\\
			& 100 & 100 & 0.0520 & 101 & 0.0476 & 102 & 0.0507 & 1601 & 0.0549 & 99 & 0.0553\\
			\noalign{\smallskip}\hline\noalign{\smallskip}
			$100$ & 50 & 49 & 0.0564 & 50 & 0.0556 & 50 & 0.0509 & 775 & 0.0599 & 50 & 0.0513\\
			& 100 & 100 & 0.0516 & 101 & 0.0518 & 101 & 0.0534 & 1650 & 0.0545 & 101 & 0.0571\\
			\noalign{\smallskip}\hline\noalign{\smallskip}
			$500$ & 50 & 51 & 0.0588 & 50 & 0.0508 & 49 & 0.0536 & 830 & 0.0543 & 49 & 0.0539\\
			& 100 &100 & 0.0500 & 100 & 0.0529 & 98 & 0.0569 & 1688 & 0.0444 & 100 & 0.0568\\
			\noalign{\smallskip}\hline
			
		\end{tabular}
	\end{center}
\end{table}

Tables  \ref{tab:proof_prob_depend_pto1_2R_app} and \ref{tab:proof_prob_depend_pto2R.app} are the expansion of the  Table \ref{tab:proof_prob_depend_pto2R_n50} to $n=100$ and $n=500$.

\begin{table}[!htb]
	\caption{Estimation of the probability of declaring as an outlier a vector such that $\left\Vert \bX \right\Vert_{\mSigma}= 1.2 C_n^d$, for several values of $n, d$ and $\mSigma$. We also show the sample means of $L_n$.}
	\label{tab:proof_prob_depend_pto1_2R_app}
	\begin{center}
		\begin{tabular}{cc|rcrcrcrcrc}
			\hline\noalign{\smallskip}
			& & \multicolumn{10}{c}{$d=50$}\\
			\cmidrule(r){2-12}
			$n$& $l_{I}^1$ & $\hat l_I^{1.2}$ & $I_d$ & $\hat l_1^{1.2}$ & $\mSigma_1^d$ & $\hat l_2^{1.2}$ & $\mSigma_2^d$ & $\hat l_3^{1.2}$ & $\mSigma_3^d$ & $\hat  l_4^{1.2}$ & $\mSigma_4^d$ \\
			\noalign{\smallskip}\hline\noalign{\smallskip}
			$ 50$ & $50$&  48 & 0.2378 & 48 & 0.2247 & 48 & 0.2338 & 163 & 0.1752 & 48 & 0.2333\\
			&$100$ & 93 & 0.2729 & 96 & 0.2412 & 95 & 0.2617 & 313 & 0.1867 & 92 & 0.2639\\
			\noalign{\smallskip}\hline\noalign{\smallskip}
			$ 100$& $50$ & 47 & 0.2618 & 48 & 0.2403 & 47 & 0.2530 & 163 & 0.1764 & 47 & 0.2626\\
			& $100$ & 89 & 0.3057 & 91 & 0.2731 & 88 & 0.2935 & 307 & 0.1968 & 88 & 0.3006\\
			\noalign{\smallskip}\hline\noalign{\smallskip}
			$ 500$ &$50$& 44 & 0.3079 & 45 & 0.2702 & 44 & 0.2897 & 152 & 0.1957 & 45 & 0.2991\\
			&$100$& 82 & 0.3471 & 87 & 0.2985 & 84 & 0.3306 & 296 & 0.2054 & 82 & 0.3412\\
			\noalign{\smallskip}\hline\noalign{\smallskip}
			&  \multicolumn{10}{c}{$d=100$}\\
			\cmidrule(r){2-12}
			$n$& $l_{I}^1$ & $\hat l_I^{1.2}$ & $I_d$ & $\hat l_1^{1.2}$ & $\mSigma_1^d$ & $\hat l_2^{1.2}$ & $\mSigma_2^d$ & $\hat l_3^{1.2}$ & $\mSigma_3^d$ & $\hat  l_4^{1.2}$ & $\mSigma_4^d$ \\
			\noalign{\smallskip}\hline\noalign{\smallskip}
			$ 50$ &$50$& 48 & 0.2235 & 49 & 0.2093 & 48 & 0.2146 & 223 & 0.1729 & 49 & 0.2191\\
			&$100$& 97 & 0.2387 & 97 & 0.2236 & 95 & 0.2320 & 460 & 0.1723 & 96 & 0.2487\\
			\noalign{\smallskip}\hline\noalign{\smallskip}
			$ 100$&$50$ & 46 & 0.2501 & 47 & 0.2331 & 47 & 0.2527 & 218 & 0.1803 & 47 & 0.2571\\
			&$100$& 90 & 0.2795 & 90 & 0.2605 & 91 & 0.2766 & 418 & 0.1832 & 90 & 0.2884\\
			\noalign{\smallskip}\hline\noalign{\smallskip}
			$ 500$&$50$ & 45 & 0.2930 & 46 & 0.2746 & 45 & 0.2863 & 209 & 0.1917 & 45 & 0.3002\\
			&$100$& 84 & 0.3319 & 85 & 0.3080 & 85 & 0.3220 & 409 & 0.2026 & 84 & 0.3296\\
			\noalign{\smallskip}\hline\noalign{\smallskip}
			&  \multicolumn{10}{c}{$d=500$}\\
			\cmidrule(r){2-12}
			$n$& $l_{I}^1$ & $\hat l_I^{1.2}$ & $I_d$ & $\hat l_1^{1.2}$ & $\mSigma_1^d$ & $\hat l_2^{1.2}$ & $\mSigma_2^d$ & $\hat l_3^{1.2}$ & $\mSigma_3^d$ & $\hat  l_4^{1.2}$ & $\mSigma_4^d$\\
			\noalign{\smallskip}\hline\noalign{\smallskip}
			$ 50$&$50$ & 50 & 0.2160 & 48 & 0.2132 & 49 & 0.2168 & 518 & 0.1711 & 50 & 0.2198\\
			&$100$& 97 & 0.2454 & 99 & 0.2375 & 96 & 0.2399 & 973 & 0.1761 & 97 & 0.2412\\
			\noalign{\smallskip}\hline\noalign{\smallskip}
			$ 100$ &$50$ & 46 & 0.2647 & 47 & 0.2575 & 47 & 0.2569 & 474 & 0.1766 & 47 & 0.2558\\
			&$100$& 91 & 0.2795 & 91 & 0.2810 & 90 & 0.2737 & 963 & 0.1821 & 91 & 0.2766\\
			\noalign{\smallskip}\hline\noalign{\smallskip}
			$ 500$&$50$ & 46 & 0.2690 & 46 & 0.2701 & 46 & 0.2831 & 483 & 0.1844 & 46 & 0.2709\\
			&$100$& 89 & 0.3196 & 88 & 0.3139 & 87 & 0.3191 & 961 & 0.1954 & 86 & 0.3104\\
			\noalign{\smallskip}\hline\noalign{\smallskip}
			&  \multicolumn{10}{c}{$d=1000$}\\
			\cmidrule(r){2-12}
			$n$& $l_{I}^1$ & $\hat l_I^{1.2}$ & $I_d$ & $\hat l_1^{1.2}$ & $\mSigma_1^d$ & $\hat l_2^{1.2}$ & $\mSigma_2^d$ & $\hat l_3^{1.2}$ & $\mSigma_3^d$ & $\hat  l_4^{1.2}$ & $\mSigma_4^d$\\
			\noalign{\smallskip}\hline\noalign{\smallskip}
			$ 50$& $50$ & 49 & 0.2202 & 51 & 0.2136 & 49 & 0.2159 & 700 & 0.1632 & 49 & 0.2156\\
			&$100$& 98 & 0.2470 & 97 & 0.2338 & 97 & 0.2429 & 1383 & 0.1616 & 96 & 0.2366\\
			\noalign{\smallskip}\hline\noalign{\smallskip}
			$ 100$&$50$ & 46 & 0.2545 & 46 & 0.2575 & 47 & 0.2536 & 651 & 0.1720 & 46 & 0.2583\\
			&$100$& 92 & 0.2883 & 92 & 0.2828 & 91 & 0.2809 & 1373 & 0.1778 & 90 & 0.2835\\
			\noalign{\smallskip}\hline\noalign{\smallskip}
			$ 500$&$50$& 45 & 0.2790 & 46 & 0.2726 & 46 & 0.2723 & 689 & 0.1823 & 45 & 0.2805\\
			&$100$& 88 & 0.3172 & 85 & 0.3210 & 84 & 0.3213 & 1333 & 0.1963 & 86 & 0.3173\\
			\noalign{\smallskip}\hline
		\end{tabular}
	\end{center}
\end{table}

\begin{table}[!htb]
	\caption{Estimation of the probability of declaring as an outlier a vector such that $\left\Vert \bX \right\Vert_{\mSigma}= 2 C_n^d$, for several values of $n, d$ and $\mSigma$. We also show the sample means of $L_n$.}
	\label{tab:proof_prob_depend_pto2R.app}
	\begin{center}
		\begin{tabular}{cc|rcrcrcrcrc}
			\hline\noalign{\smallskip}
			&  &\multicolumn{10}{c}{$d=50$}\\
			\cmidrule(r){2-12}
			$n$& $l_{I}^1$ & $\hat l_I^{2}$ & $I_d$ & $\hat l_1^{2}$ & $\mSigma_1^d$ & $\hat l_2^{2}$ & $\mSigma_2^d$ & $\hat l_3^{2}$ & $\mSigma_3^d$ & $\hat  l_4^{2}$ & $\mSigma_4^d$ \\
			\noalign{\smallskip}\hline\noalign{\smallskip}
			$ 50$ &$50$& 12 & 0.8817 & 13 & 0.8660 & 12 & 0.8830 & 47 & 0.6575 & 12 & 0.8912\\
			&$100$& 16 & 0.9259 & 19 & 0.9061 & 16 & 0.9153 & 74 & 0.6985 & 16 & 0.9229\\
			\noalign{\smallskip}\hline\noalign{\smallskip}
			$ 100$&$50$&$50$  & 0.9101 & 11 & 0.8880 & 10 & 0.9009 & 40 & 0.6801 & 10 & 0.9053\\
			&$100$ & 12 & 0.9349 & 15 & 0.9257 & 13 & 0.9372 & 62 & 0.7393 & 13 & 0.9375\\
			\noalign{\smallskip}\hline\noalign{\smallskip}
			$ 500$&$50$ & 8 & 0.9193 & 9 & 0.9116 & 9 & 0.9185 & 35 & 0.7134 & 8 & 0.9205\\
			&$100$& 10 & 0.9506 & 12 & 0.9391 & 11 & 0.9509 & 51 & 0.7658 & 10 & 0.9502\\
			\noalign{\smallskip}\hline\noalign{\smallskip}
			& & \multicolumn{10}{c}{$d=100$}\\
			\cmidrule(r){2-12}
			$n$& $l_{I}^1$ & $\hat l_I^{2}$ & $I_d$ & $\hat l_1^{2}$ & $\mSigma_1^d$ & $\hat l_2^{2}$ & $\mSigma_2^d$ & $\hat l_3^{2}$ & $\mSigma_3^d$ & $\hat  l_4^{2}$ & $\mSigma_4^d$ \\
			\noalign{\smallskip}\hline\noalign{\smallskip}
			$ 50$&$50$ & 13 & 0.8829 & 13 & 0.8678 & 13 & 0.8734 & 70 & 0.6289 & 13 & 0.8743\\
			&$100$& 18 & 0.9150 & 19 & 0.9081 & 18 & 0.9115 & 113 & 0.6738 & 18 & 0.9160\\
			\noalign{\smallskip}\hline\noalign{\smallskip}
			$ 100$ &$50$& 10 & 0.9009 & 11 & 0.8924 & 11 & 0.9019 & 56 & 0.6631 & 10 & 0.9003\\
			&$100$& 13 & 0.9369 & 15 & 0.9251 & 14 & 0.9351 & 90 & 0.7050 & 14 & 0.9353\\
			\noalign{\smallskip}\hline\noalign{\smallskip}
			$ 500$&$50$ & 9 & 0.9240 & 9 & 0.9109 & 9 & 0.9134 & 49 & 0.6858 & 9 & 0.9182\\
			&$100$& 11 & 0.9480 & 12 & 0.9379 & 11 & 0.9434 & 74 & 0.7329 & 11 & 0.9446\\
			\noalign{\smallskip}\hline\noalign{\smallskip}
			& & \multicolumn{10}{c}{$d=500$}\\
			\cmidrule(r){2-12}
			$n$& $l_{I}^1$ & $\hat l_I^{2}$ & $I_d$ & $\hat l_1^{2}$ & $\mSigma_1^d$ & $\hat l_2^{2}$ & $\mSigma_2^d$ & $\hat l_3^{2}$ & $\mSigma_3^d$ & $\hat  l_4^{2}$ & $\mSigma_4^d$  \\
			\noalign{\smallskip}\hline\noalign{\smallskip}
			$ 50$&$50$ & 13 & 0.8771 & 13 & 0.8617 & 13 & 0.8780 & 150 & 0.6139 & 13 & 0.8726\\
			&$100$& 18 & 0.9166 & 18 & 0.9185 & 18 & 0.9075 & 249 & 0.6513 & 18 & 0.9090\\
			\noalign{\smallskip}\hline\noalign{\smallskip}
			$ 100$&$50$ & 11 & 0.8985 & 11 & 0.8992 & 11 & 0.8981 & 124 & 0.6488 & 11 & 0.8949\\
			&$100$& 14 & 0.9355 & 15 & 0.9307 & 14 & 0.9273 & 204 & 0.6901 & 14 & 0.9313\\
			\noalign{\smallskip}\hline\noalign{\smallskip}
			$ 500$&$50$ & 9 & 0.9089 & 9 & 0.9133 & 9 & 0.9137 & 114 & 0.6708 & 9 & 0.9113\\
			&$100$& 12 & 0.9451 & 12 & 0.9421 & 12 & 0.9450 & 173 & 0.7119 & 12 & 0.9410\\
			\noalign{\smallskip}\hline\noalign{\smallskip}
			& & \multicolumn{10}{c}{$d=1000$}\\
			\cmidrule(r){2-12}
			$n$& $l_{I}^1$ & $\hat l_I^{2}$ & $I_d$ & $\hat l_1^{2}$ & $\mSigma_1^d$ & $\hat l_2^{2}$ & $\mSigma_2^d$ & $\hat l_3^{2}$ & $\mSigma_3^d$ & $\hat  l_4^{2}$ & $\mSigma_4^d$ \\
			\noalign{\smallskip}\hline\noalign{\smallskip}
			$ 50$&$50$ & 13 & 0.8797 & 13 & 0.8728 & 13 & 0.8729 & 214 & 0.6128 & 13 & 0.8674\\
			&$100$& 19 & 0.9116 & 19 & 0.9134 & 19 & 0.9093 & 360 & 0.6551 & 19 & 0.9124\\
			\noalign{\smallskip}\hline\noalign{\smallskip}
			$ 100$&$50$ & 11 & 0.8994 & 11 & 0.8986 & 11 & 0.8947 & 182 & 0.6508 & 10 & 0.8956\\
			&$100$& 14 & 0.9315 & 15 & 0.9300 & 14 & 0.9300 & 283 & 0.6894 & 14 & 0.9288\\
			\noalign{\smallskip}\hline\noalign{\smallskip}
			$ 500$&$50$ & 9 & 0.9143 & 9 & 0.9105 & 9 & 0.9136 & 154 & 0.6683 & 9 & 0.9080\\
			&$100$& 12 & 0.9409 & 12 & 0.9412 & 12 & 0.9477 & 238 & 0.7128 & 12 & 0.9432\\ 
			\noalign{\smallskip}\hline
		\end{tabular}
	\end{center}
\end{table}

\begin{table}
	\caption{Samples contain 2\% of outliers such that $\Vert\bX\Vert_\mSigma=rC_n^d$ for each $r=1.05,1.25,1.5,2,3$. Columns show the proportion of each type of point declared as outlier along 1000 replications, for several covariance matrices and dimensions with $n=50,100$.}
	% EK = 50
	\label{tab:masking_swamping}
	\begin{center}
		\begin{tabular}{@{\hskip .1cm}r@{\hskip .05cm}r@{\hskip .1cm}|@{\hskip .075cm}c@{\hskip .1cm}|@{\hskip .075cm}c@{\hskip .1cm}c@{\hskip .1cm}c@{\hskip .1cm}c@{\hskip .1cm}c@{\hskip .1cm}|@{\hskip .075cm}c@{\hskip .1cm}|@{\hskip .075cm}c@{\hskip .1cm}c@{\hskip .1cm}c@{\hskip .1cm}c@{\hskip .1cm}c@{\hskip .1cm}}
			\hline\noalign{\smallskip}
			& & \multicolumn{6}{c}{$n=50$} & \multicolumn{6}{@{\hskip -.1cm}|c}{$n=100$}\\
			\hline\noalign{\smallskip}
			$d$ & $\mSigma$ & $<C_n^d$ & $1.05$ & $1.25$ & $1.5$ & $2$ & $3$ & $<C_n^d$ & $1.05$ & $1.25$ & $1.5$ & $2$ & $3$\\
			\noalign{\smallskip}\hline\noalign{\smallskip}
			$50$ & $I_d$ & 0.0039 & 0.0710 & 0.2620 & 0.5860 & 0.8620 & 0.9650 & 0.0015 & 0.0700 & 0.3125 & 0.6325 & 0.8950 & 0.9790 
			\\
			& $\mSigma_1^d$ & 0.0042 & 0.0850 & 0.2470 & 0.5590 & 0.8460 & 0.9670 & 0.0016  & 0.0760 & 0.2700 & 0.5795 & 0.8800 & 0.9655 
			\\
			& $\mSigma_2^d$ & 0.0044& 0.0670 & 0.2730 & 0.5730 & 0.8510 & 0.9710 & 0.0017 & 0.0615 & 0.2810 & 0.6160 & 0.8980 & 0.9710 
			\\
			& $\mSigma_3^d$ & 0.0098 & 0.0810 & 0.1950 & 0.3860 & 0.6500 & 0.8670 & 0.0037 & 0.0635 & 0.1975 & 0.4015 & 0.6660 & 0.8795
			\\
			& $\mSigma_4^d$ & 0.0038 & 0.1030 & 0.2750 & 0.6150 & 0.8630 & 0.9690 	& 0.0017 & 0.0760 & 0.2975 & 0.6370 & 0.8945 & 0.9715 
			\\
			\noalign{\smallskip}\hline\noalign{\smallskip}
			$500$ & $I_d$ & 0.0199 & 0.0700 & 0.2590 & 0.5880 & 0.8620 & 0.9710 & 0.0167  & 0.0775 & 0.3145 & 0.6085 & 0.8955 & 0.9705 
			\\
			& $\mSigma_1^d$ & 0.0211 & 0.0530 &0.2740 & 0.5410 & 0.8640 & 0.9670 & 0.0168 & 0.0795 & 0.2795 & 0.6250 & 0.8900 & 0.9775 
			\\
			& $\mSigma_2^d$ & 0.0209& 0.0610 & 0.2880 & 0.5660 & 0.8620 & 0.9590 & 0.0160 & 0.0810 & 0.3025 & 0.6290 & 0.8930 & 0.9725 
			\\
			& $\mSigma_3^d$ & 0.0383& 0.0790 & 0.1930 & 0.3790 & 0.6080 & 0.7840 & 0.0276 & 0.0815 & 0.2025 & 0.3865 & 0.6450 & 0.8330 
			\\
			& $\mSigma_4^d$ & 0.0219& 0.0700 & 0.2700 & 0.5820 & 0.8830 & 0.9680 & 0.0166 & 0.0795 & 0.3190 & 0.6170 & 0.8890 & 0.9750 
			\\
			\noalign{\smallskip}\hline\noalign{\smallskip}
			$10^3$ & $I_d$ & 0.0257  & 0.0670 & 0.2740 & 0.5490 &0.8490 & 0.9690  & 0.0222 & 0.0835 & 0.3080 & 0.6125 & 0.8820 & 0.9670 
			\\
			& $\mSigma_1^d$ & 0.0276  & 0.0610 & 0.2620 & 0.5550 & 0.8780 & 0.9740  & 0.0212 & 0.0760 &0.2875 & 0.6265 & 0.8860 & 0.9690 
			\\
			& $\mSigma_2^d$ & 0.0272 & 0.0800 & 0.2950 & 0.5840 & 0.8600 & 0.9740   & 0.0223 & 0.0775 & 0.3105 & 0.6215 & 0.8885 & 0.9730 
			\\
			& $\mSigma_3^d$ & 0.0500  & 0.0930 & 0.2220 & 0.3580 & 0.6310 & 0.8210  &  0.0351 & 0.0845 & 0.2225 & 0.3850 & 0.6370 & 0.8160 
			\\
			& $\mSigma_4^d$ & 0.0265 & 0.0710 & 0.2490 & 0.5800 & 0.8520 & 0.9610  & 0.0217 & 0.0810 & 0.2955 & 0.6260 & 0.8800 &0.9705 
			\\
			\hline\noalign{\smallskip}
		\end{tabular}
	\end{center}
\end{table}

%

%%------------------------------------------------------------------------------------------------------------------------------

%----------------------------------------------------%
\subsubsection{Comparison with other methods}
\label{subsubsec:Comparison_MDP_PCOut}
%----------------------------------------------------%

In this subsection we present Tables \ref{tab:Comparison.Other.Procedures.App.1} and \ref{tab:Comparison.Other.Procedures.App.2} which show the results obtained with the simulations described in Subsection \ref{sec:comparison} when applied to  the families of matrices $\Sigma_i^d, i=1,\ldots, 4$ introduced  in Subsection \ref{subsec:a_b_Other}.

\begin{table}
	\caption{Proportion of outliers found in a clean data set for several covariance matrices. }
	\label{tab:Comparison.Other.Procedures.App.1}
	\begin{center}
		\begin{tabular}{r@{\hskip .12cm}r@{\hskip .12cm}|c@{\hskip .12cm}c@{\hskip .12cm}c@{\hskip .12cm}c@{\hskip .12cm}|c@{\hskip .12cm}c@{\hskip .12cm}c@{\hskip .12cm}c@{\hskip .12cm}|c@{\hskip .12cm}c@{\hskip .12cm}c@{\hskip .12cm}c@{\hskip .12cm}}
			\hline\noalign{\smallskip}
			& &\multicolumn{4}{c}{MDP}&\multicolumn{4}{|c}{PCOut} & \multicolumn{4}{|c}{RP}
			\\
			\hline\noalign{\smallskip}
			$n$& $d$ & $\mSigma_1^d$ & $\mSigma_2^d$ & $\mSigma_3^d$ & $\mSigma_4^d$ & $\mSigma_1^d$ & $\mSigma_2^d$ & $\mSigma_3^d$ & $\mSigma_4^d$ & $\mSigma_1^d$ & $\mSigma_2^d$ & $\mSigma_3^d$ & $\mSigma_4^d$
			\\
			\noalign{\smallskip}\hline\noalign{\smallskip}
			$50$ & $50$ & .2460 & .1372 & .2509 & .1348 & .1118 & .1026 & .1082 & .1043 & .1077 & .1160 & .1326 & .1095
			\\
			& $500$& .0884 & .0579 & .3748 & .1291 & .0963 & .0974 & .0964 & .0946 & .1104 & .1104 & .1368 & .1099
			\\
			& $1000$& --- & --- & --- & --- &  .0978 & .1043 & .0977 & .0970 &  .1144 & .1145 &.1462 & .1109
			\\
			\noalign{\smallskip}\hline\noalign{\smallskip}
			$100$ & $50$& .2209 & .0758 & .0797 & .0746 & .1111 & .1013 & .1056 & .1029 & .1019 & .1046 &.1088 & .1027
			\\
			& $500$& .0702 & .0552 & .2310  & .0833 &  .0803 & .0833 & .0794 & .0788 & .1069 & .1106 & .1228 & .1090
			\\
			& $1000$& --- & --- & --- & --- & .0813 & .0809 & .0806 & .0784 & .1128 & .1121 & .1249 & .1116
			\\
			\hline\noalign{\smallskip}
		\end{tabular}
	\end{center}
\end{table}

\begin{table}
	\caption{Samples contain 10\% of real outliers. Columns show the proportion of them correctly identified. }
	\label{tab:Comparison.Other.Procedures.App.2}
	\begin{center}
		\begin{tabular}{r@{\hskip .12cm}r@{\hskip .12cm}|c@{\hskip .12cm}c@{\hskip .12cm}c@{\hskip .12cm}c@{\hskip .12cm}|c@{\hskip .12cm}c@{\hskip .12cm}c@{\hskip .12cm}c@{\hskip .12cm}|c@{\hskip .12cm}c@{\hskip .12cm}c@{\hskip .12cm}c@{\hskip .12cm}}
			\hline\noalign{\smallskip}
			& &\multicolumn{4}{c}{MDP}&\multicolumn{4}{|c}{PCOut} & \multicolumn{4}{|c}{RP}
			\\
			\hline\noalign{\smallskip}
			$n$& $d$ & $\mSigma_1^d$ & $\mSigma_2^d$ & $\mSigma_3^d$ & $\mSigma_4^d$ & $\mSigma_1^d$ & $\mSigma_2^d$ & $\mSigma_3^d$ & $\mSigma_4^d$ & $\mSigma_1^d$ & $\mSigma_2^d$ & $\mSigma_3^d$ & $\mSigma_4^d$
			\\
			\noalign{\smallskip}\hline\noalign{\smallskip}
			$50$ & $50$ & .2545 & .1865 & .2886 & .1942 & .2064 & .1636 & .1724 & .1596 & .2736 & .2844 & .2412 & .2868
			\\
			& $500$& .0933 & .0803 & .1826 & .1859 & .1196 &.1224 & .1104 & .1352 & .1636 & .1680 & .1776 & .1604
			\\
			& $1000$& --- & --- & --- & --- &  .1136 & .1312 &.1184 & .1248 & .1440 & .1452 & .1612 & .1592
			\\
			\noalign{\smallskip}\hline\noalign{\smallskip}
			$100$ & $50$&.2241& .1282 &.2581 &.1320 & .2736 & .2360 & .2482 & .2376 & .2812 & .3020 & .2310 &.2986
			\\
			& $500$& .0747 & .0935 & .3330 & .1419 &.0996 & .0964 & .0990 & .0952 & .1636 & .1638 & .1584 & .1760
			\\ 
			& $1000$& --- & --- & --- & --- & .0874 &.0982 & .0892 & .0972 & .1548 & .1504 & .1598 & .1476
			\\
			\hline\noalign{\smallskip}
		\end{tabular}
	\end{center}
\end{table}


\begin{thebibliography}{}
	\bibitem{Aggarwal2015}
	Aggarwal, C.C.: Outlier {A}nalysis.
	\newblock Springer (2017)
	
	\bibitem{Anderson1955}
	Anderson, T.W.: The integral of a symmetric unimodal function over a symmetric
	convex set and some probability inequalities.
	\newblock Proc. Amer. Math. Soc. \textbf{6}(2), 170--176 (1955)
	
	\bibitem{Barnett1994}
	Barnett, V., Lewis, T.: Outliers in Statistical Data.
	\newblock Wiley Series in Probability and Statistics. Wiley (1994)
	
	\bibitem{Becker1999}
	Becker, C., Gather, U.: The masking breakdown point of multivariate outlier
	identification rules.
	\newblock J. Amer. Statist. Assoc. \textbf{94}(447), 947--955 (1999)
	
	\bibitem{Cardot2007}
	Cardot, H., Mas, A., Sarda, P.: Clt in functional linear regression models.
	\newblock Probab. Theory and Related Fields \textbf{138}(3-4), 325--361 (2007)
	
	\bibitem{Cerioli}
	Cerioli, A.: Multivariate outlier detection with high-breakdown estimators.
	\newblock J. Amer. Statist. Assoc. \textbf{105}(489), 147--156 (2010)
	
	\bibitem{Cerioli2009}
	Cerioli, A., Riani, M., Atkinson, A.C.: Controlling the size of multivariate
	outlier tests with the mcd estimator of scatter.
	\newblock Stat. Comput. \textbf{19}(3), 341--353 (2009)
	
	\bibitem{Chang}
	Chang, S., Cosman, P.C., Milstein, L.B.: Chernoff-type bounds for the gaussian
	error function.
	\newblock IEEE Trans. Commun. \textbf{59}(11), 2939--2944 (2011)
	
	\bibitem{Csorgo1983}
	Csorgo, M.: Quantile processes with statistical applications, vol.~42.
	\newblock Siam (1983)
	
	\bibitem{Cuesta2007}
	Cuesta-Albertos, J., del Barrio, E., Fraiman, R., Matr{\'a}n, C.: The random
	projection method in goodness of fit for functional data.
	\newblock Comput. Statist. Data Anal. \textbf{51}(10), 4814 -- 4831 (2007)
	
	\bibitem{Cuesta2006}
	Cuesta-Albertos, J., Fraiman, R., Ransford, T.: Random projections and
	goodness-of-fit tests in infinite-dimensional spaces.
	\newblock Bull. Braz. Math. Soc. \textbf{37}(4), 477--501 (2006)
	
	\bibitem{Cuesta2008}
	Cuesta-Albertos, J., Nieto-Reyes, A.: The random {T}ukey depth.
	\newblock Comput. Statist. Data Anal. \textbf{52}(11), 4979 -- 4988 (2008)
	
	\bibitem{Cuesta2010}
	Cuesta-Albertos, J.A., Febrero-Bande, M.: A simple multiway anova for
	functional data.
	\newblock TEST \textbf{19}(3), 537--557 (2010)
	
	\bibitem{Cuesta}
	Cuesta-Albertos, J.A., Fraiman, R., Ransford, T.: A sharp form of the
	{C}ram{\'e}r--{W}old theorem.
	\newblock J. Theor. Probab. \textbf{20}(2), 201--209 (2007)
	
	\bibitem{Cuesta2014}
	Cuesta-Albertos, J.A., Gamboa, F., Nieto-Reyes, A.: A random-projection based
	procedure to test if a stationary process is gaussian.
	\newblock Comput. Statist. Data Anal. \textbf{75}, 124--141 (2014)
	
	\bibitem{Cuesta2018}
	Cuesta-Albertos, J.A., Garc\'ia-Portugu\'es, E., Febrero-Bande, M.,
	Gonz\'alez-Manteiga, W.: Goodness-of-fit tests for the functional linear
	model based on randomly projected empirical processes.
	\newblock Ann. Stat. \textbf{47}(1), 439--467 (2019)
	
	\bibitem{Davies}
	Davies, L., Gather, U.: The identification of multiple outliers.
	\newblock J. Amer. Statist. Assoc. \textbf{88}(423), 782--792 (1993)
	
	\bibitem{Donoho}
	Donoho, D.L.: Breakdown properties of multivariate location estimators.
	\newblock Ph.D. qualifying paper  (1982)
	
	\bibitem{Esbensen}
	Esbensen, K., Guyot, D., Westad, F., Houmoller, L.: Multivariate Data Analysis
	in Practice : an Introduction to Multivariate Data Analysis and Experimental
	Design.
	\newblock CAMO (2002)
	
	\bibitem{Fang}
	Fang, K., Zhang, Y.: Generalized Multivariate Analysis.
	\newblock Berlin ; New York : Springer-Verlag (1990)
	
	\bibitem{Febrero2007}
	Febrero, M., Galeano, P., Gonz{\'a}lez-Manteiga, W.: A functional analysis of
	{NO}$_x$ levels: location and scale estimation and outlier detection.
	\newblock Comput. Statist. \textbf{22}(3), 411--427 (2007)
	
	\bibitem{Filzmoser2008}
	Filzmoser, P., Maronna, R., Werner, M.: Outlier identification in high
	dimensions.
	\newblock Comput. Statist. Data Anal. \textbf{52}(3), 1694--1711 (2008)
	
	\bibitem{Healy1968}
	Healy, M.J.R.: Multivariate normal plotting.
	\newblock J. R. Stat. Soc. Ser. C. Appl. Stat. \textbf{17}(2), 157--161 (1968)
	
	\bibitem{Huetal89}
	Hu, T., M\'oricz, F., Taylor, R.: Strong laws of large numbers for arrays of
	rowise independent random variables.
	\newblock Acta Math. Hung. \textbf{54}, 153--162 (1989)
	
	\bibitem{Hubert2015}
	Hubert, M., Rousseeuw, P.J., Segaert, P.: Multivariate functional outlier
	detection.
	\newblock Stat. Methods Appl. \textbf{24}(2), 177--202 (2015)
	
	\bibitem{Johnson}
	Johnson, W., Lindenstrauss, J.: Extensions of {L}ipschitz maps into a {H}ilbert
	space.
	\newblock Contemporary Mathematics \textbf{26}, 189--206 (1984)
	
	\bibitem{Johnstone2009}
	Johnstone, I.M., Lu, A.Y.: On consistency and sparsity for principal components
	analysis in high dimensions.
	\newblock . Amer. Statist. Assoc. \textbf{104}(486), 682--693 (2009)
	
	\bibitem{Larsen}
	Larsen, F.H., van~den Berg, F., Engelsen, S.B.: An exploratory chemometric
	study of $^1${H} {NMR} spectra of table wines.
	\newblock J. Chemom. \textbf{20}(5), 198--208 (2006)
	
	\bibitem{Laurent}
	Laurent, B., Massart, P.: Adaptive estimation of a quadratic functional by
	model selection.
	\newblock Ann. Math. Statist. \textbf{28}(5), 1302--1338 (2000)
	
	\bibitem{Maronna2019}
	Maronna, R.A., Martin, R.D., Yohai, V.J., Salibi{\'a}n-Barrera, M.: Robust
	statistics: theory and methods (with R).
	\newblock John Wiley \& Sons (2019)
	
	\bibitem{Maronna1995}
	Maronna, R.A., Yohai, V.J.: The behavior of the {S}tahel-{D}onoho robust
	multivariate estimator.
	\newblock J. Amer. Statist. Assoc. \textbf{90}(429), 330--341 (1995)
	
	\bibitem{Pan}
	Pan, J., Fung, W., Fang, K.: Multiple outlier detection in multivariate data
	using projection pursuit techniques.
	\newblock J. Statist. Plann. Inference \textbf{83}(1), 153 -- 167 (2000)
	
	\bibitem{Pena}
	Peña, D., Prieto, F.J.: Multivariate outlier detection and robust covariance
	matrix estimation.
	\newblock Technometrics \textbf{43}(3), 286--310 (2001)
	
	\bibitem{Ro2015}
	Ro, K., Zou, C., Wang, Z., Yin, G.: Outlier detection for high-dimensional
	data.
	\newblock Biometrika \textbf{102}(3), 589--599 (2015)
	
	\bibitem{Rousseeuw_2006}
	Rousseeuw, P.J., Debruyne, M., Engelen, S., Hubert, M.: Robustness and outlier
	detection in chemometrics.
	\newblock Crit. Rev. Anal. Chem. \textbf{36}(3--4), 221--242 (2006)
	
	\bibitem{Serfling}
	Serfling, R., Mazumder, S.: Computationally easy outlier detection via
	projection pursuit with finitely many directions.
	\newblock J. Nonparametr. Stat. \textbf{25}(2), 447--461 (2013)
	
	\bibitem{Stahel}
	Stahel, W.A.: Breakdown of covariance estimators.
	\newblock Fachgruppe fur Statistik  (1981)
	
	\bibitem{Tartakovsky2014}
	Tartakovsky, A., Nikiforov, I., Basseville, M.: Sequential analysis: Hypothesis
	testing and changepoint detection.
	\newblock Chapman and Hall/CRC (2014)
	
\end{thebibliography}
\end{document}